\documentclass[twocolumn]{emulateapj}

\usepackage[usenames,dvipsnames,svgnames,table]{xcolor}
\usepackage{graphicx,color,soul}
\usepackage{latexsym}
\usepackage{amsmath,amssymb}      
\usepackage[draft=false]{hyperref}
\hypersetup{
     colorlinks   = true,
     citecolor    = blue
}

\usepackage{amsmath}

\def\aap{A\&A}

\begin{document}

\title{MAGNUS: A New Resistive MHD Code with Heat Flow Terms}

\author{Anamar\'ia Navarro, F. D. Lora-Clavijo and Guillermo A. Gonz\'alez}
\affiliation{Grupo de Investigaci\'on en Relatividad y Gravitaci\'on,
Escuela de F\'isica, Universidad Industrial de Santander,
\\A. A. 678, Bucaramanga 680002, Colombia.}

\email{AN: ana.navarro1@correo.uis.edu.co ; \\ FDLC: fadulora@uis.edu.co; \\ GAG: guillego@uis.edu.co }

\date{\today}

\begin{abstract}

We present a new magnetohydrodynamic (MHD) code for the simulation of wave propagation in  the solar atmosphere, under the effects of electrical resistivity -but not dominant- and heat transference in a uniform 3D grid.  The code is based on the finite volume method combined with the HLLE and HLLC approximate Riemann solvers, which use different slope limiters like  MINMOD, MC, and  WENO5. In order to control the growth of the divergence of the magnetic field, due to numerical errors, we apply the Flux Constrained Transport method, which is described in detail to understand how the resistive terms are included in the algorithm. In our results, it is verified that this method preserves the divergence of the magnetic fields within the machine round-off error ($\sim 1\times10^{-12}$). For the validation of the accuracy and efficiency of the schemes implemented in the code, we present some numerical tests in 1D and 2D for the ideal MHD. Later, we show one test for the resistivity in a magnetic reconnection process and one for the thermal conduction, where the temperature is advected by the magnetic field lines. Moreover, we display two numerical problems associated with the MHD wave propagation. The first one corresponds to a 3D evolution of a vertical velocity pulse at the photosphere-transition-corona region, while the second one consists in a 2D simulation  of a transverse velocity pulse in a coronal loop.

\end{abstract}


\maketitle

\section{Introduction}

The theory of magnetohydrodynamics -the study of interactions between magnetic fields and conductive fluids in low frequencies- is of great importance to understand the dynamics of the plasma in the solar atmosphere \citep{Priest1991}. Since the plasma in this region is highly influenced by strong magnetic fields, some of the oscillatory phenomena observed  in its magnetic structures can be modeled  as magnetohydrodynamic waves.  For instance, some oscillations detected by TRACE in a coronal loop could be interpreted as a stationary kink mode  \citep{NakariakovOfmanDelucaRobertsDavila1999}. Moreover, in  \cite{TomczykMcIntoshKeilJudgeSchadSeeleyEdmondson2007} was proposed that Alfv\'en waves detected by CoMP (Coronal Multi-Channel Polarimeter) were a possible mechanism to heat the corona. However, \cite{DoorsselaereNakariakovVerwichte2008} discussed that these waves were better explained as kink magnetoacoustic modes. 

On the other hand, according to some data obtained from the Atmospheric Imaging Assembly, on board the Solar Dynamics Observatory,  \cite{MortonVerthMcLaughlinErdelyi2012} observed a transversal kink mode propagating in a UV/EUV solar jet. In a similar way,  \cite{PontieuEtAl2007} pointed out that some Alfv\'en waves detected by the Solar Optical Telescope, on board the Japanese Hinode satellite, could have the enough energy to accelerate solar winds. Moreover, using the data taken with this telescope, magnetohydrodynamic waves propagating along magnetic flux tubes in the solar photosphere were detected \citep{FujimuraTsuneta2009}. In addition, using the Swedish Solar Telescope (SST) and the Solar Optical Universal Polarimeter (SOUP), torsional Alfv\'en waves were detected in a bright-point group near the solar disk center \citep{JessMathioudakisErdelyi2009}. Also, from images of the Rapid Oscillations in the Solar Atmosphere instrument at the Dunn Solar Telescope, it was reported how oscillations in the pressure, in small-scale photospheric magnetic bright-points, are able to generate kink modes in the outer solar atmosphere  \citep{JessPascoeChristian2012}. Complete reviews of observations of magnetohydrodynamic waves in  solar regions like the corona, sunspots, prominences, coronal mass ejections, solar flares and solar winds can be found in  \cite{NakariakovVerwichte2005, Khomenko2015, Okamoto2007, Vrsnac2013, ShibataMagara2011, Ofman2010}, respectively.

Through numerical MHD simulations, it is possible to process and analyze the data taken by the solar missions. This is one of the reasons for which a big number of MHD codes have been developed. Many of them are very robust, with adaptive mesh refinement methods (AMR), several Riemann solvers, slope limiters and integrators. For instance: Athena \citep{athena_code}, a grid-based code for astrophysical MHD, developed primarily for studies of the interstellar medium, star formation, and accretion flows; Flash \citep{flash_code}, a high performance application code, which has evolved into a modular, extensible software system from a collection of unconnected legacy codes for diverse applications in hydrodynamics, MHD, radiation transfer, diffusion and conduction; Pluto \citep{pluto_code}, a code for the solution of hypersonic astrophysical flows; Zeus \citep{zeus_code}, a numerical code for the simulation of fluid dynamical flows in astrophysics, including a self-consistent treatment of the effects of magnetic fields and radiation transfer;  Enzo \citep{enzo_code}, a code based on block-structured AMR for modeling astrophysical fluid flows;  Nirvana \citep{nirvana_code}, a parallel computational MHD code with AMR for astrophysical problems; MAP \citep{ResistiveCode_Jiang}, a code with AMR and parallelization for astrophysics. Moreover, some codes have different numerical schemes like the one developed by \cite{MHDcodeLagrangianEulerian}, which is based on control volume averaging with a staggered grid, instead  of using approximate Riemann solvers.

 Furthermore, there are codes focused only on solar atmosphere phenomena, for instance: VAC \citep{vac_code}, a code for the simulation of the interaction of an arbitrary perturbation with a magnetohydrostatic equilibrium background, designed to reproduce physical processes in stable gravitationally-stratified plasma; Surya \citep{surya_code}, a high-resolution, well-balanced finite-volume-based massively parallel code, to study the propagation of waves in a stratified non-isothermal magnetic atmosphere; MURaM \citep{muram_code}, a 3D magnetohydrodynamics code, for applications in the solar convection zone and photosphere; LFM \citep{LFM_code}, a global magnetospheric simulation code; Bifrost \citep{bifrost_code}, a massively parallel numerical code designed for the research of the stellar atmospheres from the convection zone to the corona; CAFE \citep{CafeNewtoniano, CafeRelativista}, a 3D MHD code for the analysis of solar phenomena within the photosphere-corona region.

Multiple MHD scenarios of solar physics interest have been evolved with the codes described before, for instance,  \cite{Murawski_Musielak2010} have studied the small amplitude-Alfv\'en waves in the solar atmosphere, finding their spatial and temporal signatures in the linear regime.  \cite{KonkolMurawskiLeeWeide2010} evolved a 2D potential arcade in a gravitationally stratified corona, in order to explore the influence of thermal conduction on the attenuation of the fundamental stationary slow magnetoacoustic modes. Moreover,  \cite{Murawski_Srivastava_Zaqarashvili2011} simulated the formation of macrospicules created by a localized pulse below the transition region. On the other hand, the microflares in the corona and the chromosphere were modeled by \cite{JiangFangChen2012}, changing the bottom boundary conditions. \cite{Shelyag2011} tested the generation of small-scale vortex motions in the solar photosphere. \cite{JessShelyagMathioudakis2012}  simulated some oscillatory phenomena in the photosphere as periodic intensity fluctuations with magnetoconvection processes. In \cite{VigeeshFedun2012} a MHD wave propagation in the solar magnetic flux tubes was carried out by evolving a strong, axially symmetric magnetic flux tube in a stratified solar model atmosphere, whose lower boundary is located at photospheric levels. Later, \cite{Wedemeyer2012} reported the imprints of convectively driven vortex flows in the photosphere and the corona; \cite{Shelyag2013} also analyzed plasma motions in photospheric magnetic vortices;  \cite{JelinekMurawski2013} reproduced magnetoacoustic-gravity waves, at the gravitationally stratified corona, by using a realistic temperature profile and curved magnetic field lines. Furthermore, Kelvin-Helmholtz instabilities at the coronal mass ejection boundaries in the solar corona were found by \cite{MostlTemmerVeronig2013}.

Many other numerical studies in this direction have been developed. In \cite{BaumannGalsgaardNordlund2013} some reconnection events in the corona have been reproduced in 3D. Impulsively generated fast magneto-acoustic perturbations were presented by \cite{PascoeNakariakov2013} using 2D evolutions. Moreover, in \cite{gravity_waves} 3D simulations of magnetohydrodynamic-gravity waves and the production of vortices in a magnetized solar atmosphere were tested. Later on, \cite{Wedemeyer2013} analyzed the propagating shock waves in 3D radiation simulations of M-type dwarf stars. Impulsively generated Alfv\'en waves have been also reproduced in an isolated solar arcade, which is gravitationally stratified and magnetically confined \citep{Chmielewski_Murawski_Musielak_Srivastava2014}.  Recently, 
solar prominences embedded in sheared magnetic arcades were evolved by \cite{TerradasSolerLunaOliver2015}. Besides, the coronal heating problem was also discussed in more complex scenarios, including radiative MHD terms \citep{HansteenGuerreiro2015}. In addition, transverse MHD waves in a prominence flux tube were taken into account by \cite{AntolinOkamoto2015}. Also, in \cite{YangZhangHePeter2015}, the excitation of fast-propagating magnetosonic waves along coronal magnetic funnels was studied. In more recent works,  3D MHD models of solar flares were carried out by \cite{JanvierAulanierDemoulin2015}. \cite{MurawskiChmielewski2016} have also analyzed the response of a solar small-scale and weak magnetic flux tube with photospheric twisting motions. Finally,  the footpoint-driven transverse waves propagating in a coronal plasma with a cylindrical density structure were considered by \cite{DeMoortelPascoeWright2016}. 

Motivated by the works mentioned above, with the code described in this paper, our goal is to perform numerical simulations of the propagation of magnetohydrodynamic waves in the solar atmosphere, in scenarios with electrical resistivity and heat flow, although the amplitude of those terms will not be dominant. Specifically, this code solves the resistive MHD equations with heat flow terms and with a constant gravity in 3D. It uses the HLLE \citep{Harten_etal_1983} and HLLC \citep{Toro_hllc} Riemann solvers.  The HLLC method is implemented following the adaptation proposed by \cite{hllc_mhd_Li}.  The slope limiters are second-order MINMOD \citep{Roe1986}, MC \citep{mc} and fifth-order WENO schemes \citep{TitarevToro2004}. In addition, in regard of preserving the  magnetic field divergence-free, $\nabla\cdot\vec{B} = 0$, we follow the Flux Constrained Transport method \citep{CT_Balsara, CT_evans, CT_toth}, which was slightly modified in order to deal with the resistivity terms, and thus keep the divergence of the magnetic field to the order of the machine round-off error. 

This paper is organized as follows: in section \ref{sec:equations}, we describe the equations to be solved numerically, that is, the magnetohydrodynamic equations with resistivity, heat flow, and constant gravity. Then, these equations are written in a conservative form, which is ideal for the numerical used here. In section \ref{sec: Numerical Methods}, we present the numerical methods to solve these equations. Since the purpose of the code is to study wave phenomena in the solar atmosphere  with resistive and heat flow terms, in section \ref{sec:tests},  we show the numerical tests to validate the accuracy and the efficiency of the algorithms.  Finally, in section \ref{sec:conclusions}, we discuss the results and present our main conclusions.

\section{Resistive Magnetohydrodynamic Equations with Heat Flux}
\label{sec:equations}

The MHD equations for a resistive medium with heat flux and a constant gravitational field in the negative direction of the z-axis are the continuity of mass, the  Newton's  second law for a fluid element, the conservation of energy and the Faraday's law, given respectively by 
\begin{eqnarray}
& & \partial_t \rho + \partial_i  \left( \rho v^i  \right) = 0  \label{eqMHD1} \, , \\
& & \partial_t \left( \rho v^i \right) + \partial_j \left( \rho v^i v^j - \frac{B^i B^j}{\mu_0}  + p_T \delta^{i j}   \right) = -\rho g \delta^{i z} \label{eqMHD2} \, , \\
& & \partial_t E + \partial_j \left[ \left( E + p_T \right)v^j - \frac{B^j (\vec{B}\cdot \vec{v})}{\mu_0}  \right] \label{eqMHD3}  \\ 
& & \hspace{2cm} =  \partial_j\left\lbrace q^j - \left[ \frac{\eta}{\mu_0}  \vec{J}  \times \vec{B} \right]^j    \right\rbrace - \rho v^z g   \, , \nonumber \\ 
& & \partial_t B^i + \partial_j\left( v^i B^j - B^i v^j \right) = - \left[ \nabla \times  \eta \vec{J} \right]^i \label{eqMHD4} \, , 
\end{eqnarray}
with the Gauss's law for the magnetic field 
\begin{eqnarray}
\nabla \cdot \vec{B} = 0 \, , \label{divergencia}
\end{eqnarray}
where $\rho$ is the mass density of the fluid, $p$ the pressure,  $\vec{v}$ the velocity vector field, $\vec{B}$ the magnetic field, $\vec{J}$ the current density vector, $\eta$ the electric resistivity, $\vec{q}$ the heat flux vector, $g$ the acceleration due to gravity, $\mu_0$ the permeability of free space and $E$ the total energy density
\begin{eqnarray}
E = \frac{\rho v^2}{2} + \rho e + \frac{B^2}{2\mu_0} \, ,
\end{eqnarray}
which is expressed as the sum of the kinetic, internal ($\rho e$) and magnetic energy densities. In the equation of energy \eqref{eqMHD3}, it has been assumed that the fluid obeys an equation of state for an ideal gas, that is,
\begin{eqnarray}
p = (\Gamma - 1)\rho e \, ,
\end{eqnarray}
where $\Gamma$ is the adiabatic index. 

Furthermore, the heat flow vector $\vec{q}$ for the thermal conduction is modeled in two ways, the first one corresponds to the isotropic dependence with the temperature, given by  
\begin{eqnarray}
\vec{q} = \kappa\nabla T \, , \label{heat}
\end{eqnarray}
 where $\kappa$ is the thermal conductivity, and $T$  is the temperature. The second model for the heat flux vector is the classical model for a  magnetized plasma \citep{Spitzer2006}
\begin{eqnarray}
\vec{q} = \kappa T^{5/2}\left(\vec{B}\cdot\nabla T \right) \vec{B}/B^2 \, , \label{heat2}
\end{eqnarray}
where the thermal conduction is transferred only along the magnetic field lines. This approximation is suitable for solar applications, since the conduction in the perpendicular direction is $2 \times 10^{-31} n^2 T^{-3} B^{-2}$ (in SI units and $n$  the total number of particles per unit volume) times smaller \citep{Priest2014} due to the strong magnetic fields. It is important to notice that with the inclusion of the heat flux vector, the equations have now a diffusive term, which makes more difficult the numerical calculations, so in regards to the stability of the solution, we choose a time step proportional to the square of the spatial resolutions.

The system of equations \eqref{eqMHD1}-\eqref{eqMHD4} can be  written in the compact form
\begin{eqnarray}
\partial_t \vec{U} + \partial_i \vec{F^i} = \vec{S} \, , \label{MHD_eq_ConservativeForm}
\end{eqnarray}
where $\vec{U}$ is a state vector of conservative variables defined as
\begin{eqnarray}
\vec{U} &=& \left[
\rho ,
\rho \vec{v},
 E , 
\vec{B}
  \right] ^T \, ,
\end{eqnarray}
$\vec{F^i}$ are the fluxes along each axis
\begin{eqnarray}
\vec{F^i} &=& \left[ \begin{array}{c} \rho v^i \\  
 \rho v^i v_j - \frac{B^i B_j}{\mu_0} + p_T\delta^i_j \\
( E + p_T)v^i - \frac{B^i({\bf B}\cdot {\bf v})}{\mu_0} \\
v^i B_k - v_k B^i   \end{array}  \right] \, , \label{flujos}
\end{eqnarray}
and $\vec{S}$ is a source vector
\begin{eqnarray} 
\vec{S} = \left[ \begin{array}{c}
0 \\
- \rho \vec{g} \\ 
-\rho \vec{v}\cdot \vec{g} - \nabla \cdot \left( \vec{q} +  \frac{\eta}{\mu_0} \vec{J} \times \vec{B} \right) \\
-\nabla \times \eta \vec{J}
 \end{array} \right] .  \label{S_vector}
\end{eqnarray}
With the purpose to achieve an optimal performance in the numerical processes of the code, the equations are adimensionalized with the following conversions
\begin{eqnarray}
\arraycolsep=0.1cm\def\arraystretch{2.0}
\begin{array}{llll}
\partial_i \rightarrow \frac{1}{l_0}\partial_i \, , &
\partial_t \rightarrow \frac{1}{t_0} \partial_t \, ,  &
l \rightarrow \frac{l}{l_0} \, , &
t \rightarrow \frac{t}{t_0} \, , \\  
\rho \rightarrow \frac{\rho}{\rho_0} \, , & 
\eta \rightarrow \frac{\eta}{l_0 v_0 \mu_0}  \, , &
\vec{q} \rightarrow \frac{\vec{q}}{\rho_0 v_0^3} \, , &
\vec{B} \rightarrow \frac{\vec{B}}{B_0} \, , \\
\vec{v} \rightarrow \frac{\vec{v}}{v_0} \, , &
p \rightarrow \frac{p}{\rho v_0^2} \, , & 
p_T \rightarrow \frac{p_T}{\rho v_0^2} \, , & 
E \rightarrow \frac{E}{\rho v_0^2} \, , \label{adim_relations}
\end{array}
\end{eqnarray}
where $v_0 = l_0/t_0 $ and  $B_0 = v_0^2 \mu_0 \rho_0 $. With \eqref{adim_relations}, the system of equations \eqref{MHD_eq_ConservativeForm} can be re-written in the same form except for the term of the magnetic permeability $\mu_0$, which disappears.

\section{Numerical methods}\label{sec: Numerical Methods}

This code solves the time-dependent system of partial differential equations  \eqref{MHD_eq_ConservativeForm} in an  uniform grid.  The MHD equations are solved in time by using the method of lines, which is equipped with different time integrators. Specifically, it has implemented second and third order total variation diminishing Runge-Kutta, fourth order regular Runge-Kutta and iterative Cranck-Nicholson time integrators. These equations are then discretized following the finite volume approximation
\begin{eqnarray}
& & \frac{\mathrm{d} \vec{U}_{(i,j,k)} }{\mathrm{d}t} = 
 - \frac{ \vec{F}^x_{(i+1/2,j,k)}  - \vec{F}^x_{(i-1/2,j,k)}}{\Delta x}  \label{finite_volumes}  \\
& & - \frac{ \vec{F}^y_{(i,j+1/2,k)}  - \vec{F}^y_{(i,j+1/2,k)}}{\Delta y} 
 - \frac{ \vec{F}^z_{(i,j,k+1/2)}  - \vec{F}^z_{(i,j,k-1/2)}}{\Delta z} \nonumber \\
& &   + \vec{S}_{(i,j,k)} \nonumber \, , 
\end{eqnarray}
where $\vec{F^x}_{(i\pm1/2,j,k)}$, $\vec{F^y}_{(i,j\pm1/2,k)}$ and $\vec{F^z}_{(i,j,k\pm1/2)}$ are the numerical fluxes at the interfaces of a cell and will be calculated using the High Resolution Shock Capturing methods (HRSC). Specifically they are HLLE,  HLLC and ROE (only for hydrodynamics) schemes which are coupled with the slope limiters of second-order MINMOD \citep{Roe1986}, MC \citep{mc} and the fifth-order weighted Essentially Non Oscillatory WENO5 \citep{TitarevToro2004}. 

In order to obtain stable evolutions and to calculate the convergence, the time step is chosen following the Courant-Friedrichs-Levy condition \citep{2005JCoPh.204..715T}
\begin{equation}
\Delta t = C_\text{cfl} \times \text{min}_{ijk}\left(\frac{\mathrm{\Delta}x}{| \lambda_{ijk}^{n,x} |}, \frac{\mathrm{\Delta}y}{| \lambda_{ijk}^{n,y} |}, \frac{\mathrm{\Delta}z}{| \lambda_{ijk}^{n,z} |}\right), 
\end{equation}
where $C_\text{cfl}$ stands for the Courant number  and $| \lambda_{ijk}^{n,d} |$ is the speed of the fastest wave present at time level $n$ traveling in the $d$ direction. It is worth mentioning that the Courant number is chosen depending on the space dimension, that is, $0<C_\text{cfl}\leq1/2$ for 2D and $0<C_\text{cfl}\leq1/3$ for 3D. Furthermore, in the cases where the resistivity and the heat flux are considered, the time step is chosen using the same formula but in this case the spatial resolutions are squared, since in those cases the equations have now a diffusive component.

On the other hand, the default four boundary conditions (BC) implemented in the code are outflow, periodic, reflecting and fixed in time. For instance, for the boundaries at the face $x = x_\text{min}$ and $x=x_\text{max}$, they are calculated as
\begin{eqnarray}
 \left. \begin{array}{c}
 \vec{U}_{(0,j,k)} = \vec{U}_{(1,j,k)} \\ 
 \vec{U}_{(N_x,j,k)} = \vec{U}_{(N_x-1,j,k)} 
\end{array} \right\rbrace \text{for \ outflow \ BC}, \\
 \left. \begin{array}{c}
\vec{U}_{(0,j,k)} = \vec{U}_{(N_x-1,j,k)} \\
\vec{U}_{(N_x,j,k)} = \vec{U}_{(1,j,k)}
\end{array} \right\rbrace \text{for \ periodic \ BC}, 
\end{eqnarray}
and for the reflecting we use outflow conditions in all the variables except for the velocity in the x-direction, which reflects as
\begin{eqnarray}
 \left. \begin{array}{c}
v^x_{(0,j,k)} = -v^x_{(1,j,k)} \\
v^x_{(N_x,j,k)} = -v^x_{(N_x-1,j,k)} 
\end{array} \right\rbrace \text{for \ reflecting \ BC}.
\end{eqnarray}
For the fixed in time conditions, we reset the values at the boundary with the initial values. 
Finally, we parallelized the code with OpenMP \citep{openmp} to improve the computational times of the calculations. 

\subsection{Finite differences for the current, resistive and heat flux terms}\label{subsec: Current terms}

For the numerical calculations, it is necessary to compute the terms related to the resistivity and heat flux with finite differences. The components of the current density vector  $\vec{J}$ are obtained from Amp\'ere's law 
\begin{eqnarray}
J_x &=& \partial_y B_z - \partial_z B_y \, , \\
J_y &=& \partial_z B_x - \partial_x B_z \, , \\
J_z &=& \partial_x B_y - \partial_y B_x \, ,
\end{eqnarray}
and the partial derivatives are computed by using simple second-order finite differences. It is worth mentioning that we have dropped the term $\mu_0$ since the equations implemented are dimensionless. The fifth component of the source vector  \eqref{S_vector} and the heat flux vector $\vec{q}$ from equation \eqref{heat} are calculated following the schemes presented in \cite{ResistiveCode_Jiang}, which are given by
\begin{eqnarray}
& & \nabla \cdot \left( \eta \vec{J}\times \vec{B} \right)_{(i,j,k)} = \\
 & & \frac{[\eta( J_z B_y - J_y B_z )] _{(i+1,j,k)}   - [\eta( J_z B_y - J_y B_z )]_{(i-1,j,k)}   }{2 \Delta x}   \nonumber \\ 
 & & + \frac{ [\eta( J_x B_z - J_z B_x )]_{(i,j+1,k)}  - [\eta( J_x B_z - J_z B_x )]_{(i,j-1,k)}   }{2 \Delta y}   \nonumber \\
 & & + \frac{ [\eta( J_y B_x - J_x B_y )]_{(i,j,k+1)}  - [\eta( J_y B_x - J_x B_y )]_{(i,j,k-1)}   }{2 \Delta z}  \, ,  \nonumber
 \end{eqnarray}
 \begin{eqnarray}
 \left( \nabla \cdot \vec{q} \right)_{(i,j,k)}  &=& \\ 
& & \frac{ \sqrt{\kappa_{(i,j,k)}\kappa_{(i+1,j,k)}}\left( T_{(i+1,j,k)} - T_{(i,j,k)}  \right) }{\Delta x^2} \nonumber \\ 
&  -& \frac{\sqrt{\kappa_{(i,j,k)}\kappa_{(i-1,j,k)}}\left(T_{(i,j,k)} - T_{(i-1,j,k)} \right) }{\Delta x^2}  \nonumber \\
 &  +& \frac{ \sqrt{\kappa_{(i,j,k)}\kappa_{(i,j+1,k)}}\left( T_{(i,j+1,k)} - T_{(i,j,k)}  \right) }{\Delta y^2} \nonumber \\
 &  -& \frac{\sqrt{\kappa_{(i,j,k)}\kappa_{(i,j-1,k)}}\left(T_{(i,j,k)} - T_{(i,j-1,k)} \right) }{\Delta y^2} \nonumber \\
 & +& \frac{ \sqrt{\kappa_{(i,j,k)}\kappa_{(i,j,k+1)}}\left( T_{(i,j,k+1)} - T_{(i,j,k)}  \right) }{\Delta z^2} \nonumber \\
&  -&  \frac{\sqrt{\kappa_{(i,j,k)}\kappa_{(i,j,k-1)}}\left(T_{(i,j,k)} - T_{(i,j,k-1)} \right) }{\Delta z^2} \, . \nonumber
\end{eqnarray}
For the second model of the  heat flux vector \eqref{heat2}, simple second-order centered finite differences are used. Finally, the three last components of the source vector \eqref{S_vector}, $\nabla\times\eta \vec{J}$, are not written in a finite differences scheme, since the equations \eqref{eqMHD4} will be solved through the Flux Constrained Transport technique to preserve the magnetic field divergence-free and will be discussed in section \ref{subsec: CT}.

\subsection{Approximate Riemann Solver}\label{subsec: Riemann Solvers}

Since the purpose of the code is to study scenarios with low resistivity and low heat transport, that is, small amplitudes of the resistivity and thermal conductivity, we have assumed that the eigenvalue structure of the system of equations \eqref{eqMHD1}-\eqref{eqMHD4} is not altered by the inclusion of the resistivity and heat flux terms. This means the eigenvalues are the same that those obtained in \cite{Powell} for the ideal MHD.

The HRSC methods compute the fluxes in equation \eqref{finite_volumes} by solving a local Riemann problem on both sides of every inter-cell  (with the slope limiters MINMOD, MC, etc...).  Let us summarize both methods, the  HLLE and HLLC for the 1D case (x-direction) since the 3D version can be easily extended by doing a cyclic permutation of the variables. 

The HLLE scheme \citep{Harten_etal_1983} is based on the eigenvalues $\lambda_i$ associated with the Jacobian matrix of the system of equations \eqref{MHD_eq_ConservativeForm}.  The first one is an entropy wave traveling with speed $\lambda_{1} = v_x $, the second and third ones are two Alfv\'en waves traveling with speeds $\lambda_{2,3} = v_x \pm c_a$, and the other four are magneto-acoustic waves, two slow 
$ \lambda_{4,5} = v_x \pm c_s$, and two fast $ \lambda_{6,7} = v_x \pm c_f$, where $c_a = \frac{B_x}{\sqrt{\rho}}$ is the Alfv\'en velocity, $c_f$ and $c_s$ are the fast and slow magnetosonic speeds, 
\begin{eqnarray}
& & c_{f,s} = \frac{1}{2\rho}\left( \Gamma p + \vec{B}\cdot\vec{B} 
\pm \sqrt{ \left(\Gamma p + \vec{B}\cdot\vec{B} \right)^2 - 4\Gamma p B_x^2  } \right) \nonumber \, . \\ 
\end{eqnarray} 
The solution is approximated by three constant states separated by two waves, one at the left, one intermediate  and one at the right side, given by
\begin{eqnarray}
& & \vec{U}_L \quad  \text{if} \quad S_L > 0 \, , \nonumber \\
 & & \vec{U}^\text{HLLE}= \frac{S_R \vec{U}_R - S_L \vec{U}_L + \vec{F}_L - \vec{F}_R}{S_R - S_L} \quad \text{if} \quad S_L \leq 0 \leq S_R \, , \nonumber \\ 
& &\vec{U}_R \quad \text{if} \quad  S_R < 0 \, , \nonumber 
\end{eqnarray}
where $\vec{F}_L = \vec{F}^x(\vec{U}_L)$ and $\vec{F}_R = \vec{F}^x(\vec{U}_R)$ are the fluxes evaluated in the left and right states respectively.  $S_L$ and $S_R$ are the velocities of the fastest waves moving to the left and to the right, that is, $S_L = \text{min}(0,\lambda_i^R, \lambda_i^L) $ and $S_R = \text{max}(0,\lambda_i^R, \lambda_i^L) $.
Finally, the fluxes for the intermediate state are given by
\begin{eqnarray}
\vec{F^x}^* = \frac{S_R \vec{F}^x_L - S_L \vec{F}^x_R - S_L S_R ( \vec{U}_R - \vec{U}_L )}{ S_R - S_L} \, .
\end{eqnarray}
 An important advantage of the HLLE method is that it uses only the eigenvalue information of the system of equations, so it is easier to implement compared to others. However, this scheme dissipates more than others, since it ignores the contact discontinuity of the characteristic waves.

 On the other hand, the HLLC scheme \citep{Toro_hllc} for hydrodynamics, is a modification of HLLE method, where the contact discontinuity is solved. This method estimates an intermediate contact wave with velocity $S^*$, so the solution is divided into the following substates:
\begin{eqnarray}
& & \vec{U}_L \quad \text{if} \quad 0<S_L \, ,  \nonumber \\
& &  \vec{U}^*_L \quad \text{if} \quad S_L\leq 0 \leq S^*  \, , \nonumber \\
& &  \vec{U}^*_R \quad \text{if}  \quad  S^*\leq 0 \leq S_R \, , \nonumber \\
 & &  \vec{U}^R \quad \text{if} \quad 0>S_R \, . \nonumber
\end{eqnarray}
Now, by applying the Rankine-Hugoniot conditions across $S_L$ and $S_R$, the fluxes can be written in terms of the state variables $\vec{U}_L^*$ and $\vec{U}_R^*$ as 
\begin{eqnarray}
\vec{F}^*_L = \vec{F}_L + S_L(\vec{U}^*_L - \vec{U}_L) \, ,& \ \text{if} \ \ S_L \leq 0 \leq S^* \, , \\
\vec{F}^*_R = \vec{F}_R + S_R(\vec{U}^*_R - \vec{U}_R) \, ,& \ \text{if} \ \ S^* \leq 0 \leq S_R \, .
\end{eqnarray}
where it has been assumed that the pressure and velocities remain unchanged across the contact discontinuity wave
\begin{eqnarray}
& & p^*_L = p^*_R = p^*  \, , \\
& & {v_x}_L^* = {v_x}_R^* = {v_x}^* = S^* \, .  
\end{eqnarray}

The extension of this method to the MHD case is straight forward,  but in the work of \cite{hllc_mhd_Li} it was pointed out that  the consistency condition (the integral form of the conservation laws \cite{Toro}) 
\begin{eqnarray}
& & \frac{S^*-S_l}{S_R-S_L}\vec{U}_L^* +  \frac{S_R-S^*}{S_R-S_L}\vec{U}_R^* \\ 
& & \hspace{3cm} = \frac{S_R \vec{U}_R - S_L\vec{U}_L -(\vec{F}_R - \vec{F}_L ) }{S_R - S_L} \nonumber \, , 
\end{eqnarray}
is not satisfied under the MHD extension.  However, \cite{hllc_mhd_Li} has shown that the preceding condition is satisfied with the following restriction for the magnetic fields
\begin{eqnarray}
 & &{B_{i}}^*_L = {B_{i}}^*_R = {B_{i}}^* = B_i^{\text{HLLE}} \, , \\
 & & (\vec{B}\cdot  \vec{v})^*_L = (\vec{B}\cdot \vec{v})^*_R = (\vec{B}\cdot  \vec{v})^* = \vec{B}^{\, \text{HLLE}}\cdot \vec{v}^{ \, \text{HLLE}} \, , \nonumber \\
\end{eqnarray}
in such a way that the intermediate state $\vec{U}_L^*$ is
\begin{eqnarray}
& & \rho^*_L = \rho_L \frac{S_L-v_x}{S_L-S^*} \, , \\
& & (\rho v_x)_L^* = \rho^*_L S^* \, , \\
& & (\rho v_y)_L^* = (\rho v_y)_L \frac{S_L-v_x}{S_L-S^*} - \frac{(B_x^*B_y^*-B_x B_y)}{S_L-S^*} \, , \\
& & (\rho v_z)_L^* = (\rho v_z)_L\frac{S_L-v_x}{S_L-S^*} - \frac{(B_x^*B_z^*-B_x B_z)}{S_L-S^*} \, , \\
& & E^*_L =  \frac{1}{S_L-S^*} \left\lbrace E(S_L-v_x)  + \right.    \\ 
& &  \hspace{1cm} \left.  (p^* S^* - p v_x) - \left[ B_x^*(\vec{B} \cdot \vec{v})^* - B_x(\vec{B}\cdot\vec{v})  \right] \right\rbrace \, .  \nonumber
\end{eqnarray}
where
\begin{eqnarray}
S^* &=&   \frac{\rho_R {v_x}_R(S_R-{v_x}_R) - \rho_L {v_x}_L(S_L-{v_x})_L }{\rho_R(S_R-{v_x}_R) - \rho_R(S_R-{v_x}_L) }      \\
& & \hspace{2cm} + \frac{p_L - p_R - {B_x}_L^2 + {B_x}_R^2  }{\rho_R(S_R-{v_x}_R) - \rho_R(S_R-{v_x}_L) } \nonumber \, , \\
p^* &=& \rho_L (S_L-v_x)(S^*-v_x) + p_L - B_x^2 + (B_x^*)^2  \, , 
\end{eqnarray}
and a similar solution for the right state vector $\vec{U}_R^*$ can be found.

\subsection{Divergence of magnetic field: Flux Constrained Transport}\label{subsec: CT}

When the MHD equations are solved numerically, the accumulation of errors makes the divergence of the magnetic field grow in time. Therefore, several techniques have been implemented to prevent this issue \citep{CT_toth}. In our code, the Flux Constrained Transport method \citep{CT_Balsara,CT_evans} is applied, which consists in a special discretization of Faraday's equation. 
Now, from Maxwell equations, Faraday's law can be re-written in terms of a vector $\vec{\Omega}$ 
\begin{eqnarray}
\partial_t \vec{B} = \nabla \times \vec{\Omega}  \, , \label{eq_Bt}
\end{eqnarray}
that according to the Ohm's law, can be expressed as
\begin{eqnarray}
\vec{\Omega} = \vec{v}\times \vec{B} -\eta \vec{J} \, . \label{eq_Omega}
\end{eqnarray}
Applying finite central differences in \eqref{eq_Bt}, the discretized evolution equations for the area-average components of the magnetic field on each face of the numerical cell, centered in $(i,j,k)$, are
\begin{eqnarray}
\frac{\mathrm{d}B^x_{(i+\frac{1}{2},j,k)}}{\mathrm{d}t}  &=&  
 \frac{ \Omega^z_{(i+\frac{1}{2},j+\frac{1}{2},k)}  - \Omega^z_{(i+\frac{1}{2},j-\frac{1}{2},k)} }{\Delta y} \label{eq_B_caras_x} \\   
& &  - \frac{\Omega^y_{(i+\frac{1}{2},j,k+\frac{1}{2})} - \Omega^y_{(i+\frac{1}{2},j,k-\frac{1}{2})} }{\Delta z} \, ,  \nonumber \end{eqnarray}
\begin{eqnarray}
 \frac{\mathrm{d}B^y_{(i,j+\frac{1}{2},k)}}{\mathrm{d}t} &=& 
 \frac{ \Omega^x_{(i,j+\frac{1}{2},k+\frac{1}{2})} - \Omega^x_{(i,j+\frac{1}{2},k-\frac{1}{2})} }{\Delta z}  \label{eq_B_caras_y} \\ 
 & &  - \frac{ \Omega^z_{(i+\frac{1}{2},j+\frac{1}{2},k)} - \Omega^z_{(i-\frac{1}{2},j+\frac{1}{2},k)} }{\Delta x}   \, , 
\nonumber 
\end{eqnarray}
\begin{eqnarray}
\frac{\mathrm{d}B^z_{(i,j,k+\frac{1}{2})}}{\mathrm{d}t} &=&  
\frac{ \Omega^y_{(i+\frac{1}{2},j,k+\frac{1}{2})} - \Omega^y_{(i-\frac{1}{2},j,k+\frac{1}{2})} }{\Delta x} \label{eq_B_caras_z} \\ 
& & -  \frac{ \Omega^x_{(i,j+\frac{1}{2},k+\frac{1}{2})} - \Omega^x_{(i,j-\frac{1}{2},k+\frac{1}{2})} }{\Delta y}  \, . \nonumber
 \end{eqnarray} 
where the values of $\Omega$ are defined at the cell vertex and computed, at the inter-cells,  as simple averages of the numerical fluxes $F^{ij} = v^i B^j - v^j B^i $ ,
\begin{eqnarray}
& & \Omega^x_{(i,j+\frac{1}{2},k+\frac{1}{2})} =  
 \frac{1}{4} \left( F^{yz}_{(i,j+\frac{1}{2},k)} + F^{yz}_{(i,j+\frac{1}{2},k+1)} - F^{zy}_{(i,j,k+\frac{1}{2})} \nonumber \right. \\   
& & \left.  - F^{zy}_{(i,j+1,k+\frac{1}{2})} \right)- 
  \frac{1}{4} \left( \eta J^x_{(i,j+\frac{1}{2},k)}  + \eta  J^x_{(i,j+\frac{1}{2},k+1)} \right. \nonumber \\
& & \left.   + \eta  J^x_{(i,j,k+\frac{1}{2})} + \eta  J^x_{(i,j+1,k+\frac{1}{2})} \right)   \, ,
\end{eqnarray}
\begin{eqnarray}
& & \Omega^y_{(i+\frac{1}{2},j,k+\frac{1}{2})} = 
 \frac{1}{4} \left( F^{zx}_{(i,j,k+\frac{1}{2})} + F^{zx}_{(i+1,j,k+\frac{1}{2})} - F^{xz}_{(i+\frac{1}{2},j,k)}
\nonumber \right. \\
& & \left. - F^{xz}_{(i+\frac{1}{2},j,k+1)} \right) - 
 \frac{1}{4} \left(\eta J^y_{(i,j,k+\frac{1}{2})}  + \eta  J^y_{(i+1,j,k+\frac{1}{2})} \right. \nonumber \\
& & \left. + \eta  J^y_{(i+\frac{1}{2},j,k)} + \eta  J^y_{(i+\frac{1}{2},j,k+1)} \right)  \, ,
\end{eqnarray}
\begin{eqnarray}
& & \Omega^z_{(i+\frac{1}{2},j+\frac{1}{2},k)} = 
 \frac{1}{4} \left( F^{xy}_{(i+\frac{1}{2},j,k)} + F^{xy}_{(i+\frac{1}{2},j+1,k)}   - F^{yx}_{(i,j+\frac{1}{2},k)} 
 \nonumber \right. \\
 & & \left. - F^{yx}_{(i+1,j+\frac{1}{2},k)} \right) - 
 \frac{1}{4} \left( \eta J^z_{(i+\frac{1}{2},j,k)} + \eta  J^z_{(i+\frac{1}{2},j+1,k)} \right.\nonumber \\
& & \left. + \eta  J^z_{(i,j+\frac{1}{2},k)} + \eta  J^z_{(i+1,j+\frac{1}{2},k)} \right) \,  ,
\end{eqnarray}
with the fluxes $F^{ij}$ computed with the HLLE or HLLC schemes along each spatial direction. Furthermore, in the last equations, we have included the average of the second term in the right-hand-side of \eqref{eq_Omega}, $\eta \vec{J}$, since it is not presented in the definition of the fluxes. It is worth mentioning that we have chosen to express those terms in that way, so the method could be used with any approximate Riemann solver and preserve the $\nabla\cdot\vec{B}$ to the order of the machine round-off-error, as it will be shown in the next section for all the tests considered in this work.

The last step is to calculate the magnetic field in the center of the cell by averaging the values obtained in equations \eqref{eq_B_caras_x} to \eqref{eq_B_caras_z}, that is 
\begin{eqnarray}
B^x_{(i,j,k)} = \frac{1}{2}\left( B^x_{(i-\frac{1}{2},j,k)} + B^x_{(i+\frac{1}{2},j,k)} \right) \, , \\
B^y_{(i,j,k)} = \frac{1}{2}\left( B^y_{(i,j-\frac{1}{2},k)} + B^y_{(i,j+\frac{1}{2},k)} \right) \, , \\
B^z_{(i,j,k)} = \frac{1}{2}\left( B^z_{(i,j,k-\frac{1}{2})} + B^z_{(i,j,k+\frac{1}{2})} \right)\, .
\end{eqnarray}
Finally, we compute $\nabla\cdot\vec{B}$ in each step of the evolution using the cell corner centered definition, since the centered one is not conserved by the Constrainted Transport Method \citep{CT_toth}
\begin{eqnarray}
& & \left( \nabla \cdot \vec{B} \right)_{(i+\frac{1}{2},j+\frac{1}{2},k+\frac{1}{2})} =  
 \frac{\partial B^x_{(i+\frac{1}{2},j+\frac{1}{2},k+\frac{1}{2})} }{\partial x} \\
& & \qquad \qquad \qquad + \frac{\partial B^y_{(i+\frac{1}{2},j+\frac{1}{2},k+\frac{1}{2})} }{\partial y} + \frac{\partial B^z_{(i+\frac{1}{2},j+\frac{1}{2},k+\frac{1}{2})} }{\partial z} \, ,\nonumber 
\end{eqnarray}
where the averages of the derivatives are given by
\begin{eqnarray}
& & \frac{\partial B^x_{(i+\frac{1}{2},j+\frac{1}{2},k+\frac{1}{2})} }{\partial x} =  \frac{1}{4 \ \Delta x} \left[ 
 B^x_{(i+1,j,k)}   - B^x_{(i,j,k)}  \nonumber \right. \\
& &  + B^x_{(i+1,j+1,k)}  -B^x_{(i,j+1,k)} + B^x_{(i+1,j,k+1)} \nonumber \\
& & \left.-B^x_{(i,j,k+1)} + B^x_{(i+1,j+1,k+1)} -B^x_{(i,j+1,k+1)}\right] \, , \\  
& & \frac{\partial B^y_{(i+\frac{1}{2},j+\frac{1}{2},k+\frac{1}{2})} }{\partial y} =  \frac{1}{4 \ \Delta y} \left[ 
 B^y_{(i,j+1,k)} - B^y_{(i,j,k)}  \nonumber \right. \\
& & +  B^y_{(i+1,j+1,k)} - B^y_{(i+1,j,k)}  + B^y_{(i,j+1,k+1)} - B^y_{(i,j,k+1)} \nonumber \\
& & \left. +  B^y_{(i+1,j+1,k+1)} - B^y_{(i+1,j,k+1)}\right] \, ,  \\                         
 & & \frac{\partial B^z_{(i+\frac{1}{2},j+\frac{1}{2},k+\frac{1}{2})} }{\partial z} =  \frac{1}{4 \ \Delta z} \left[
B^z_{(i,j,k+1)} - B^z_{(i,j,k)} \right. \nonumber  \\ 
& & + B^z_{(i+1,j,k+1)}  - B^z_{(i+1,j,k)}  + B^z_{(i,j+1,k+1)} - B^z_{(i,j+1,k)} \nonumber \\
& &\left.  + B^z_{(i+1,j+1,k+1)} - B^z_{(i+1,j+1,k)}\right]   \, .                         
\end{eqnarray}

\begin{table*} 
\centering
\def\arraystretch{1.5}
\begin{tabular}{|llllll|}
 \hline 
 Test & Domain & Grid points  & $C_\text{cfl}$ & $\Gamma$ & Slope limiter  \\ 
\hline
Brio Wu & $[-0.5, 0.5]\times[-0.5, 0.5]\times [-0.5, 0.5]$ & $800\times 4 \times 4$ & 0.25 & 5/3 & All \\ 
\hline
Current Sheet & $[-0.5, 0.5]\times[-0.5, 0.5]\times [-0.5, 0.5]$ & $128\times 128 \times 4$ & 0.25 & 5/3 & WENO5\\
\hline
MHD rotor & $[0.0, 1.0]\times[0.0, 1.0]\times [0.0, 1.0]$ & $200\times 200 \times 4$ & 0.1 & 1.4 & MINMOD\\
\hline
Cloud Shock Interaction & $[0.0, 1.0]\times[0.0, 1.0]\times [0.0, 1.0]$ & $400\times 400 \times 4$ & 0.05 & 5/3 & WENO5\\
\hline
Magnetic Reconnection & $[-0.5,0.5]\times[-2.0,2.0]\times [0.0, 1.0]$ & $200\times 800 \times 4$ & 0.05 & 5/3 & WENO5\\
\hline
Thermal Conduction & $[-0.5,0.5]\times[-0.5,0.5]\times [0.0, 1.0]$ & $200\times 200 \times 4$ & 0.05 & 5/3 & MINMOD\\
\hline
Transversal Oscillations & $[-25.0,25.0]\times[0.0,1.0]\times [0.0, 50.0]$ & $400\times 4 \times 400$ & 0.25 & 5/3 & MINMOD\\
\hline
MHD Gravity Waves & $[-0.75,0.75]\times[-0.75,0.75]\times [-0.25, 5.75]$ & $105\times 105 \times 210$ & 0.25 & 1.4 & MINMOD\\
\hline    
\end{tabular}\caption{Information of the numerical parameters used in the simulations }  \label{tabla_info}  
\end{table*}

\begin{figure}
\centering
\includegraphics[width=0.45\textwidth]{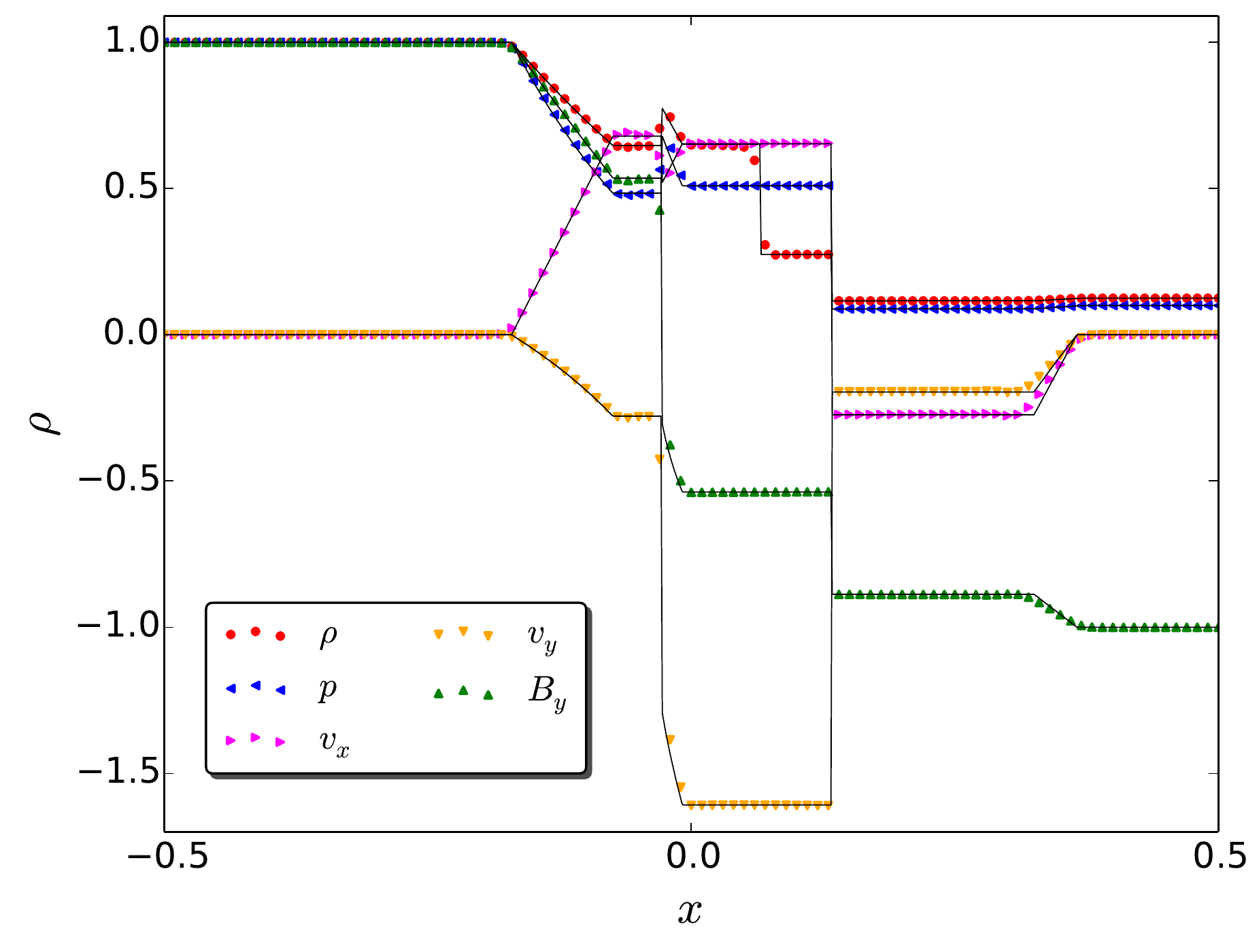} 
\caption{Comparison of the density, pressure, velocity components $v_x$, $v_y$ and magnetic field component $B_y$ in the Brio-Wu test in time $0.1$. In the same plot, the solid line is the non-regular exact solution taken from \cite{takahashi2013}. }\label{fig_BW}
\end{figure}

\begin{table} 
\centering
\begin{tabular}{|lccc|}
 \hline 
 Method & N & $L^1$  & $L^1_\text{order}$  \\ 
\hline
          &  50 & 2.46$\times 10^{-2}$ &   -  \\
          & 100 & 1.46$\times 10^{-2}$ & 0.75   \\    
HLLC/     & 200 & 9.33$\times 10^{-3}$ & 0.65  \\    
MM       & 400 & 5.31$\times 10^{-3}$ & 0.81  \\    
          & 800 & 3.22$\times 10^{-3}$ & 0.72   \\     
          &1600 & 1.88$\times 10^{-3}$ & 0.78   \\    
\hline      
        & 50 & 2.12$\times 10^{-2}$   &  -  \\
        & 100 &  1.27$\times 10^{-2}$ & 0.74 \\     
HLLC/   & 200 &  6.92$\times 10^{-3}$ & 0.87  \\   
MC        & 400 &  3.59$\times 10^{-3}$ & 0.95  \\   
        & 800 &  2.19$\times 10^{-3}$ & 0.72  \\     
        & 1600 & 1.27$\times 10^{-3}$ & 0.79 \\ 
\hline 
          & 50  &  2.10$\times 10^{-2}$ &  - \\
          & 100 &  1.22$\times 10^{-2}$ & 0.79  \\    
          & 200 &  7.64$\times 10^{-3}$ & 0.67  \\    
HLLC/     & 400 &  3.95$\times 10^{-3}$ & 0.95  \\    
WENO5     & 800 &  2.28$\times 10^{-3}$ & 0.80  \\    
          & 1600 &  1.27$\times 10^{-3}$& 0.85 \\  
\hline              
          &  50 & 2.56$\times 10^{-2}$  & -  \\
          & 100  & 1.68$\times 10^{-2}$ & 0.60   \\    
HLLE/     & 200  & 1.11$\times 10^{-2}$ & 0.60   \\    
MM        & 400  & 6.54$\times 10^{-3}$ &  0.77  \\    
          & 800  & 4.018$\times 10^{-3}$ &  0.70  \\    
          & 1600 &  2.37$\times 10^{-3}$ &  0.76  \\    
\hline
        &  50 &  2.50$\times 10^{-2}$ &-\\
        & 100 &   1.44$\times 10^{-2}$ & 0.80  \\     
HLLE/   & 200 &  8.21$\times 10^{-3}$  & 0.81  \\    
MC      & 400 &  4.07$\times 10^{-3}$ &  1.01  \\    
        & 800 &  2.17$\times 10^{-3}$  & 0.90  \\    
        & 1600 &  1.19$\times 10^{-3}$ & 0.87  \\    
\hline 
          & 50 &  2.04$\times 10^{-2}$   &  -     \\
          & 100 &  1.22$\times 10^{-2}$  &  0.74  \\     
HLLE/     & 200 &   8.24$\times 10^{-3}$ &  0.56  \\    
WENO5     & 400 &  4.43$\times 10^{-3}$  &  0.90  \\    
          & 800 &  2.61$\times 10^{-3}$  &  0.76   \\   
          & 1600 &  1.52$\times 10^{-3}$ &  0.78  \\    
\hline    
\end{tabular}\caption{Error $L^1$ for the density in the Brio Wu test }  \label{tabla_BW}  
\end{table}

\section{Tests}\label{sec:tests}

With regard to analyzing the accuracy of the code, in this section, we present some tests to validate the implementation of the algorithms. We start with the 1D test Brio-Wu \citep{BrioWu} in the regime of ideal MHD, which is compared with the exact solution to calculate its convergence.  Later we repeat the same test with resistivity and with isotropic heat flux.  Secondly, we depict three ideal MHD tests in 2D,   the Current Sheet test \citep{current_sheet_chapter3_4}, the MHD rotor \citep{Balsara1999} and  the Cloud Shock Interaction test \citep{Woodward1998}.  Also, to test the resistivity terms and the second model for the heat flux \eqref{heat2} we run the 2D magnetic reconnection test and the thermal conduction test from \cite{ResistiveCode_Jiang}.  Finally, we implement two ideal MHD evolutions for wave propagation in the solar atmosphere, the first one is the 2D Transverse Oscillations in Solar Coronal Loops \citep{transverse_waves}, and the second one is a 3D simulation of gravity waves in the solar atmosphere using a realistic temperature profile \citep{gravity_waves}. All of those evolutions were done using a uniform grid in Cartesian coordinates.   The results were obtained with the HLLC Riemann solver with different slope limiters, as indicated in each case in table \ref{tabla_info}, along with the information of the numerical domain, the number of points in the grid, the Courant number and the adiabatic index $\Gamma$. For the integration in time, we use the third order TVD Runge-Kutta. Finally, to check the accuracy and to verify that there are not non-physical phenomena, we present the maximum value of the divergence of the magnetic field in each of the 2D and 3D simulations.

\begin{figure}
\centering
\includegraphics[width=0.45\textwidth]{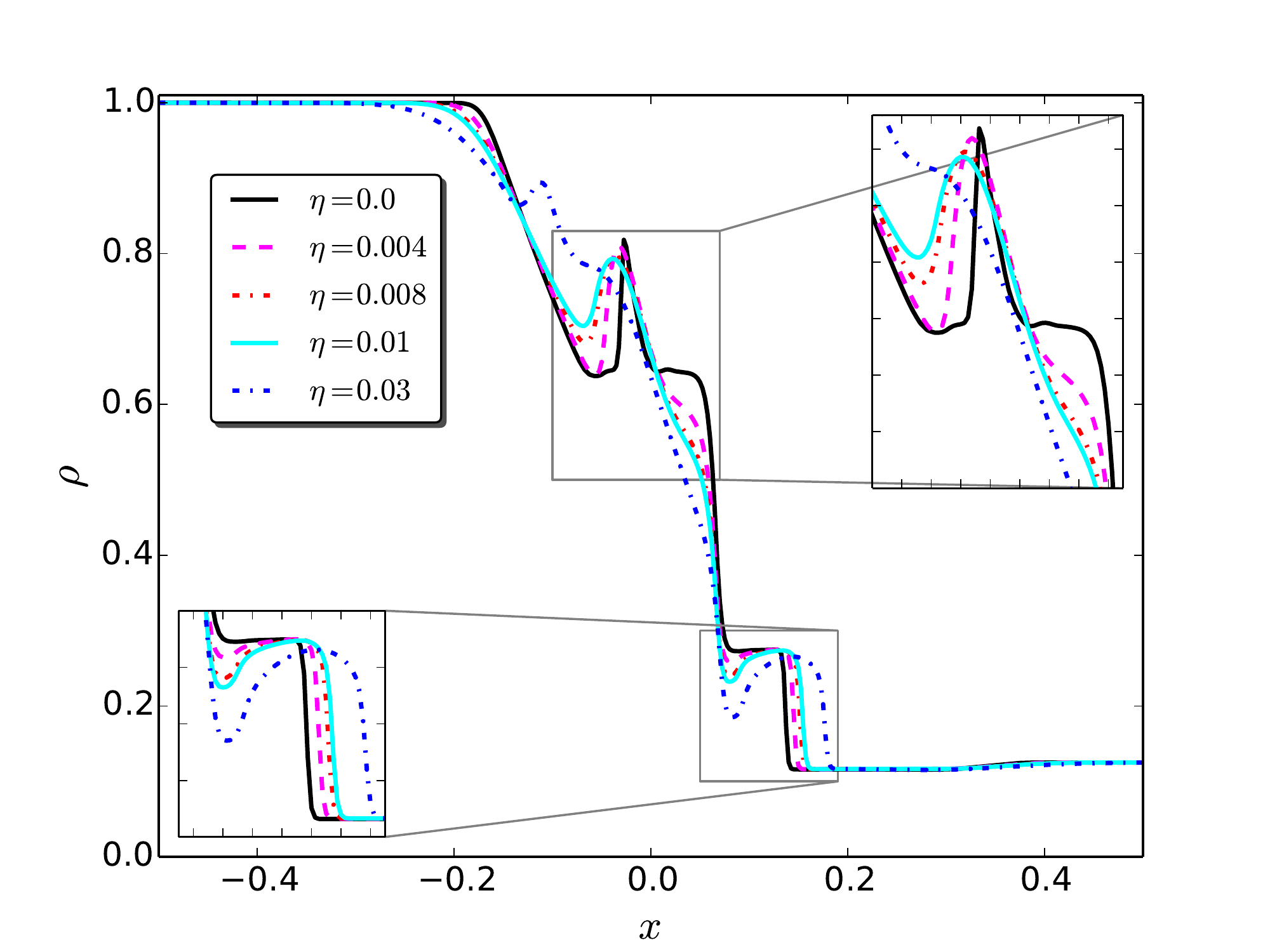} 
\caption{Comparison of the density in the Brio-Wu test in time $0.1$ for different values of the resistivity.}\label{fig_BW_Resistive}
\end{figure}
\begin{table} 
\centering
\begin{tabular}{|lccc|}
 \hline 
  & $L^1$ error 1 & $L^1$ error 2 & $L^1$ order  \\ 
\hline
$\eta = 0.01 $ & 1.41$\times 10^{-3}$ &  9.70$\times 10^{-4}$ & 0.54  \\   
$\eta = 0.004 $ & 1.45$\times 10^{-3}$ &  8.09$\times 10^{-4}$ & 0.84  \\   
$\eta = 0.007 $ &  1.36$\times 10^{-3}$ &  8.27$\times 10^{-4}$ & 0.71 \\   
\hline    
\end{tabular} \caption{Errors of the Brio Wu test with Resistivity. } \label{tabla_BW_Resistive}   
\end{table}

\begin{figure}
\centering
\includegraphics[width=0.45\textwidth]{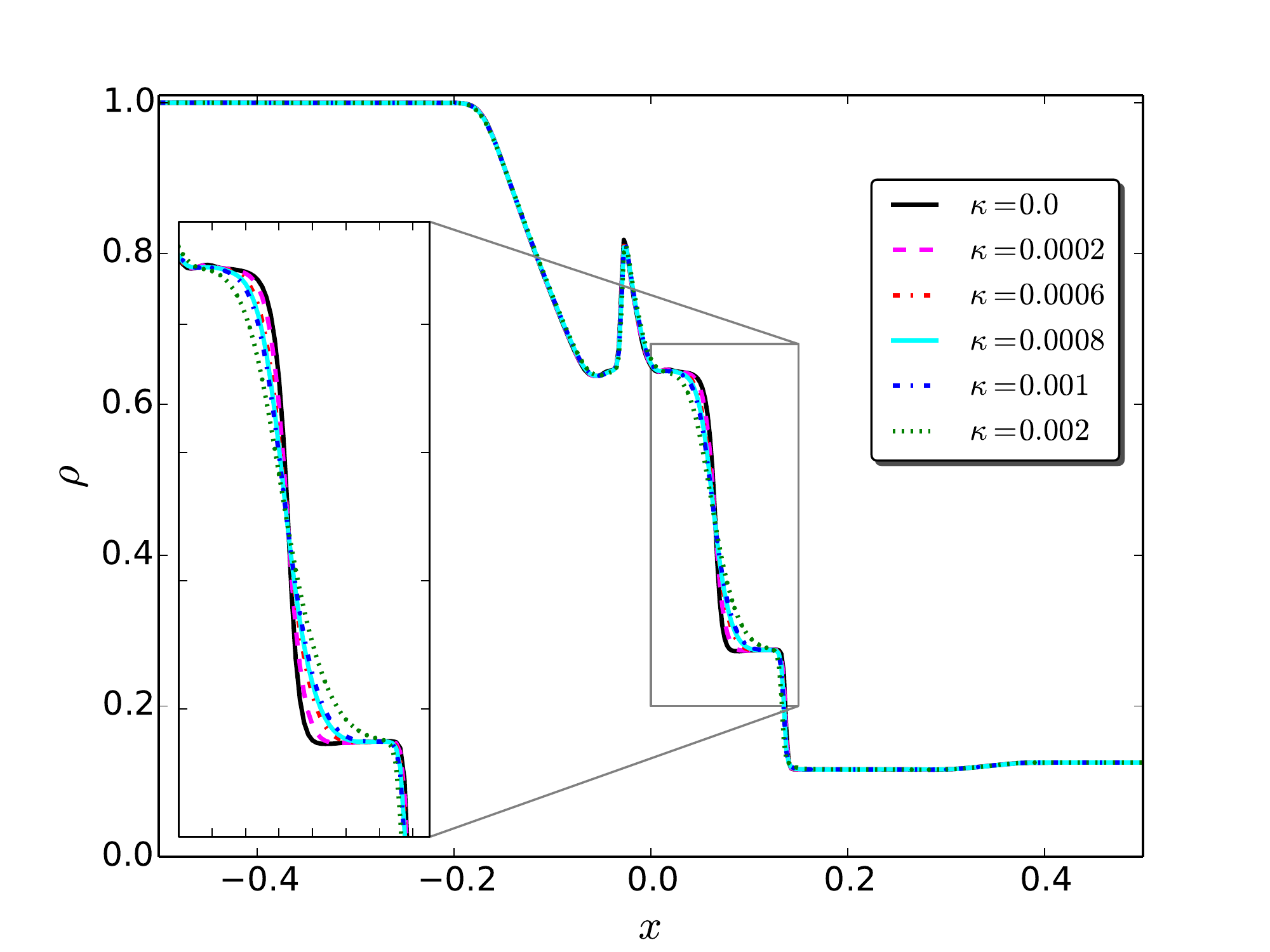} 
\caption{Comparison of the density in the Brio-Wu test in time $0.1$ for different values of the thermal conductivity.}\label{fig_BW_heat}
\end{figure}

\begin{figure*}
\centering
\includegraphics[height=0.23\textheight]{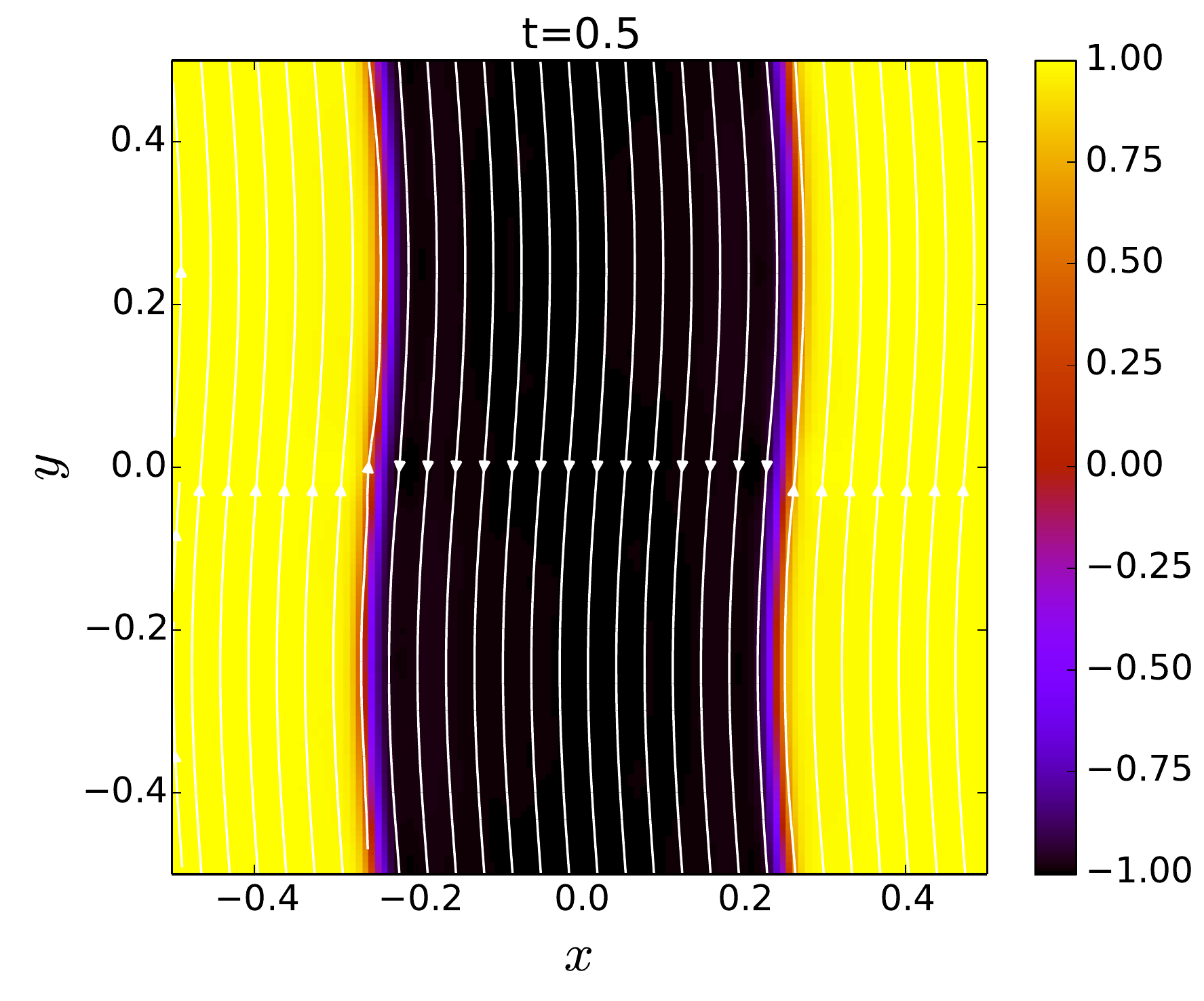} 
\includegraphics[height=0.23\textheight]{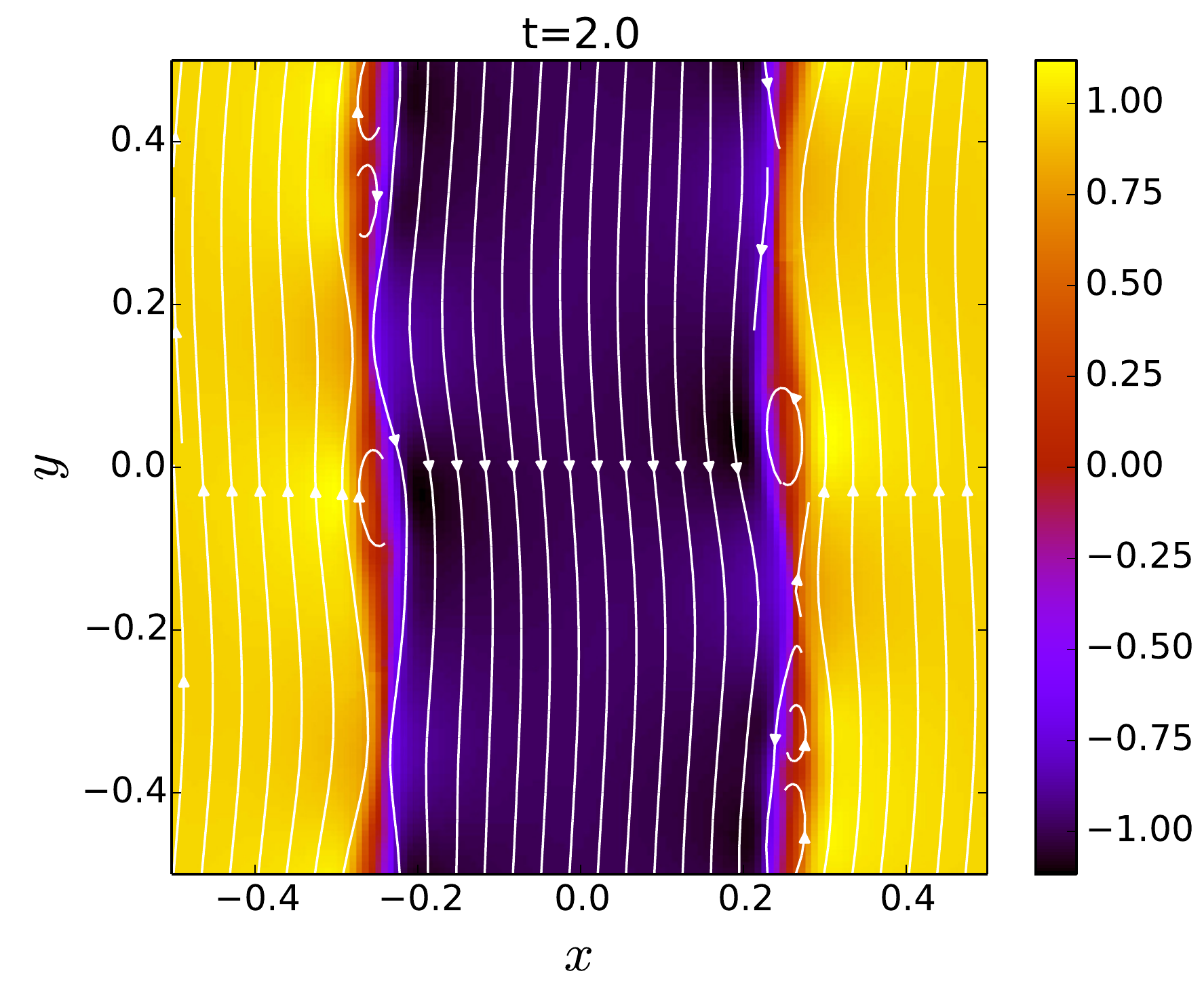}
\includegraphics[height=0.23\textheight]{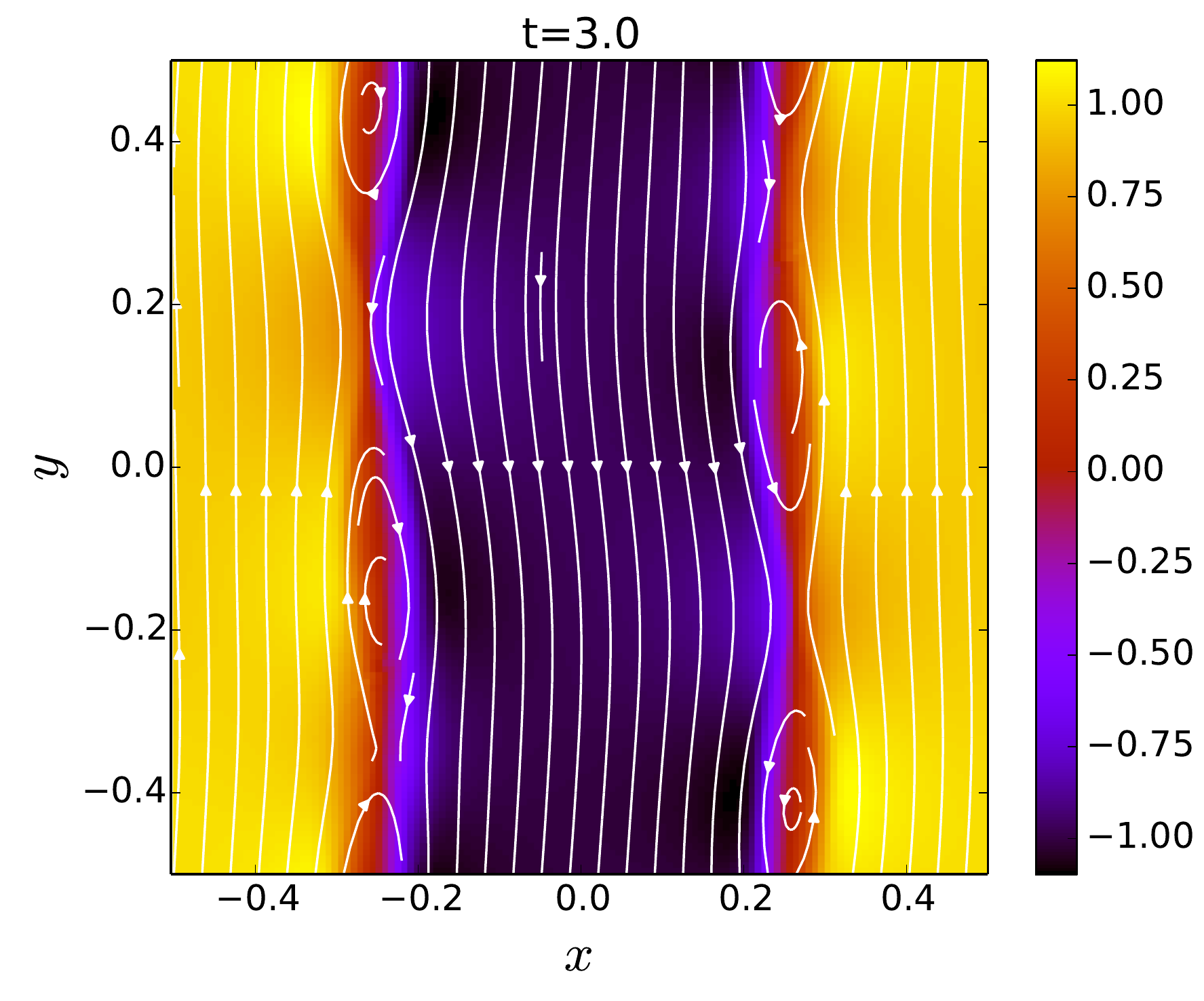} 
\includegraphics[height=0.23\textheight]{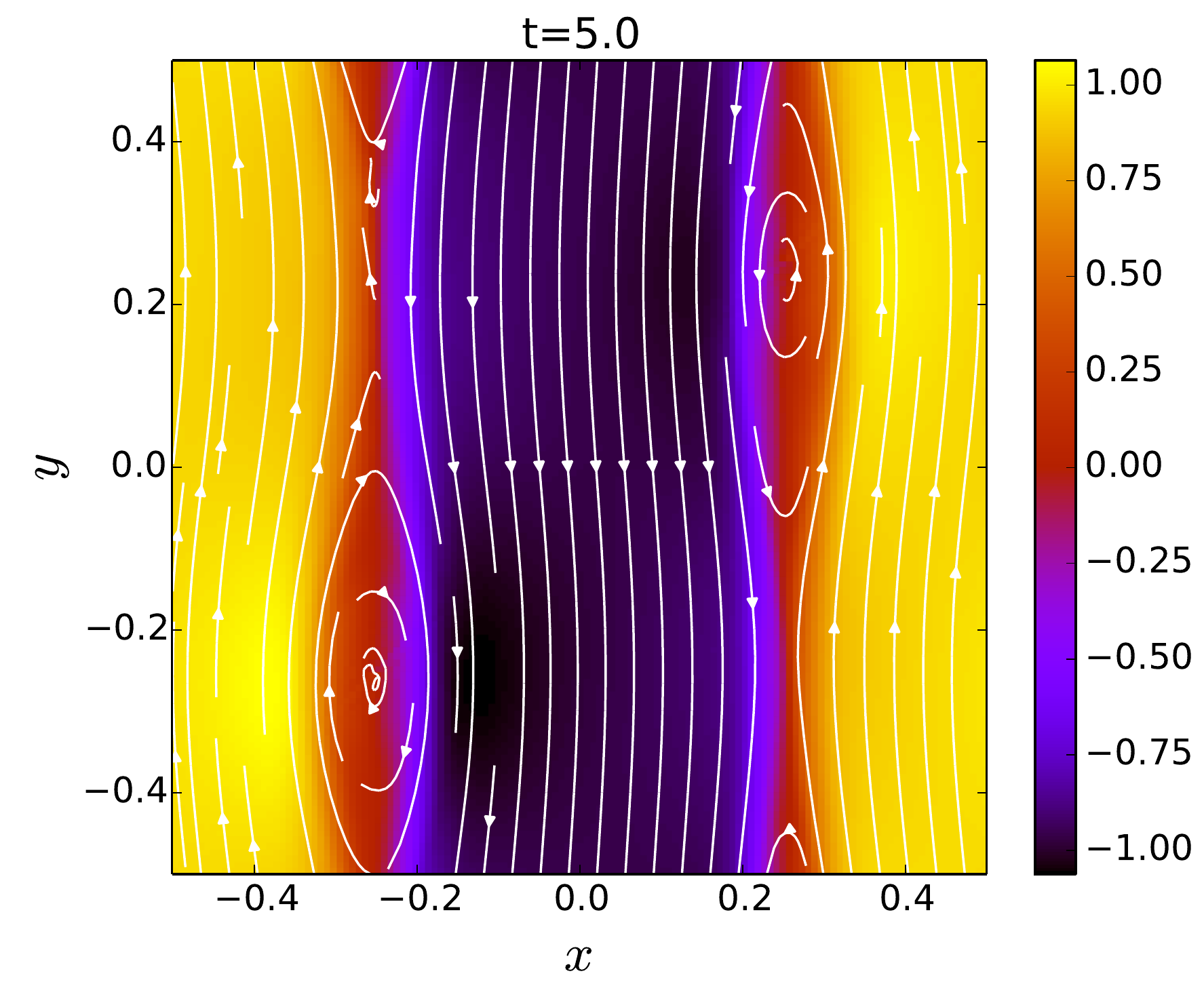}   
\caption{Magnetic field lines and $B_y$ component in the Current Sheet Test in the $z=0$ plane at times $t=0.5$, $t=2.0$, $t=3.0$ and $t=5.0$.} \label{fig_current}
\end{figure*}

\begin{table}
\centering
\begin{tabular}{|lccc|}
 \hline 
  & $L^1$ error 1 & $L^1$ error 2 &  $L^1$ order \\ 
\hline
$\kappa = 0.001 $ & 1.68$\times 10^{-3}$ &  8.63$\times 10^{-4}$ & 0.96   \\   
$\kappa = 0.0002 $ & 2.07$\times 10^{-3}$ &  1.06$\times 10^{-3}$ & 0.96   \\   
$\kappa = 0.0006 $ & 1.86$\times 10^{-3}$ &  8.58$\times 10^{-4}$ & 1.08   \\   
\hline    
\end{tabular} \caption{$L^1$ self-convergence error of the Brio Wu test with thermal conductivity }\label{tabla_BW_heat}   
\end{table}

\subsection{Test 1: Brio-Wu} \label{subsec: BW}

The \cite{BrioWu} test is the extension of the Sod's shock-tube problem with magnetic fields, which can verify the ability of the code to capture the characteristic waves of a MHD problem, that is, the shocks, rarefactions and contact discontinuity waves. The initial state, given by $p = 1$, $v_x = v_y = v_z = B_z= 0$, $B_x = 0.75$ and 
\begin{eqnarray}
\rho &=& \left\lbrace \begin{array}{ll}
1   & \quad \text{for} \quad  x<0.5  \, ,  \\
0.125 & \quad \text{for} \quad  x\geq 0.5  \, ,
\end{array} \right. \\
B_y &=& \left\lbrace \begin{array}{rl}
1.0 & \quad \text{for} \quad x<0.5 \, , \\
-1.0 & \quad \text{for} \quad x\geq 0.5 \, ,
\end{array} \right.   
\end{eqnarray}
consists of a left and right constant states. The parameters for the numerical simulation are listed in table \ref{tabla_info}. The results are presented in figure \ref{fig_BW}, where we display the density, pressure, velocity components $v_x$ and $v_y$ and the magnetic field component $B_y$ at the time $t=0.1$,  using the MINMOD slope limiter. In this figure, we also plot the non-regular exact solution from \cite{takahashi2013} in order to compare the results.  Moreover, with the exact solution for the density, we calculate the $L^1$ norm of the error 
\begin{eqnarray}
L^1_{N} &=& \frac{1}{N}\sum_{i=1, N} |\rho_i - \rho^\text{exact}_i| \, , \label{eq_L1_error} 
\end{eqnarray}
where $N$ is the number of grid points. To compute the order of the error, $L^1$ needs to be calculated with $2N$ points, then with them, the order is given by
\begin{eqnarray}
L^1_\text{order} &=& \log_2\left(\frac{L^1_N}{L^1_{2N}}\right) \, . \label{eq_L1_order}   
\end{eqnarray}
These two values are obtained for each one of the Riemann solvers and slope limiters implemented in the code. The results are displayed in table \ref{tabla_BW} for 50, 100, 200, 400, 800 and 1600 grid points.

\subsection{Test 2: Brio-Wu with Resistivity} \label{subsec: BW_Resistive}

A version of the Brio-Wu Test with a uniform resistivity has been implemented. Using the same configuration than before, we run the test using different resistivities. In figure \ref{fig_BW_Resistive}, we plot the density for each value of $\eta$, in order to visualize the effect of this term. It can be seen that for small values of the resistivity, the effect is a dissipation of the discontinuities, and for bigger values, the contact discontinuity is displaced to the left, meanwhile, the shock discontinuity moves to the right, as it is shown in the inset plots of figure \ref{fig_BW_Resistive}.  

 Since the analytical solution, in this case, does not exist, we calculate a self-convergence of the  $L^1$ norm of the error. To do this,  we first calculate the error between the solutions with $N$ and $2N$ points, secondly, we estimate the error comparing the solutions with $2N$ and $4N$ points, and with both values, we calculate the order of the error with \eqref{eq_L1_order}. This process is repeated for the resistivities, $0.01$, $0.004$ and $0.007$, which are depicted in table \ref{tabla_BW_Resistive}. 

\subsection{Test 3: Brio-Wu with Isotropic Thermal Conduction} \label{subsec: BW_heat}
To conclude with the Brio-Wu tests, we have implemented a new version of this problem by adding the effect of the thermal conduction model \eqref{heat}, where the heat flux is modeled for an isotropic medium, proportional to the gradient of the temperature . Repeating the simulation with the same values than before, we obtain the results of figure \ref{fig_BW_heat}, where we plot the density for six different values of the thermal conductivity $\kappa$,  $0.0$, $0.0002$, $0.0006$, $0.0008$, $0.001$ and $0.002$. We can note that the effect of the heat flux is a dissipation and attenuation of the discontinuities, which is more notorious at the contact wave, as it is shown in the inset plot; however no changes are presented in the global topology of the solution, as it was the case of the test with resistivity. Also, the self-convergence norms of the error are  calculated for three different values of the thermal conductivity $\kappa = 0.001, 0.0002, 0.0006$ and are depicted in table \ref{tabla_BW_heat}. 

\begin{figure}
\centering
\includegraphics[height=0.23\textheight]{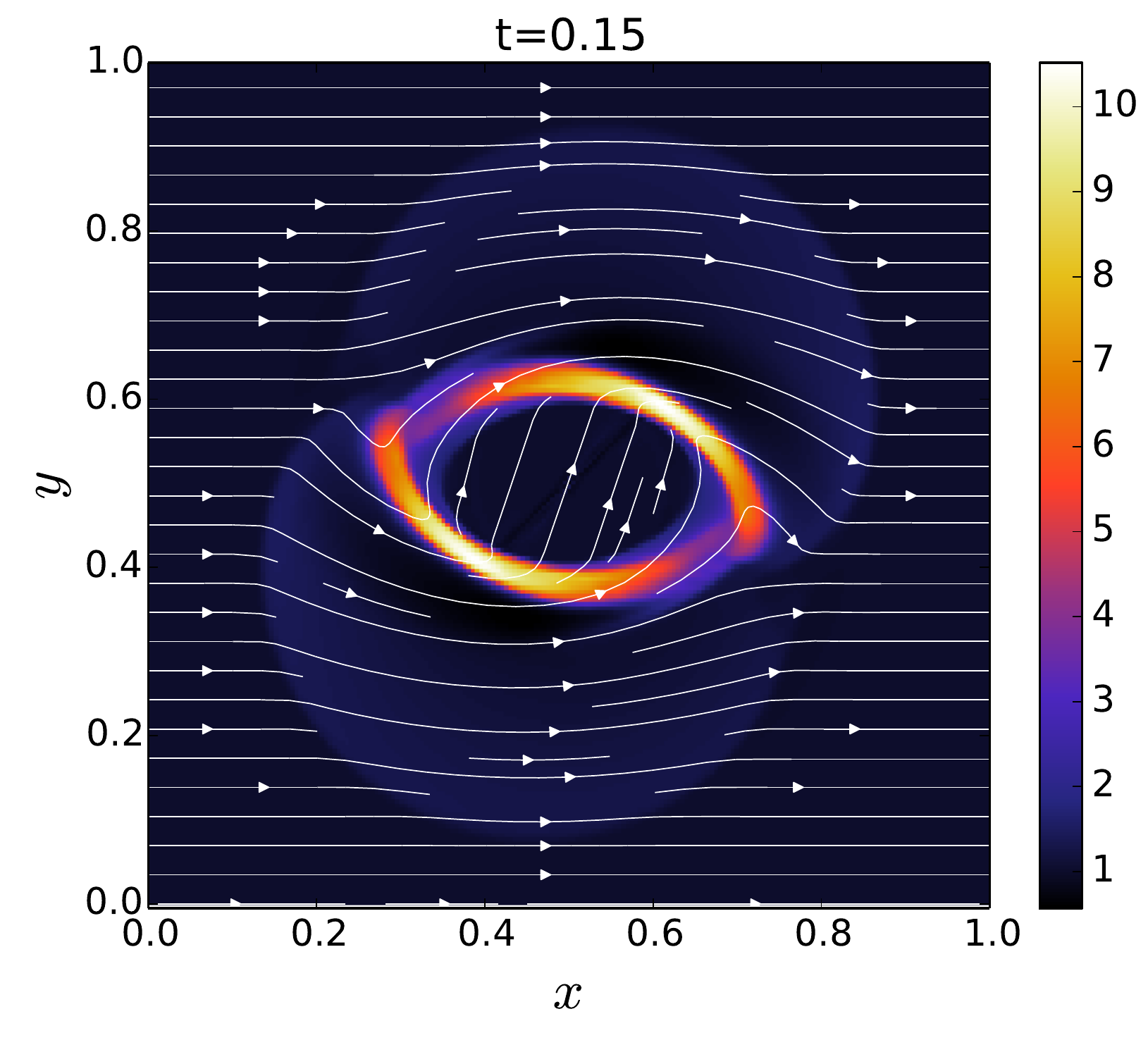}\\
\includegraphics[height=0.23\textheight]{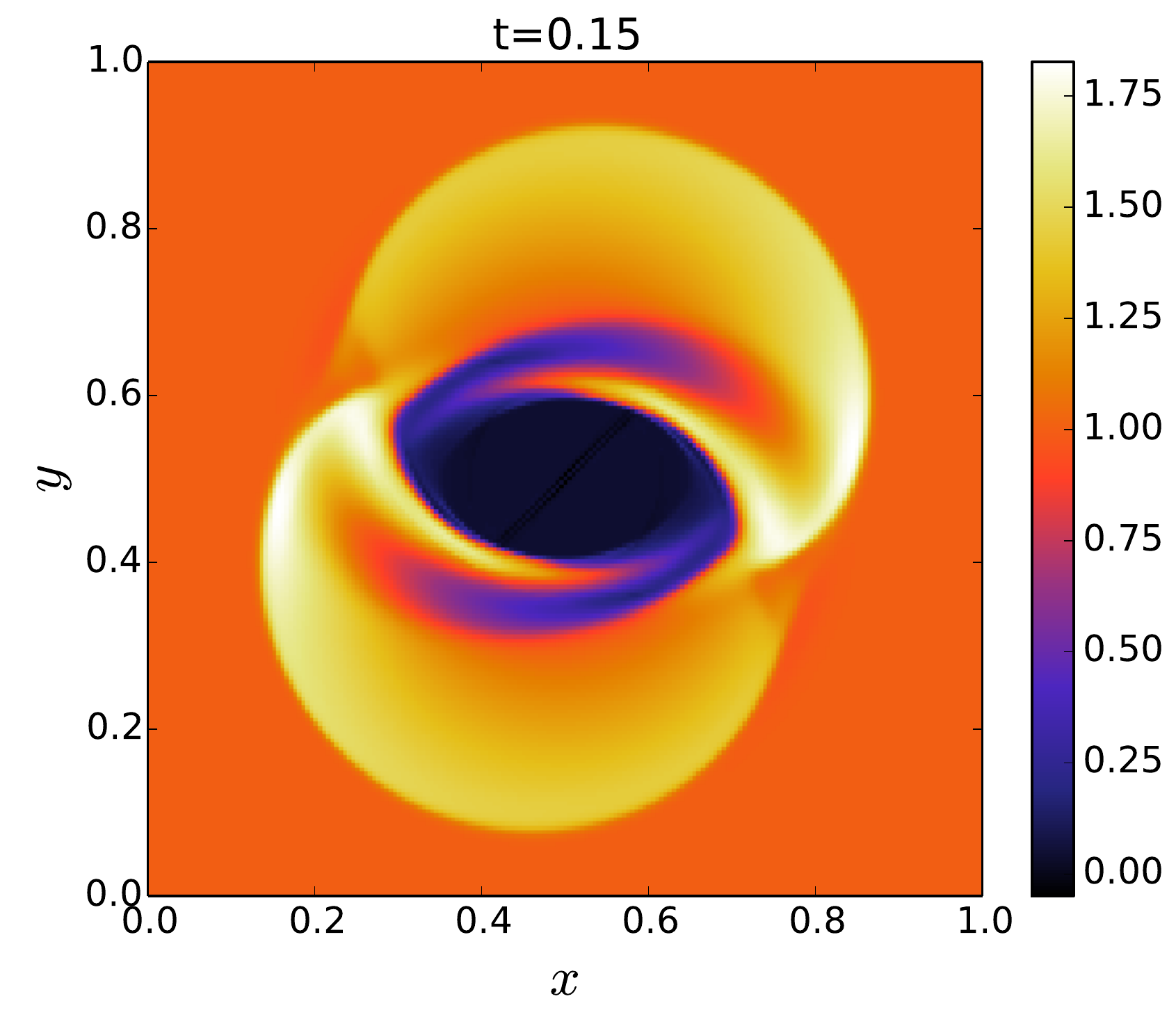} 
\caption{Gas density with the magnetic field lines and thermic pressure in the MHD Rotor Test at the time $t=0.15$.}
 \label{fig_rotor}
\end{figure}

\subsection{Test 4: Current Sheet} \label{subsec: test current sheet}
The first 2D test of ideal MHD is the Current Sheet problem \citep{current_sheet_chapter3_4}. This test is implemented in order to prove the robustness of the algorithms of integration in the code. The initial state consists in two current sheets with density $\rho = 1$, pressure $p = 0.3$, velocity components $v_x = 0.1 \sin{2\pi y}$, $v_y = v_z = 0$ and the  magnetic field  components $ B_x = B_z = 0$,
\begin{eqnarray}
B_y &=& \left\lbrace \begin{array}{rl}
1 & \quad \text{for} \quad |x|<0.25 \, , \\
-1 & \quad \text{for} \quad |x| \geq 0.25 \, . \\
\end{array}
 \right.
\end{eqnarray}
The parameters for the numerical simulation are listed in table \ref{tabla_info}. The boundary conditions are periodic in the planes $|x|=0.5$, $|y|=0.5$, and outflow in  $|z|=0.5$. In figure \ref{fig_current}, we plot a color map of the $B_y$ component and the magnetic field lines at times $t=0.5$, $t=2.0$, $t=3.0$ and $t=5.0$, where the magnetic reconnection can be tracked in time with the formation of magnetic islands that grow in time, causing a loss in the magnetic energy, which is converted into heat. From this results with low resolution, we can prove that the numerical methods implemented are good enough to describe standard non linear dynamics as the compressions and rarefactions raised in this test. The maximum divergence of the magnetic field in each step of this evolution is shown in figure \ref{fig_DivBmax}(a).

\subsection{Test 5: MHD Rotor 2D} \label{subsec:_test_mhd_rotor}

The second 2D test is the ideal MHD Rotor, proposed by \cite{Balsara1999}. It describes a rapidly spinning cylinder in a light fluid, with the following initial set up $p = 1$, $B_x = 5/\sqrt{4\pi}$, $B_y = B_z = v_z = 0$,  
\begin{eqnarray}
\rho &=& \left\lbrace \begin{array}{ll}
10 &  \quad \text{for} \quad r<r_0 \, , \\
1 + 9 f(r) & \quad  \text{for} \quad   r_0 < r <r_1 \, , \\
1  & \quad  \text{for} \quad r\geq r_1 \, ,
\end{array}   \right. \\
v_x &=& \left\lbrace \begin{array}{ll}
 -f(r)u_0(y-0.5)/r_0  & \quad \text{for} \quad r\leq r_0 \, , \\
 -f(r)u_0(y-0.5)/r  & \quad  \text{for} \quad r_0 < r \leq r_1 \, , \\
 0 & \quad  \text{for} \quad r\geq r_1 \, , \\
\end{array}  \right. \\ 
v_y &=& \left\lbrace \begin{array}{ll}
 f(r)u_0(x-0.5)/r_0  & \quad  \text{for} \quad r\leq r_0 \, , \\
 f(r)u_0(x-0.5)/r  & \quad  \text{for} \quad r_0 < r \leq r_1 \, , \\
 0 & \quad  \text{for} \quad r\geq r_1 \,  ,
\end{array}  \right.  \\
f(r) &=& (r_1-r)/(r-r_0) \, , \, r_1= 0.115 \, ,  r_0 = 0.1 \, ,
\end{eqnarray}
where $f(r)$ is a smooth function connecting the two regions, $r_0$ and $r_1$ are the inner and outer radius. 

For the evolution, we have used the information given in table \ref{tabla_info} and outflow boundary conditions in all directions. The results of the simulation are presented in figure \ref{fig_rotor},  they correspond to the color map of the density with the magnetic field lines (top plot), and the color map of the pressure (bottom plot), both in the time $t=0.15$. Here we find a very similar behavior than the one obtained by other codes, see for instance \cite{pluto_code}, finding that the initial torsional Alfv\'en waves, sent by the rotor into the fluid around, reduce their angular momentum, followed by the compression of the magnetic field, which distorts its initial circular shape into an oval one. In addition in figure \ref{fig_DivBmax}(b), we plot the maximum divergence of the magnetic field as a function of time.

\begin{figure}[h]
\centering
\includegraphics[height=0.23\textheight]{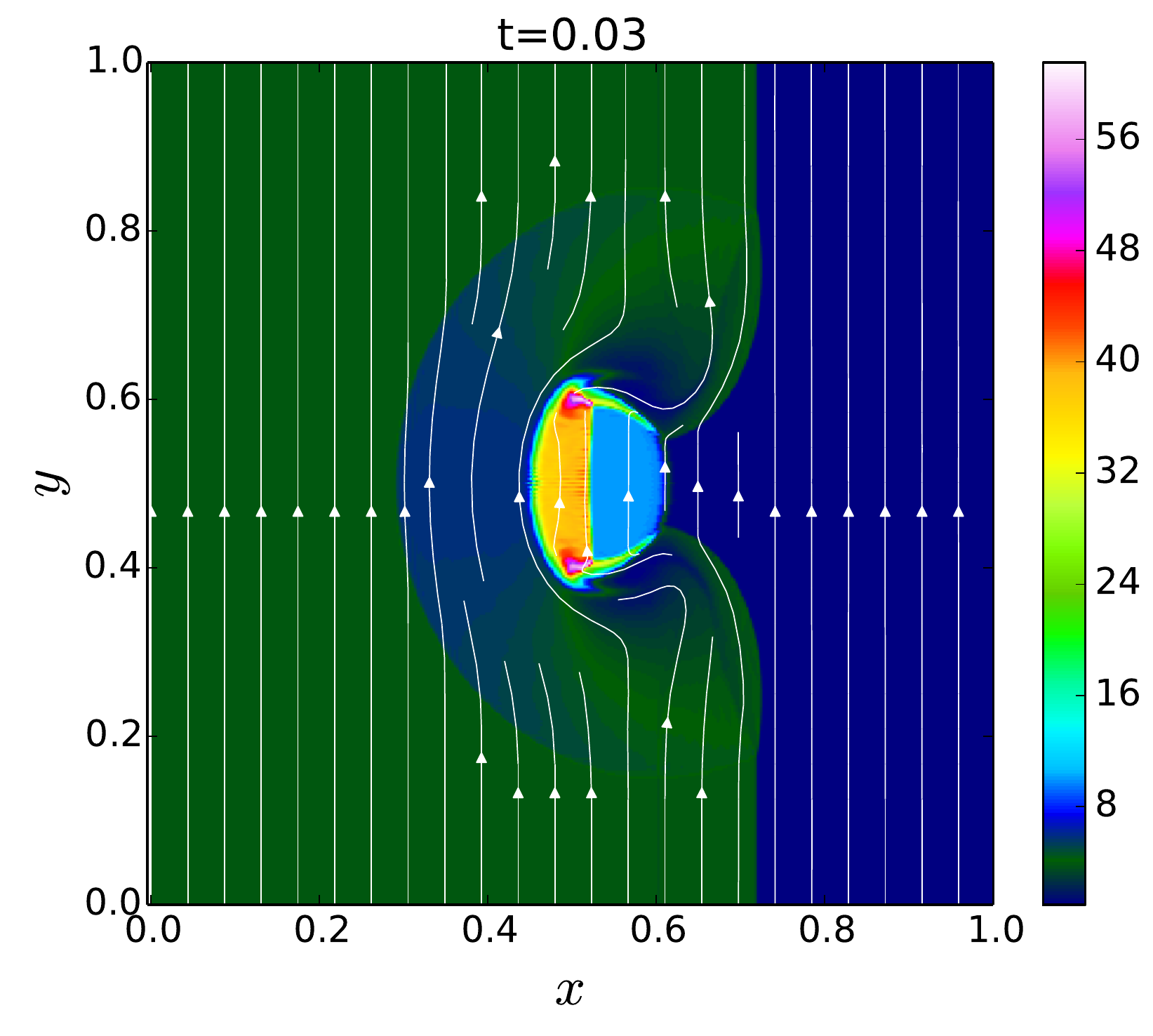}\\ \includegraphics[height=0.23\textheight]{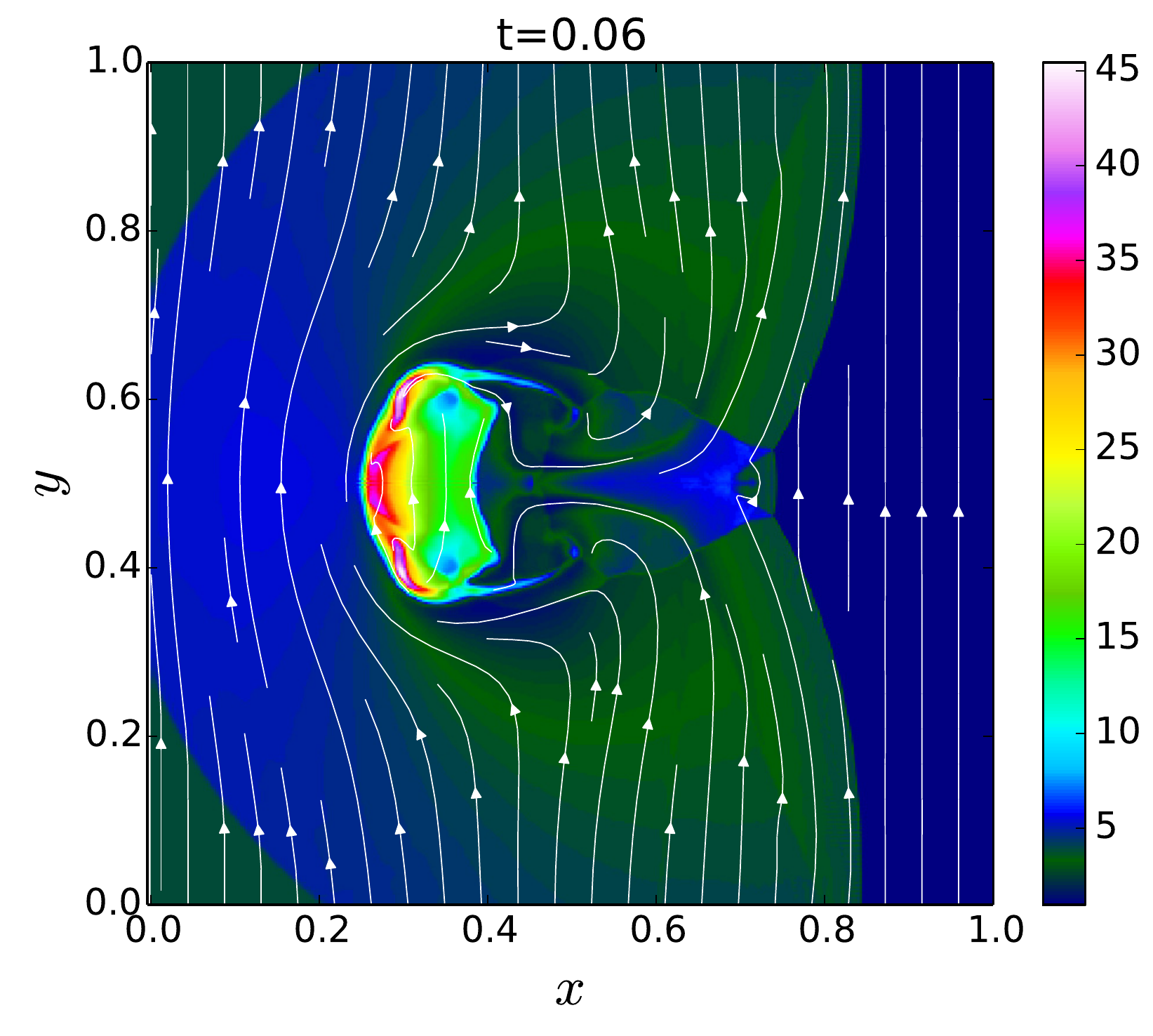}
\caption{Gas density with the magnetic field lines in the Cloud Shock Interaction Test at the times $t=0.03$ and $t=0.06$.}
 \label{fig_cloud}
\end{figure}

\begin{figure*}
\centering
\includegraphics[width=0.24\textwidth]{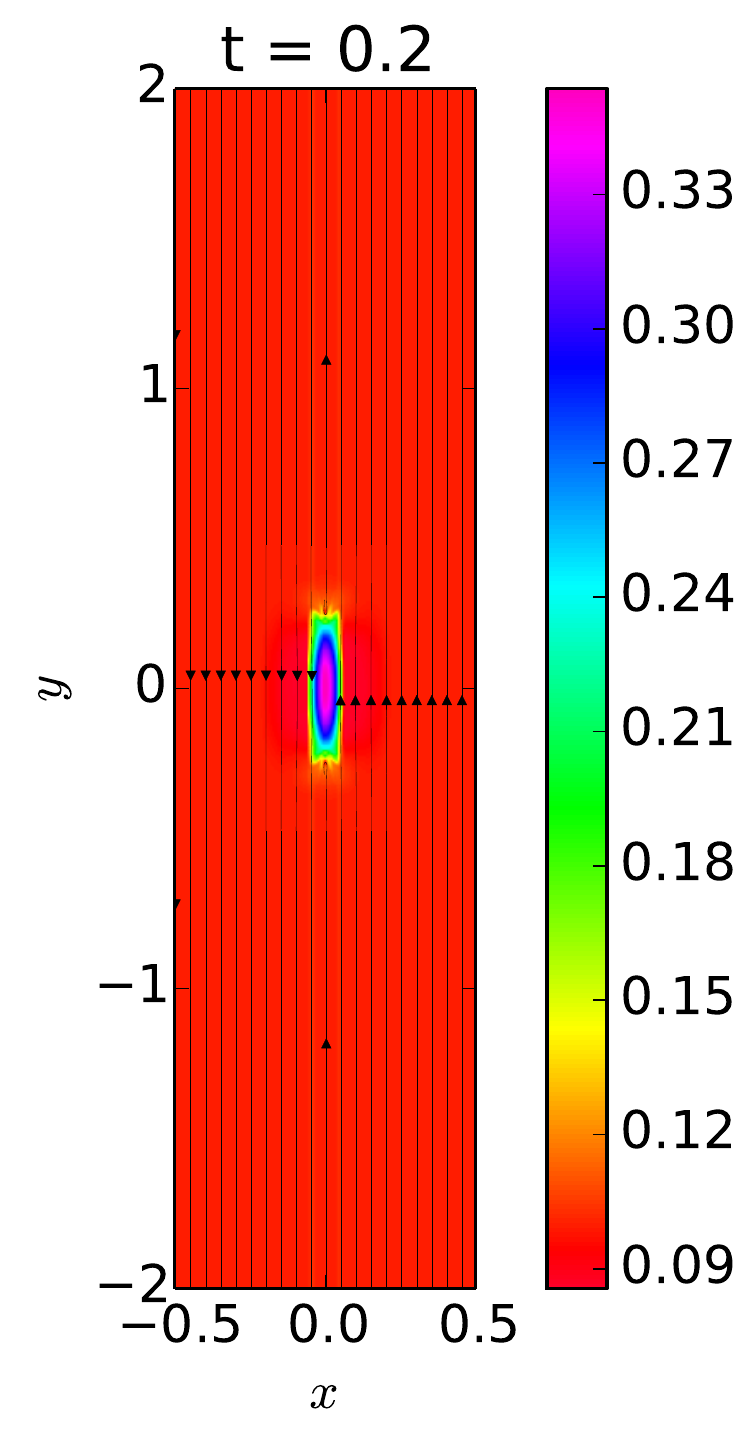} 
\includegraphics[width=0.24\textwidth]{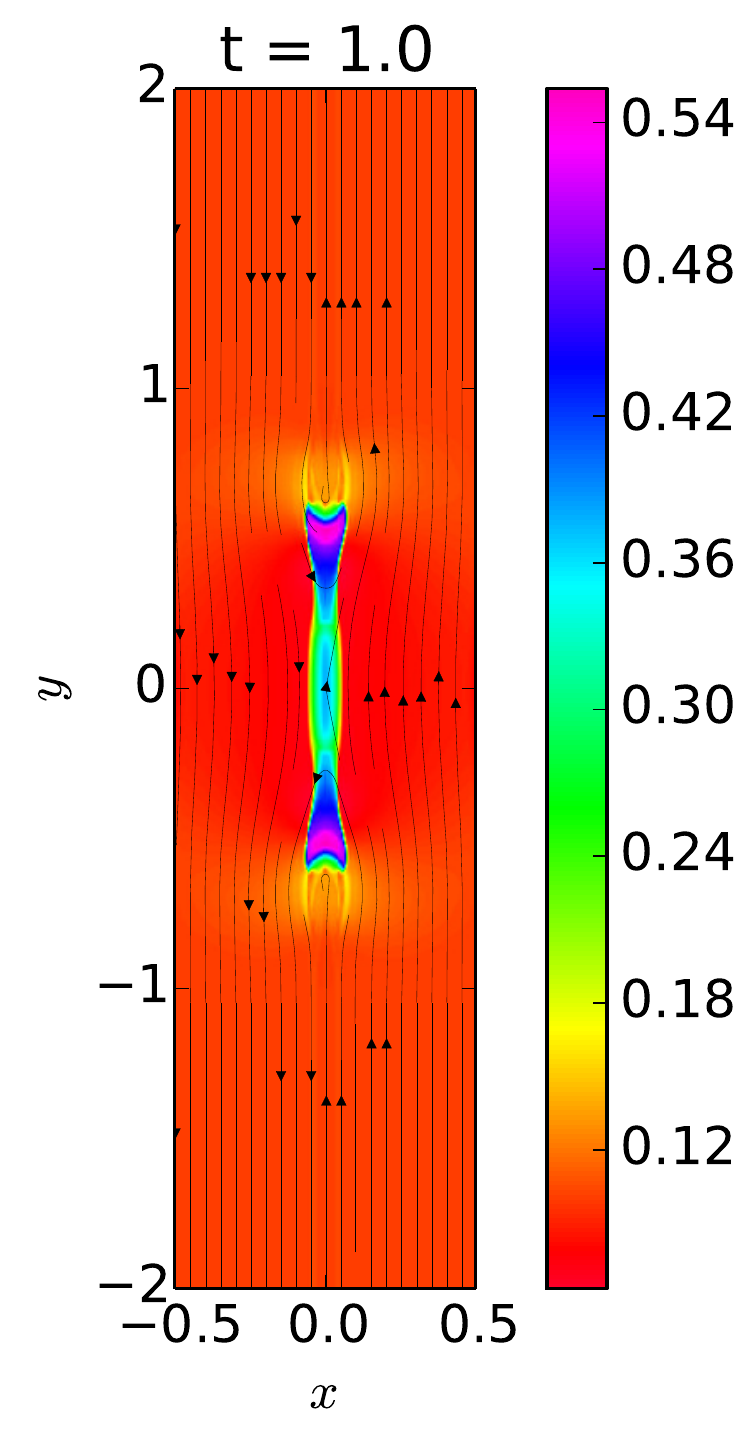}
\includegraphics[width=0.24\textwidth]{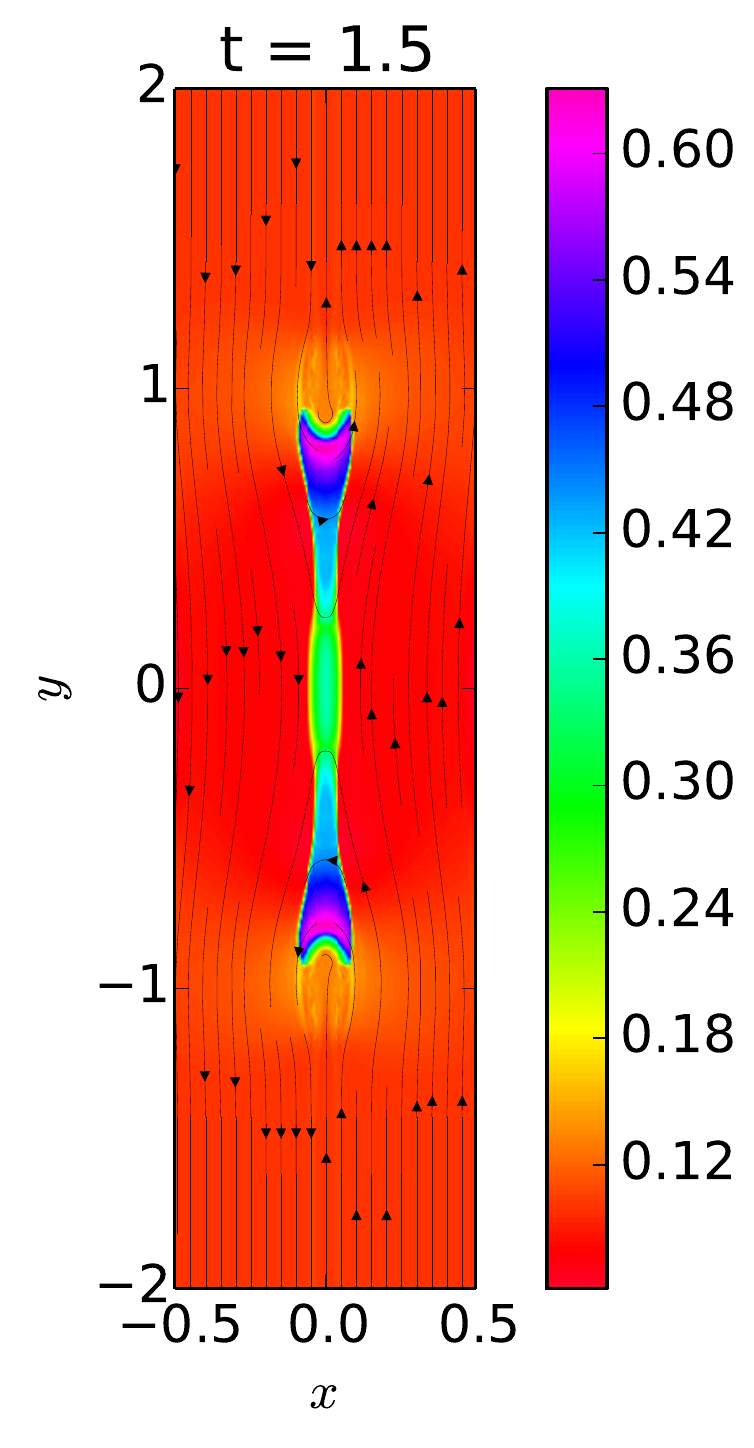} \\
\caption{Thermal pressure and magnetic field lines in the Magnetic Reconnection Test at the times $t=0.16$, $t=0.4$ and $t=1.2$.}
 \label{fig_reconnection}
\end{figure*}

\begin{figure}
\centering
\includegraphics[width=0.3\textwidth]{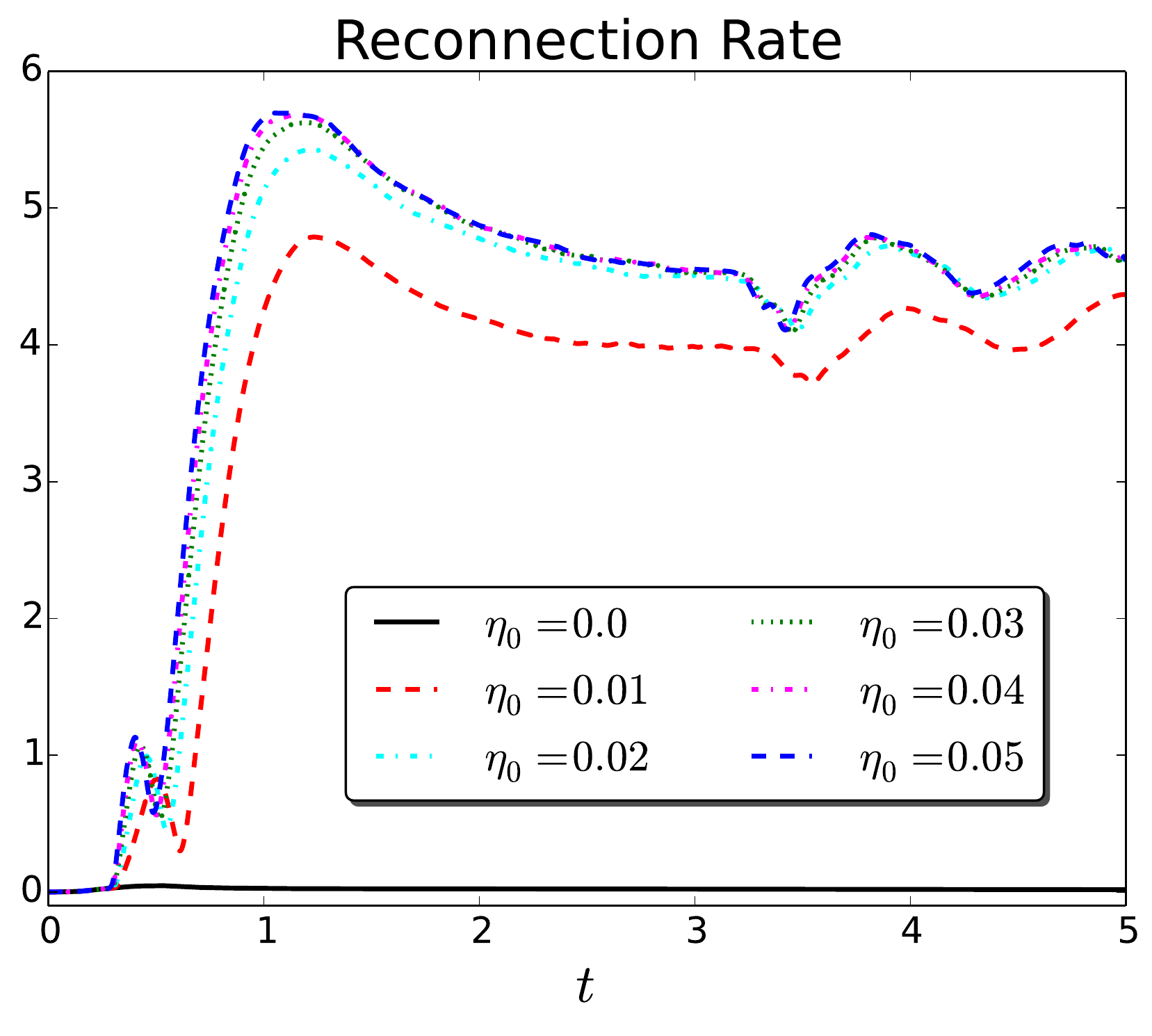} 
\caption{Magnetic Reconnection rate of Petschek for different values of the resistivity.}\label{fig_rec_rate}
\end{figure}

\begin{figure*}
\centering
\includegraphics[height=0.2\textheight]{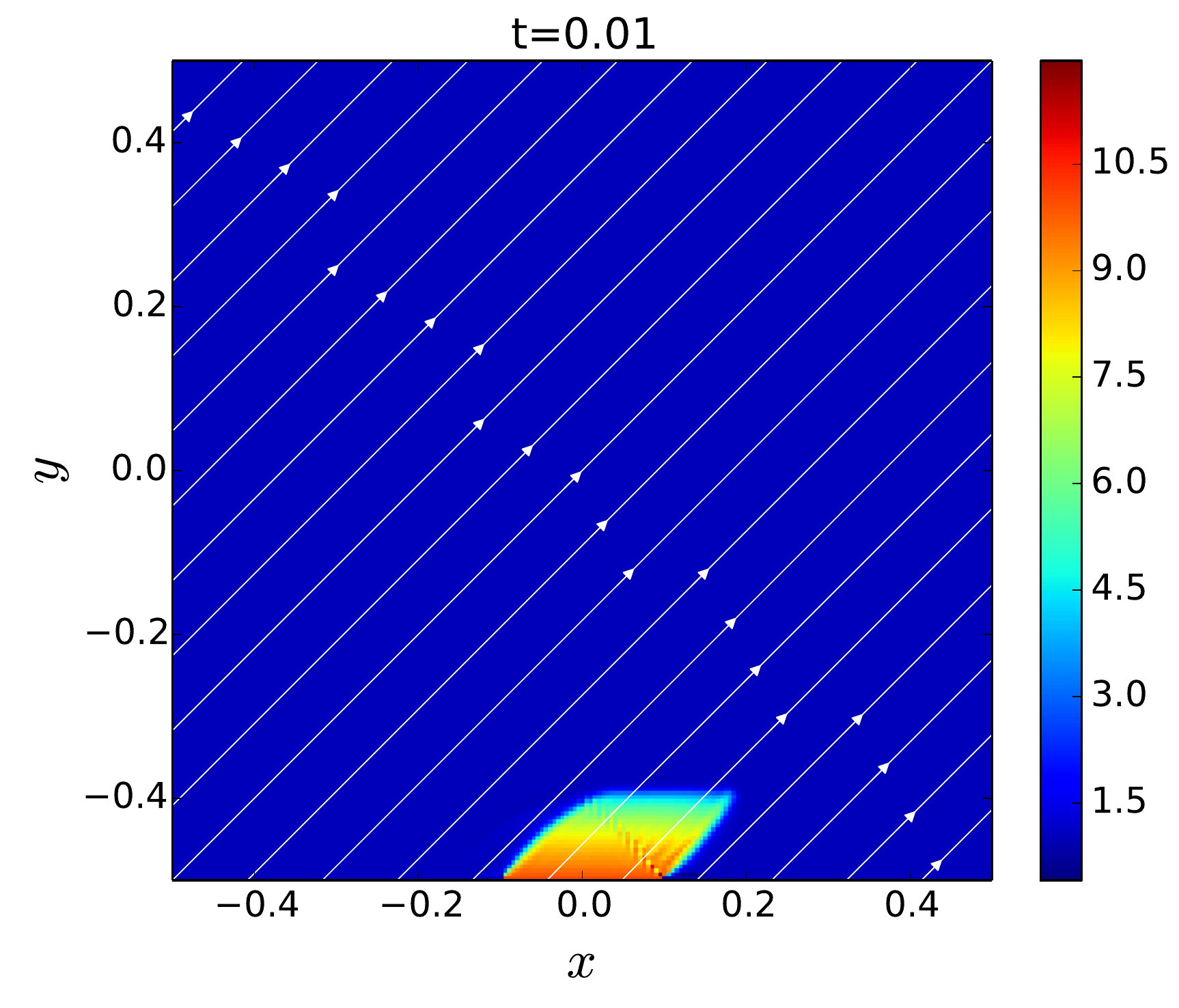}
\includegraphics[height=0.2\textheight]{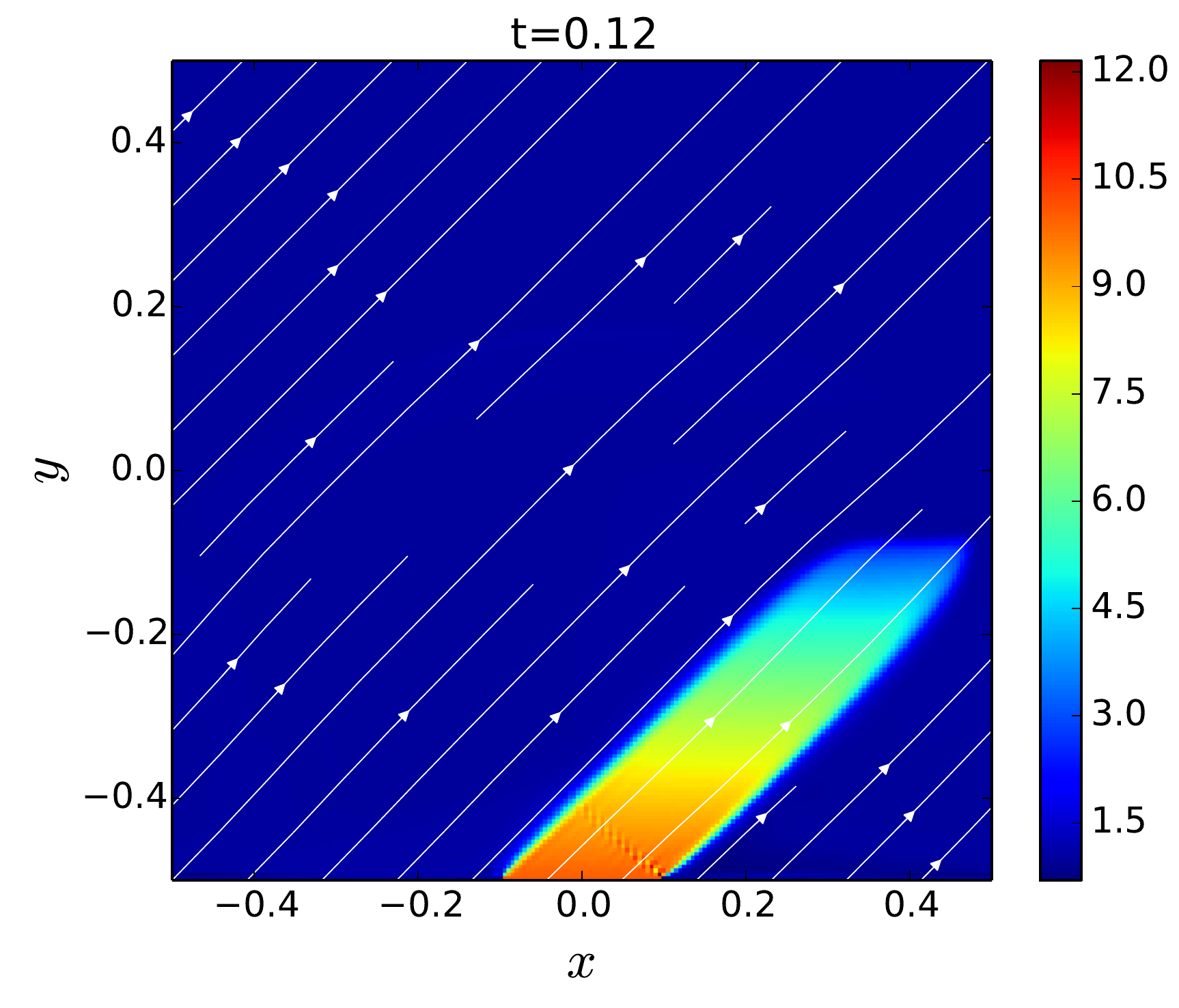}  
\includegraphics[height=0.2\textheight]{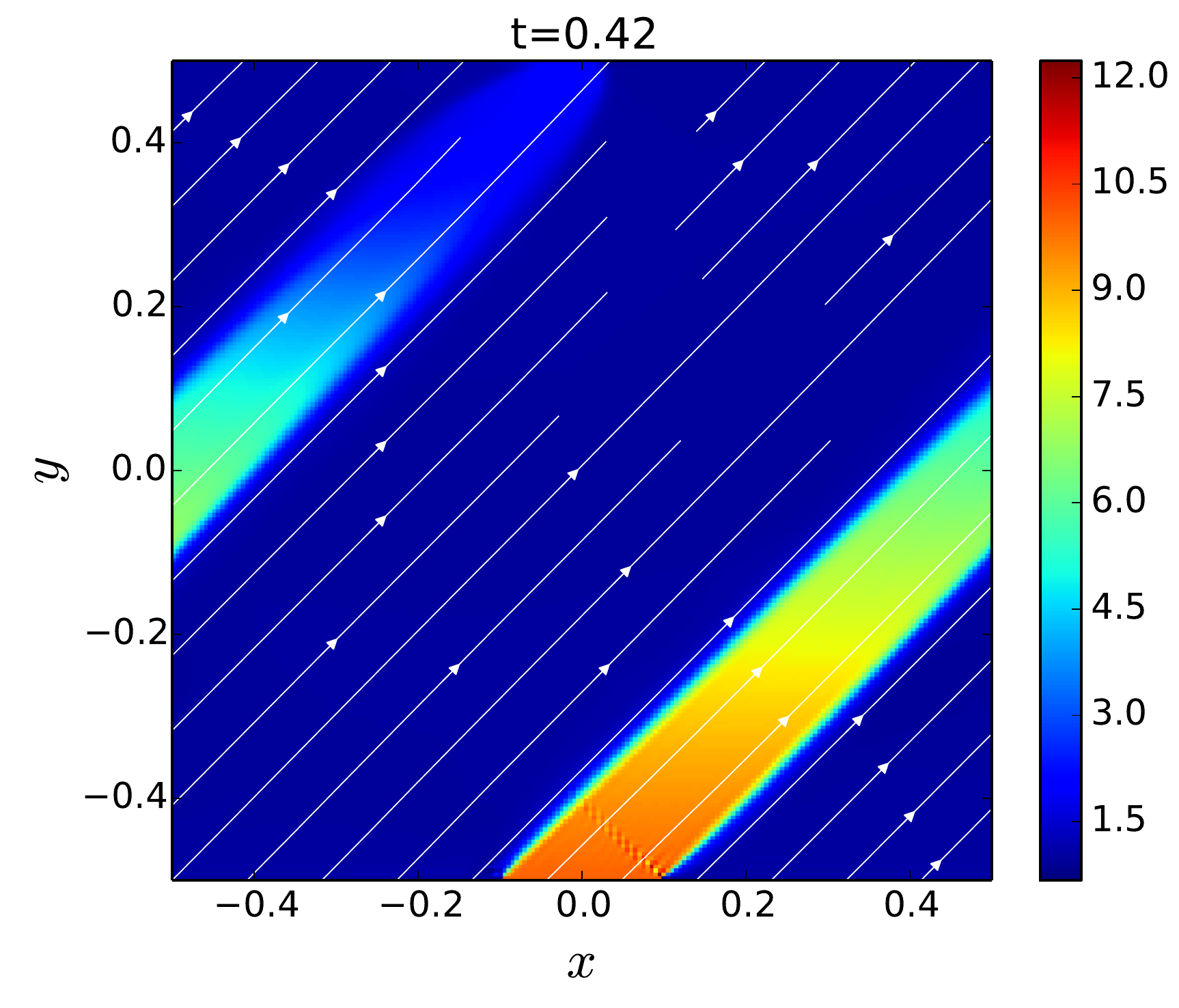}  
\caption{Temperature and magnetic field lines in the Thermal Conduction Test at the times $t=0.01 $, $t=0.12$ and $t=0.42$.}
 \label{fig_Thermic}
\end{figure*}

\subsection{Test 6: Cloud Shock Interaction} \label{subsec:_test_cloud}

The third 2D test, is very interesting from the astrophysical point of view, since it describes the disruption of a cloud with high density with a strong shock wave. It was proposed by \cite{Woodward1998} and
has been widely used to check the algorithms behavior with superfast flows. Here we have implemented the version of \cite{CT_toth}, given by the initial state  $B_x = 0 = v_y =v_z = 0$, 
\begin{eqnarray}
\rho &=& \left\lbrace \begin{array}{ll}
3.86859 & \quad \text{for} \quad x<0.6 \, ,  \\
1 & \quad \text{for} \quad x>0.6  \, , \\
10 & \quad \text{for} \quad r<0.15 \, ,
\end{array} \right. 
\end{eqnarray}
\begin{eqnarray}
v_x &=& \left\lbrace \begin{array}{ll}
0 & \quad \text{for} \quad x<0.6 \, , \\
-11.2536 & \quad \text{for} \quad x>0.6 \, , \\
\end{array} \right. 
\end{eqnarray}
\begin{eqnarray}
p &=& \left\lbrace \begin{array}{ll}
167.345 & \quad \text{for} \quad x<0.6 \, , \\
1 & \quad \text{for} \quad x>0.6 \, , \\ 
\end{array}\right. 
\end{eqnarray}
\begin{eqnarray}
B_y &=& \left\lbrace \begin{array}{ll}
2.1826182 & \quad \text{for} \quad x<0.6 \, , \\
0.56418958 & \quad \text{for} \quad x>0.6 \, , \\ 
\end{array}\right. 
\end{eqnarray}
\begin{eqnarray}
B_z &=& \left\lbrace \begin{array}{ll}
-2.1826182 & \quad \text{for} \quad x<0.6 \, , \\
0.56418958 & \quad \text{for} \quad x>0.6 \, , \\ 
\end{array}\right. 
\end{eqnarray}
where the cloud is centered in $x=0.8$, $y=0.5$ and has a radius $0.15$. The  discontinuity consists in a fast shock wave and a rotational discontinuity in the $B_z$ component.  For the results presented here, we show the numerical information in table \ref{tabla_info} and outflow boundary conditions in all the faces. In the figure \ref{fig_cloud}, we plot the color map of the density with the magnetic field lines for the times  $t=0.03$ and $t=0.06$. The maximum violation in the divergence of the magnetic field is plotted in figure \ref{fig_DivBmax}(c).

\subsection{Test 7: Magnetic Reconnection} \label{subsec:reconnection}

In order to evaluate the implementation of the resistive terms in the algorithms, we reproduced a 2D test taken from  \cite{ResistiveCode_Jiang}. It consists in an application for the magnetic reconnection, given by an initial state at rest, $v_x = v_y = v_z = 0$, with a density and pressure with constant values $\rho=1$ and $p=0.1$, and the magnetic field $B_x = 0$,
\begin{eqnarray}
B_y &=& \left\lbrace \begin{array}{ll}
-1 & \ \text{for} \quad x< -L_r \, , \\
\sin (\pi x L_r /2  ) & \ \text{for} \quad |x| \leq -L_r \, , \\
1 & \ \text{for} \quad x> L_r \, ,
\end{array} \right. 
\end{eqnarray}
\begin{eqnarray}
B_z &=& \left\lbrace \begin{array}{ll}
0 & \ \text{for} \quad x< -L_r \, , \\
\cos (\pi x  L_r /2) & \ \text{for} \quad |x| \leq -L_r \, , \\
1 & \ \text{for} \quad x> L_r \, ,
\end{array} \right.
\end{eqnarray}
where $L_r = 0.05$ is a characteristic length of the region with resistivity, which is modeled as
\begin{eqnarray}
\eta = \frac{\eta_0}{4} (\cos(10\pi x)+1)(\cos(40\pi y)+1) \, ,
\end{eqnarray}
with an amplitude $\eta_0=0.05$ in the domain $(-L_r, L_r)\times(-4L_r, 4L_r)\times (0.0, 1.0)$. For the simulation, it was used the information listed in table \ref{tabla_info} and outflow boundary conditions everywhere. In figure \ref{fig_reconnection}, we plot a density map of the thermal pressure with the magnetic field lines at times $t=0.2$, $t=0.5$ and $t=1.5$, where the phenomena of magnetic reconnection due to localized resistivity can be tracked in time. In early times the phenomena is hardly noticeable but it is intensified later. To evaluate the reconnection rate, we plot in figure \ref{fig_rec_rate} the ratio between the inflow velocity $v_{\text{in}}$ and the Alfv\'en speed, that is $v_\text{in}/c_a$, which is the reconnection rate \cite{Petschek1964}, calculated along the line at $ y=2.5, |x|<0.25, z = 0 $.  Here it was found that the rate tends to have a maximum value around $t=1$ as the resistivity is increased. On the other hand, once the system reaches a steady state, we can see that the bigger the resistivity, the bigger the reconnection rate. As it is expected, without resistivity there is no reconnection rate (solid black line). However, as the magnitude of the resistivity increases, the rate is fairly the same. Finally, to verify the violation of the Gauss's Law for the magnetic field, we plot the maximum divergence of the magnetic field in figure \ref{fig_DivBmax}(d).

\begin{figure*}
\centering
\includegraphics[height=0.23\textheight]{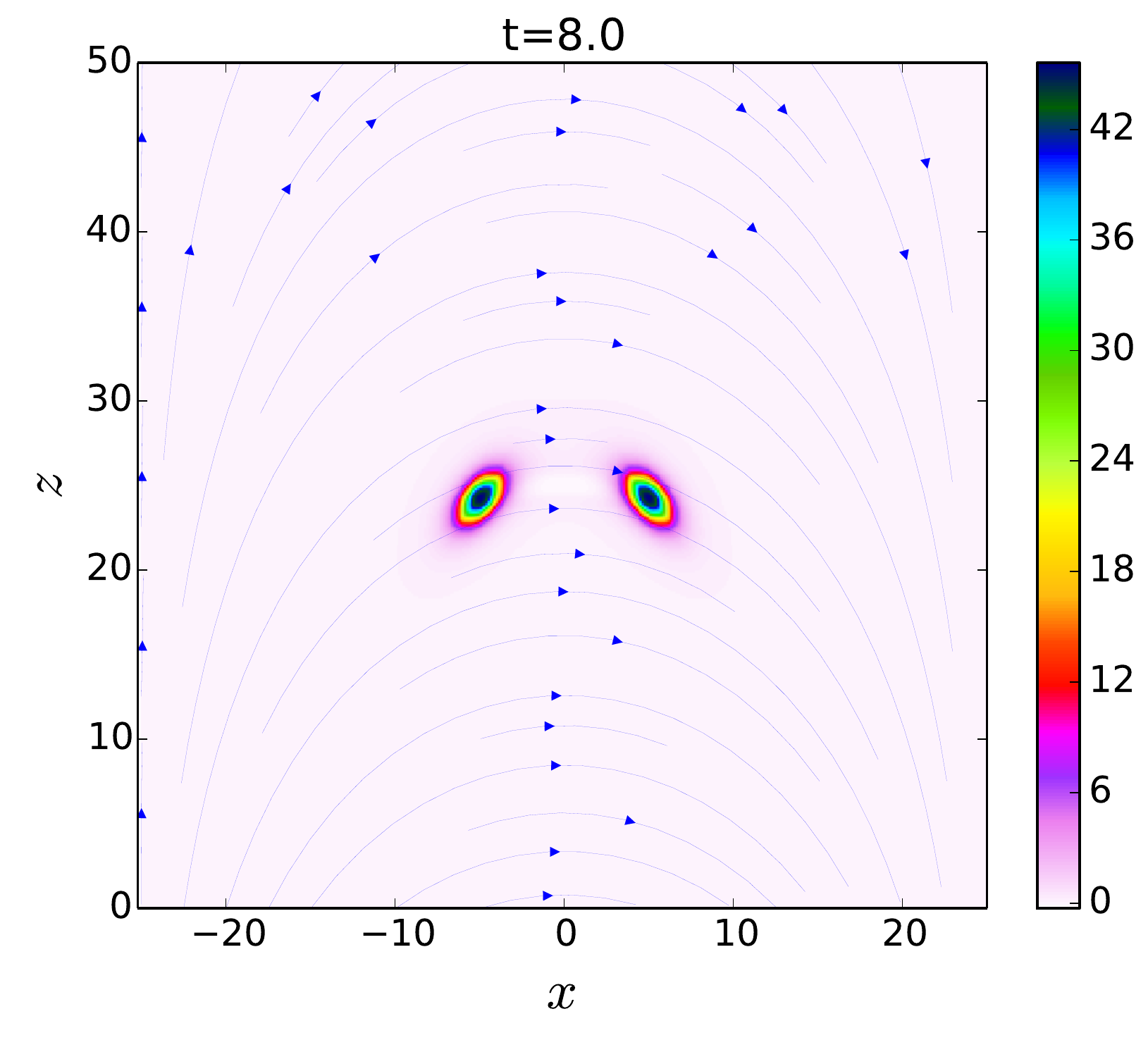} 
\includegraphics[height=0.23\textheight]{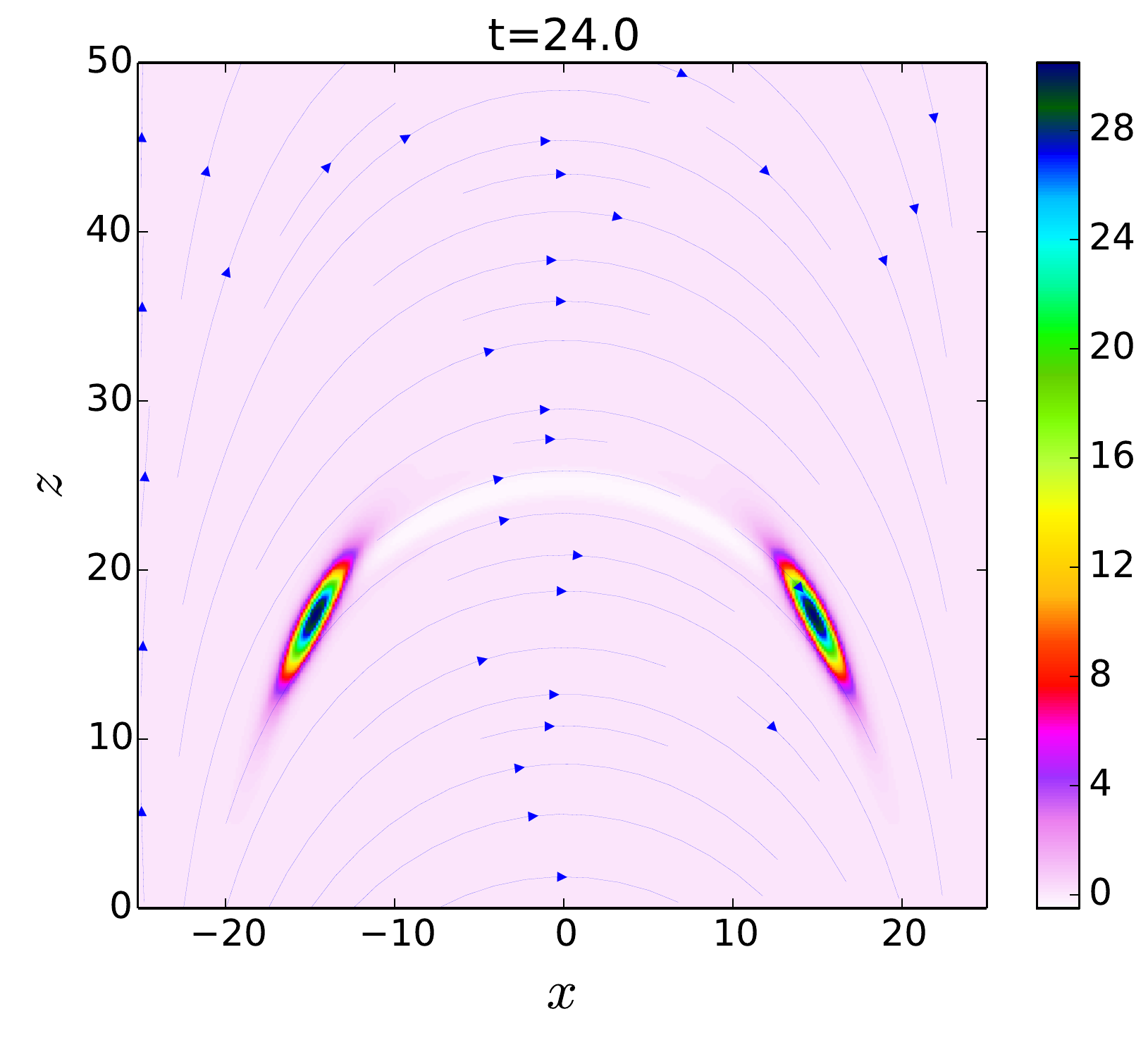}
\includegraphics[height=0.23\textheight]{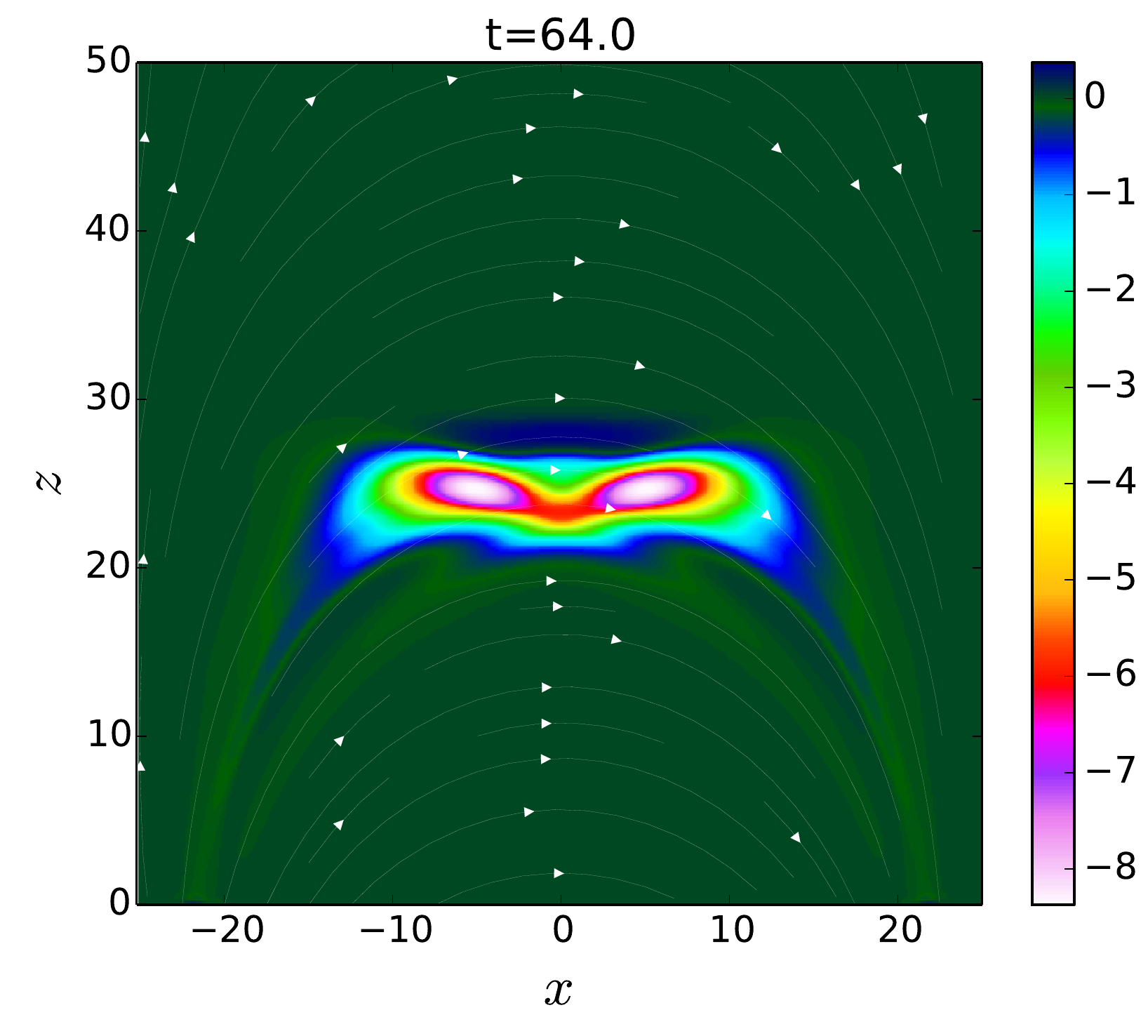} 
\includegraphics[height=0.23\textheight]{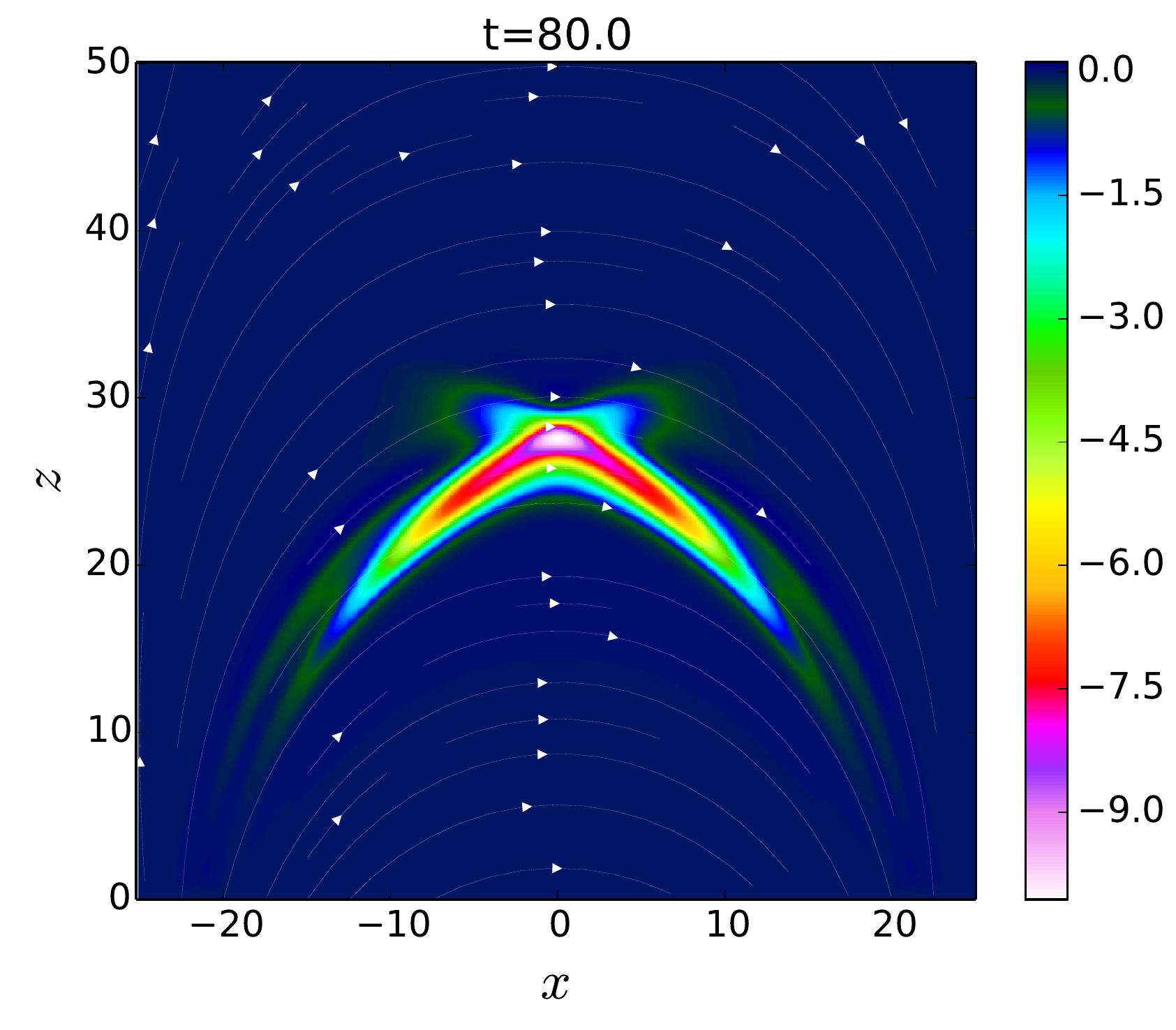} 
\caption{ Transverse velocity $v_y$ (km/s) and magnetic field lines in the Transverse Oscillations of the solar loops problem  at the times $t=8$s, $t=24$s, $t=64$s and $t=128$s.} \label{fig_transverse}
\end{figure*}

\subsection{Test 8: Thermal Conduction} \label{subsec:thermic test}

For the evaluation of the heat flux model \eqref{heat2}, we have implemented the thermal conduction evolution proposed by \cite{JiangFangChen2012}, where a small hot region in the bottom of the domain is advected by the magnetic field lines.  The initial state has a constant pressure $p = 0.1$, is at rest $v_x = v_y = v_z = 0$, and with the magnetic field $B_x = B_y = 1 $, $B_z = 0$, which is oriented forming a $45^\circ$ angle. The density is 
\begin{eqnarray}
\rho &=& \left\lbrace \begin{array}{ll}
0.01 \ \text{for} \ |x|<0.1 \ \text{and} \ y=-0.5 \, , \\
0.1 \quad \text{elsewhere}  \ , 
\end{array} \right.  
\end{eqnarray}
and describes a small hot region located in the bottom. The information for the simulation is given in table \ref{tabla_info}. The boundary conditions are periodic at $|x|=0.5$, inflow in the bottom surface $y=-0.5$ and outflow in the rest. The results are presented in figure \ref{fig_Thermic}, where the temperature is depicted in times $t=0.01 $, $t=0.12$ and $t=0.4$. It can be seen that the heat flux advects the temperature gradient along the diagonal magnetic field lines. In addition, the maximum violation in the divergence of the magnetic field is plotted in figure \ref{fig_DivBmax}(e).

\subsection{Test 9: 2D - Transverse Oscillations in Solar Coronal Loops}  \label{subsec: transverse}

Here we present the evolution of MHD wave propagation in a solar phenomenon, in order to evaluate if the code is able to reproduce this type of simulations performed by other authors. Following \cite{transverse_waves}, we test the propagation of Alfv\'enic pulses in the solar coronal arcades with the initial magnetic field $B_y = 0$,   
\begin{eqnarray}
B_x &=& B_0 \cos{(kx)}\exp{(-kz)} \, ,  \label{Transverse_Bx} \\
B_z &=& -B_0 \sin{(kx)}\exp{(-kz)} \, ,  \label{Transverse_By} 
\end{eqnarray}
where $B_0 = 40 $ G is the magnitude of the magnetic field at the photospheric level in the footpoints $x=\pm L/2$. The angular frequency is $k=\pi/L$ and $L$ is a characteristic length, $L = 50 $ Mm $=50\times 10^3 $ km. On the other hand,  the equation of hydrostatic equilibrium is given by 
\begin{eqnarray}
\frac{\mathrm{d}p}{\mathrm{d}z} + \rho g = 0 \, , \label{eq_equilibrio}
\end{eqnarray}
and through the equation of state, the pressure can be written as $
p = 2 k_\beta \rho T/m_p $, where $m_p$ is the mass of the proton and $k_\beta$ is the Boltzmann constant.  The pressure can be solved from \eqref{eq_equilibrio} 
\begin{eqnarray}
p(z) = p(z_0) \exp{\left[  -\frac{m_p g}{2 k_B} \int_{z_0}^z \frac{\mathrm{d}z'}{T(z')}  \right]} \, , 
\end{eqnarray}
where the temperature profile from the photosphere to the corona is modeled  by the step function
\begin{eqnarray}
T(z) = \frac{1}{2} (T_\text{cor} + T_\text{phot}) + \frac{1}{2}(T_\text{cor} - T_\text{phot})\tanh{\left(\frac{z-z_t}{z_w}\right)} \, , \nonumber \\
\end{eqnarray}
which locates the transition region at a height of $z_t = 2 $ Mm and with a width of $z_w = 0.2 $ Mm. The temperatures in the corona and photosphere are $T_{\text{cor}} = 1.2\times 10^6 $ K and $T_{\text{phot}} = 6000$ K. The value of the density in the base of the corona is $\rho = 10^{-15}$ g cm$^{-3}$ and the magnitude of the gravity acceleration is $g=274$ m/s. The fluid is at rest initially $v_x = v_z = 0$, but its transverse velocity is perturbed by a pulse located in $x=0$ and $z=L/2$ as follows
\begin{eqnarray}
v_y = \frac{0.1 v_0}{1 + \left\lbrace \left[ x^2 + (z-L/2)^2 \right]^2/r_0^4 \right\rbrace } \, ,
\end{eqnarray}
where $r_0 = 1 $ Mm and  the Alfv\'en speed at the corona is $v_0 = 1 $ Mm/s.

\begin{figure*}
\centering
\includegraphics[height=0.26\textheight]{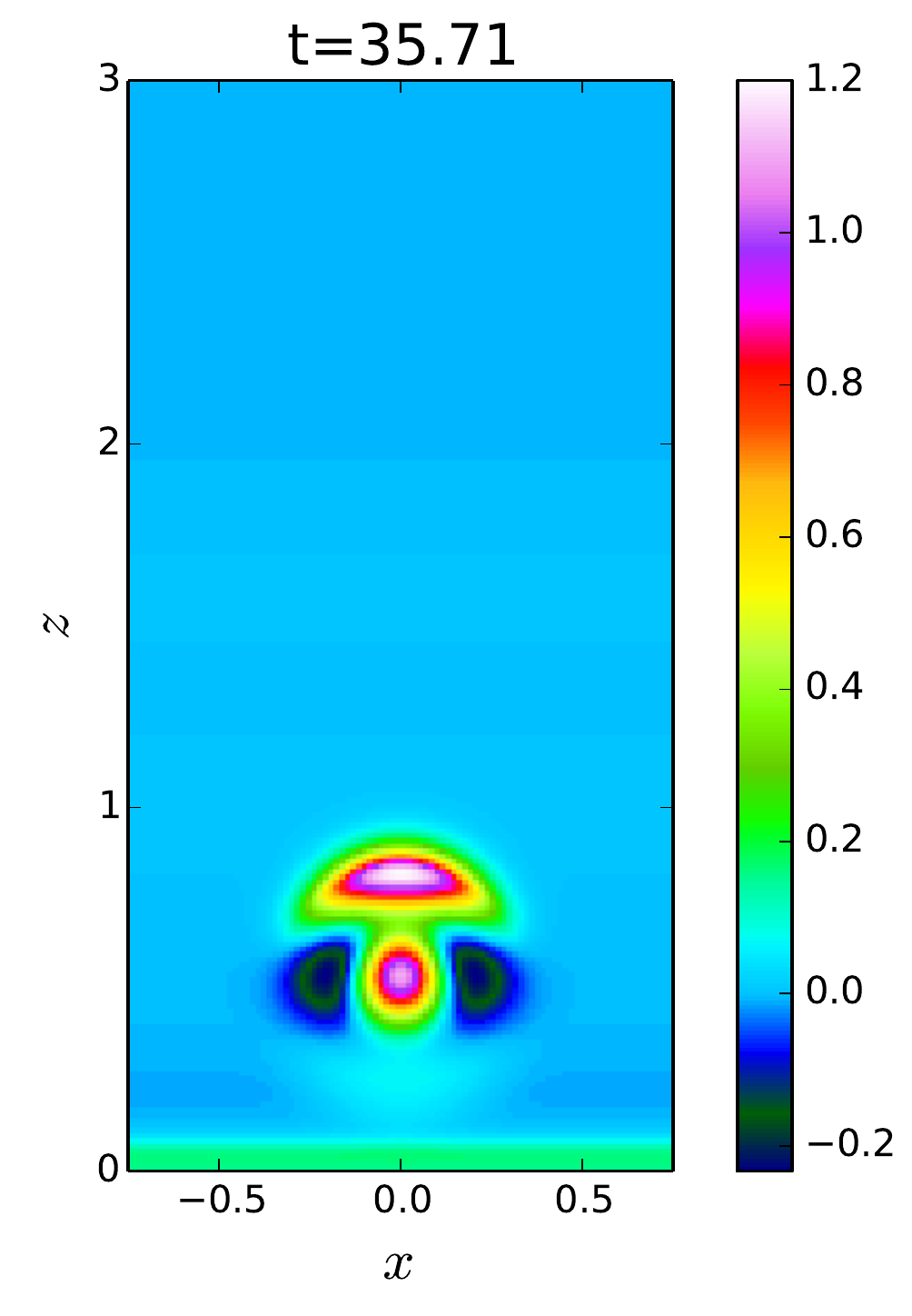}
\includegraphics[height=0.26\textheight]{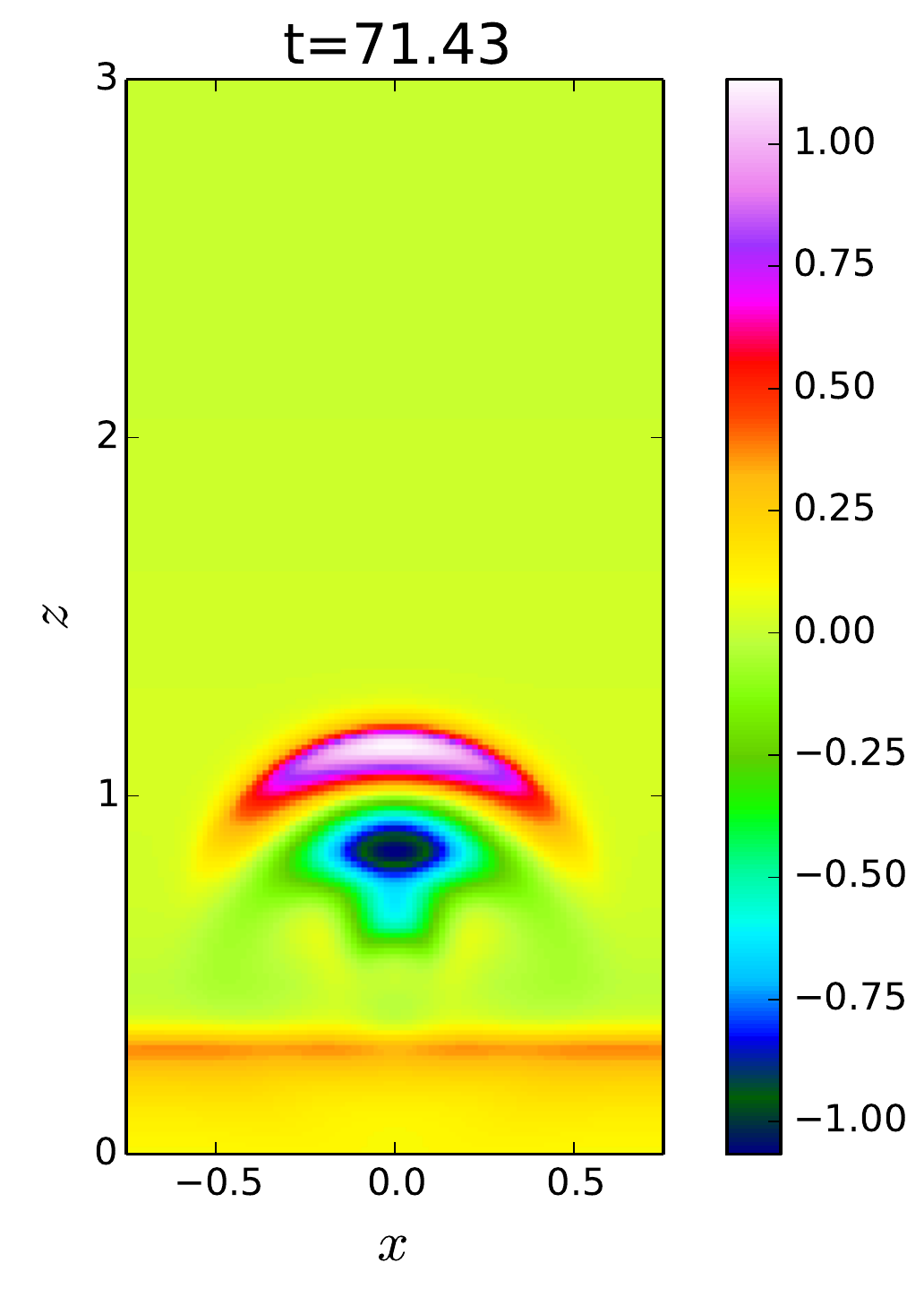}
\includegraphics[height=0.26\textheight]{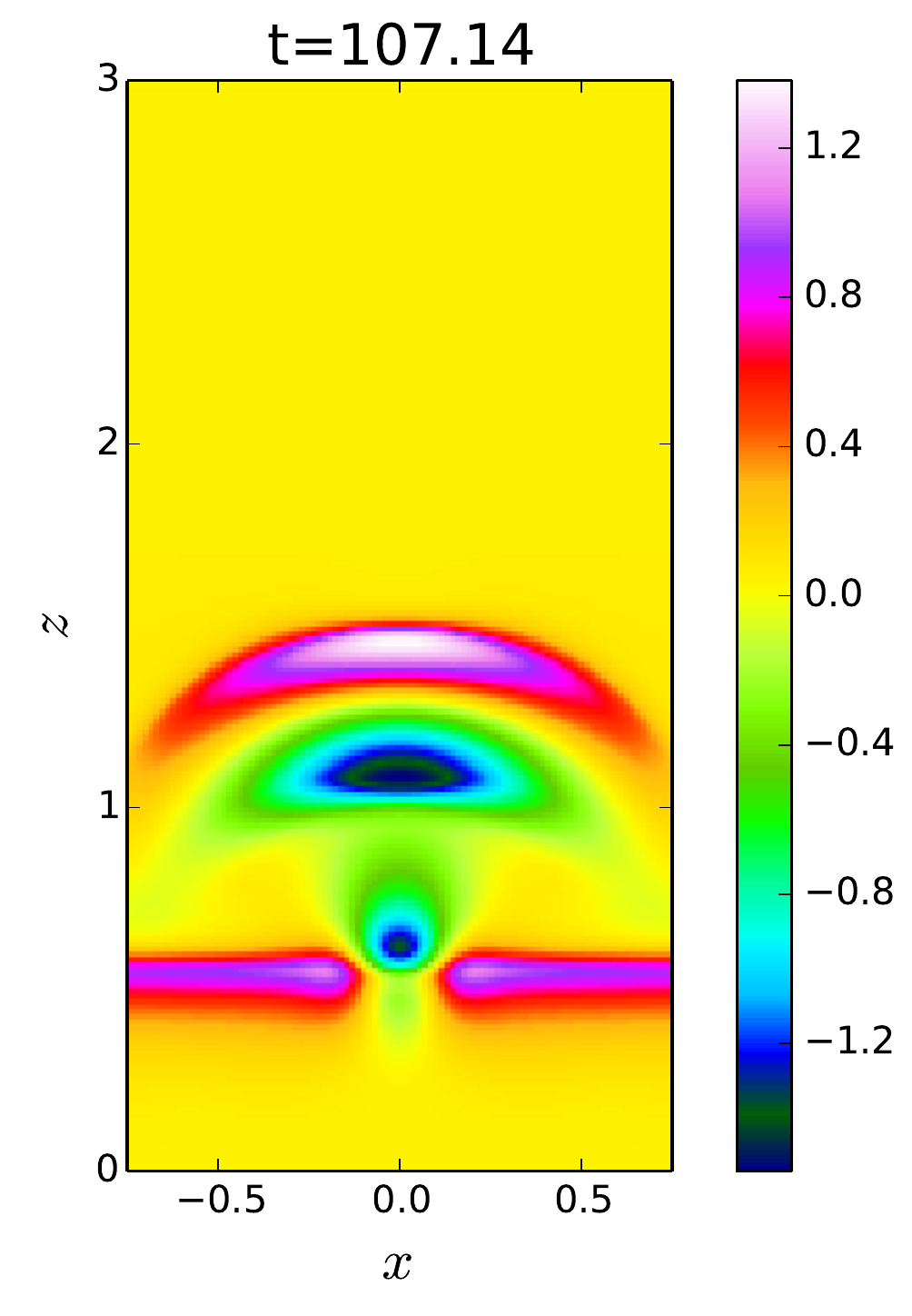}
\includegraphics[height=0.26\textheight]{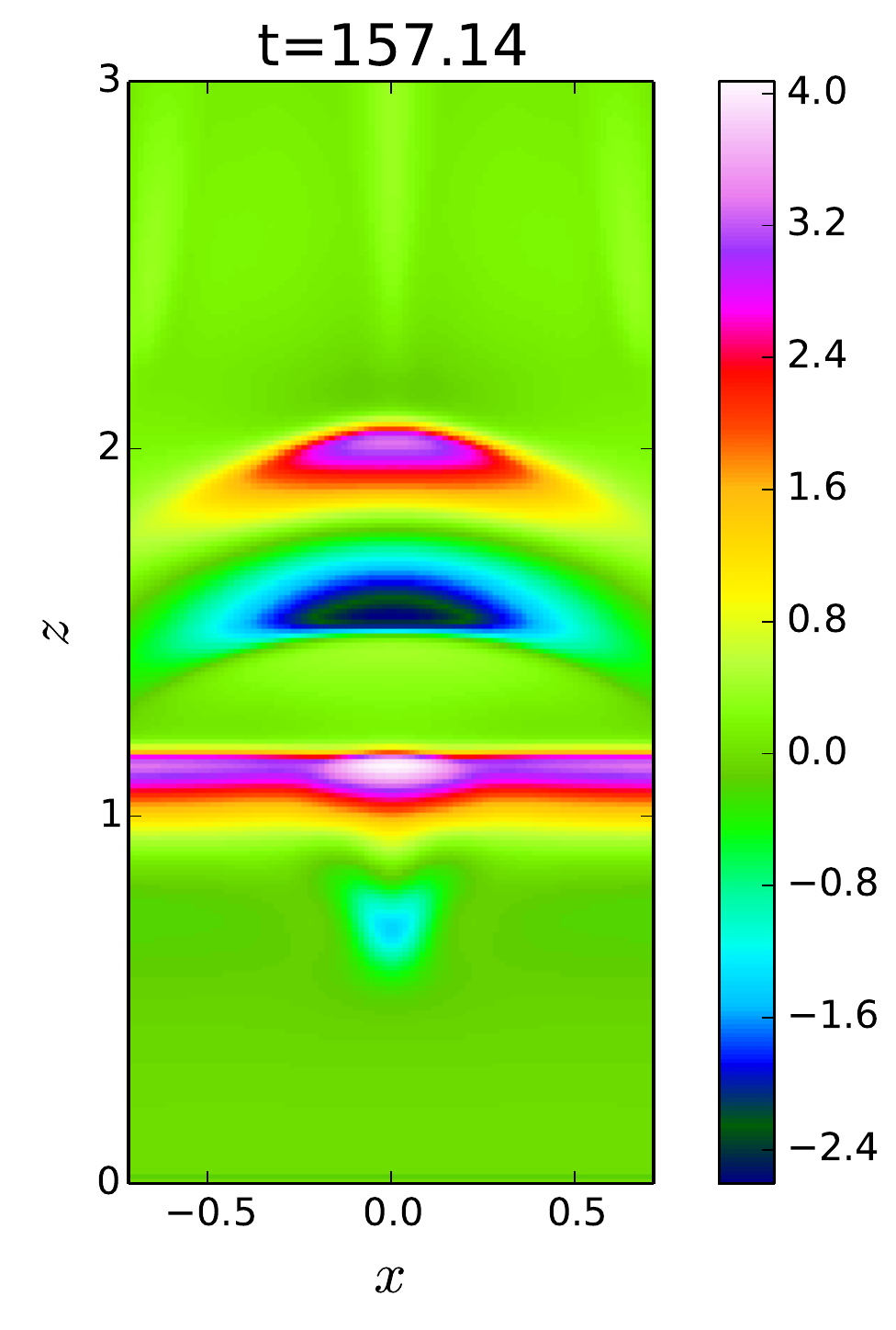}
\caption{ Vertical velocity $v_z$ in the magnetohydrodynamic-gravity waves problem at the times $t=35.71$ s, $t=71.43$ s, $t=107.14$ s and $t=157.14$ s in the plane $y=0$.}
 \label{fig_gravity_waves}
\end{figure*}

For the simulation, we use the information listed in table \ref{tabla_info} and outflow boundary conditions were implemented in all the faces. In figure \ref{fig_transverse}, we present the transverse velocity $v_y$ (km/s) and the magnetic field lines at times $t=8$ s, $t=24$ s, $t=64$ s and $t=80$ s. In this plots, we can follow the propagation of the twin Alfv\'enic pulses along the magnetic field lines and the fast and slow magneto-acoustic waves in the outer and inner arcs. Furthermore, we observe the reflection with the bottom surface as was described in \cite{transverse_waves}.  The maximum violation of $\nabla \cdot \vec{B}$ (in Tesla/km) is plotted in figure \ref{fig_DivBmax}(f), where we can see how its initial value is not absolute zero, since it is calculated with the approximation of finite differences for the initial values of the magnetic fields \eqref{Transverse_Bx} and \eqref{Transverse_By}.

\subsection{Test 10: 3D - Gravity Waves}\label{subsec:gravitywaves}

Until this point we have only shown evolutions in 1D and 2D, therefore we present here the 3D simulation of gravity waves in the solar atmosphere, which was performed using the Flash code \citep{flash_code} by \cite{gravity_waves}. The initial state is given by a constant magnetic field in the $z-$ direction 
\begin{eqnarray}
\vec{B} = B_0 \hat{e}_z \, ,
\end{eqnarray}
with a magnitude of $B_0 = 23$ G. The equation of equilibrium is given by \eqref{eq_equilibrio} and through the equation of state, for this particular configuration, the pressure can be written as $p = k_\beta \rho T/(1.24 m_p)$. In a similar way as the Transverse Oscillations test, the pressure can be solved integrating the temperature
\begin{eqnarray}
p(z) = p(z_0) \exp{\left[  -\frac{m_p g}{k_B} \int_{z_0}^z \frac{\mathrm{d}z'}{T(z')}  \right]} \, , 
\end{eqnarray}
where $p_0$ is the pressure at the level of reference in $z_0=10$ Mm at the solar corona. Assuming a temperature profile similar to VAL-IIC \citep{Vernazza_temperature_profile}, the pressure and density can be obtained for the equilibrium state. 
The perturbation of the velocity is a vertical Gaussian pulse 
\begin{eqnarray}
v_z = A_v \exp( [x^2 + y^2 + (z-z_0)^2] / \omega^2) \, ,
\end{eqnarray}
with amplitude $A_v$ = 3 km/s,  width $\omega = 100$ km and is located in $x=0$ and $z_0 = 500$ km. 

\begin{figure*}
  \centering
  \begin{tabular}[b]{@{}p{0.3\textwidth}@{}}
    \centering\includegraphics[height=0.15\textheight]{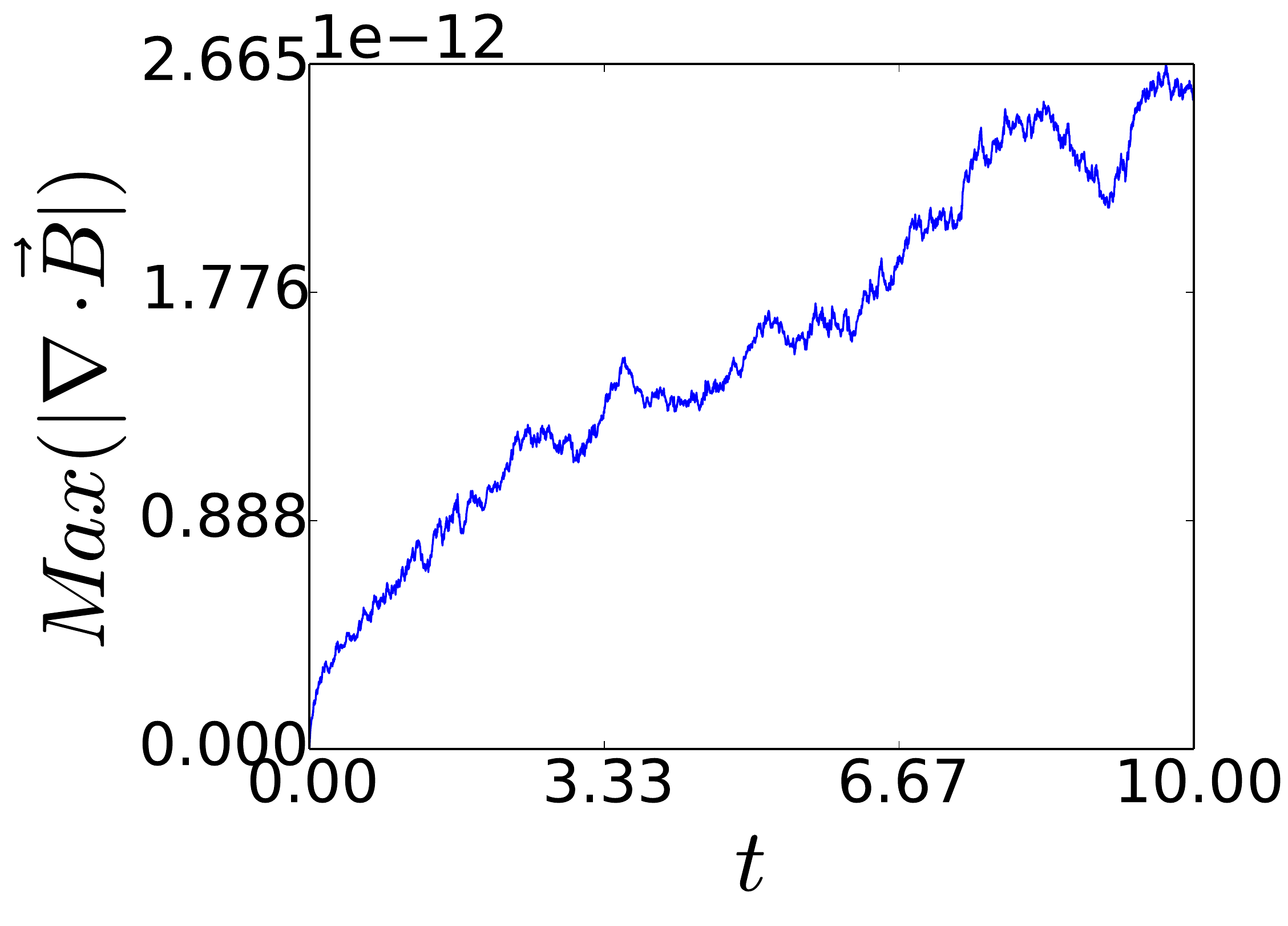}  \\
    \centering\small (a) Test 4: Current Sheet 
  \end{tabular} \begin{tabular}[b]{@{}p{0.3\textwidth}@{}}
    \centering\includegraphics[height=0.15\textheight]{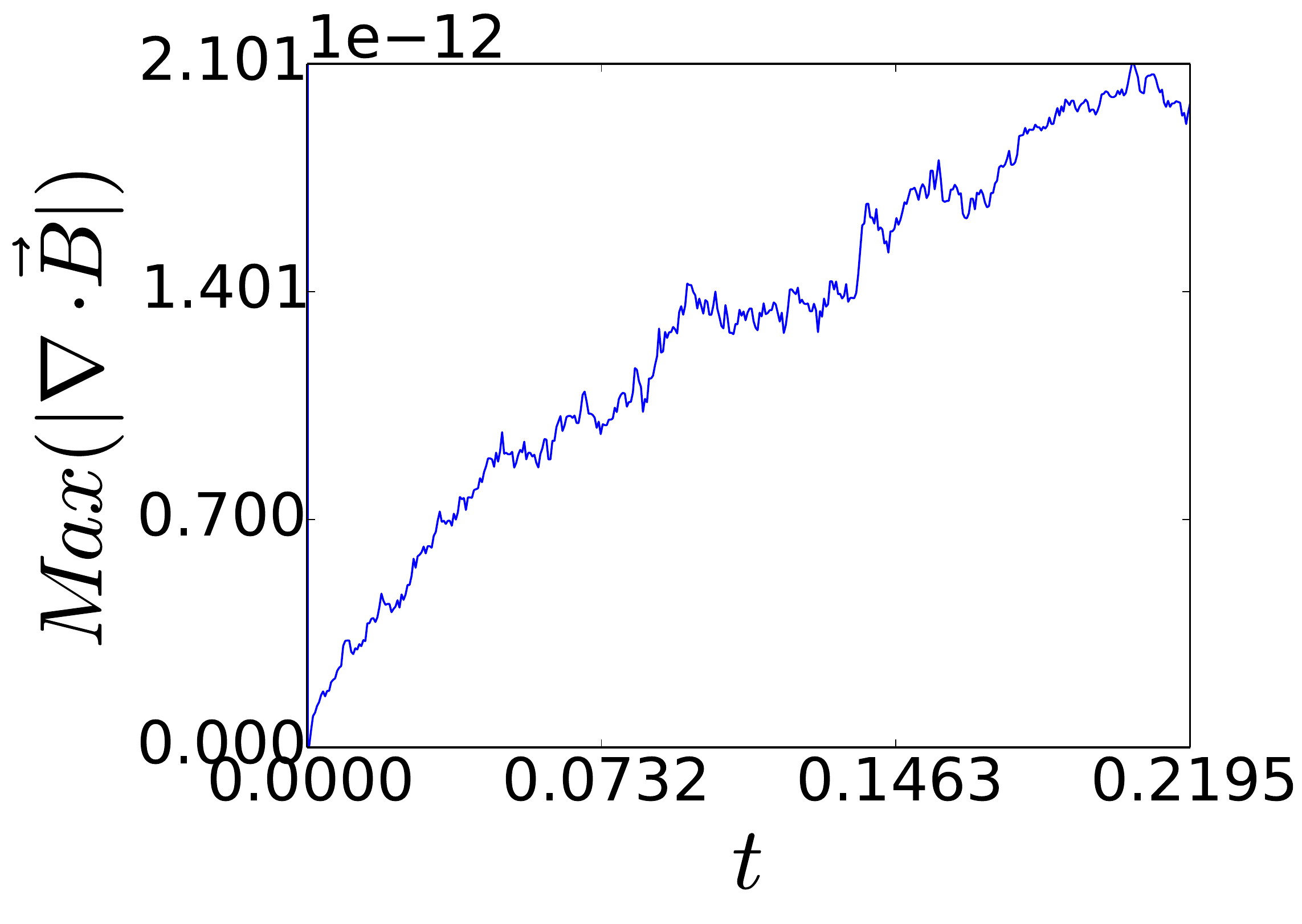} \\
    \centering\small (b) Test 5: MHD Rotor
  \end{tabular}  \begin{tabular}[b]{@{}p{0.3\textwidth}@{}}
    \centering\includegraphics[height=0.15\textheight]{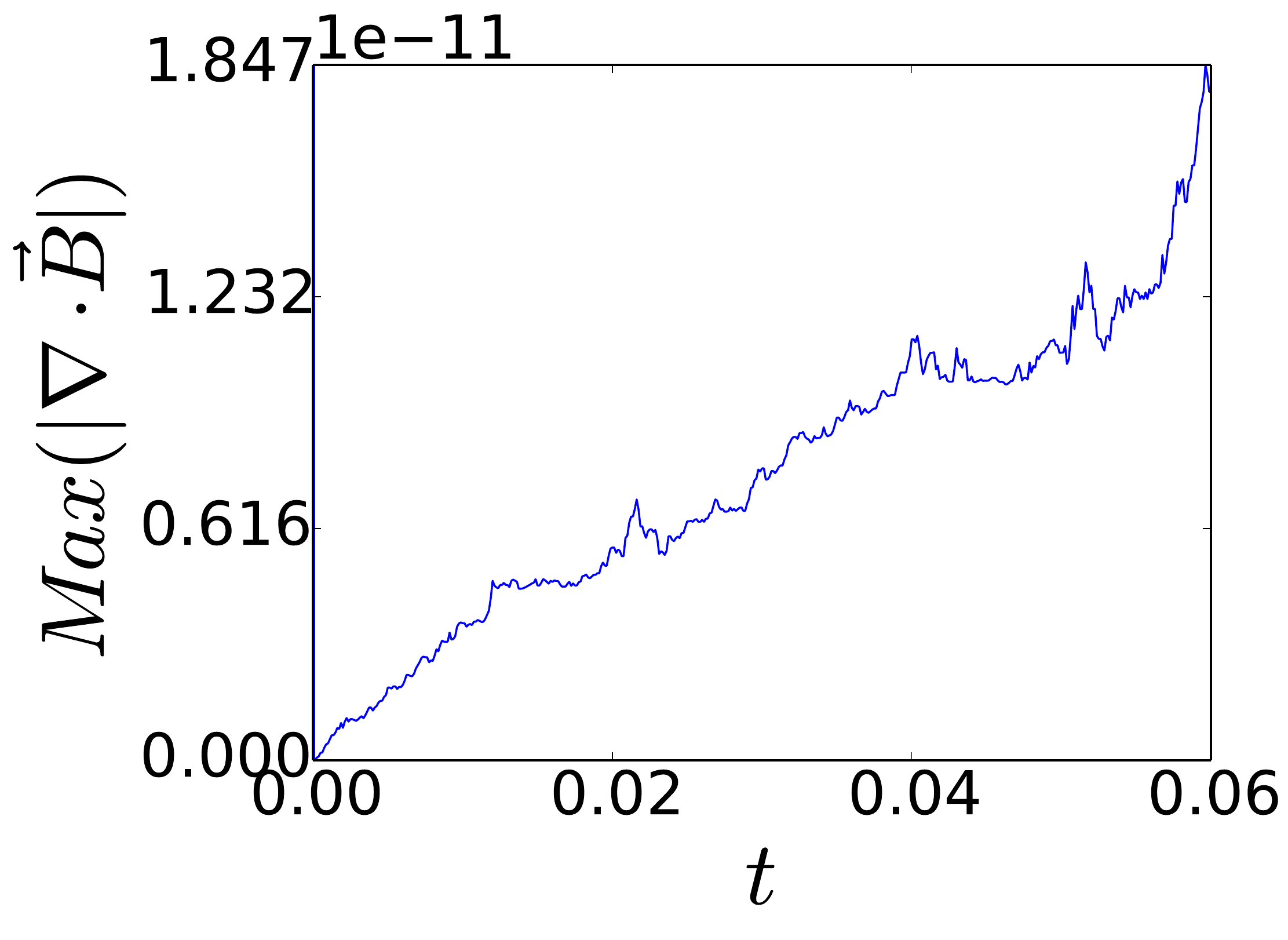}  \\
    \centering\small (c) Test 6:  Cloud Shock Interaction 
  \end{tabular}
  
  \bigskip
    \bigskip

   \begin{tabular}[b]{@{}p{0.3\textwidth}@{}}
    \centering\includegraphics[height=0.15\textheight]{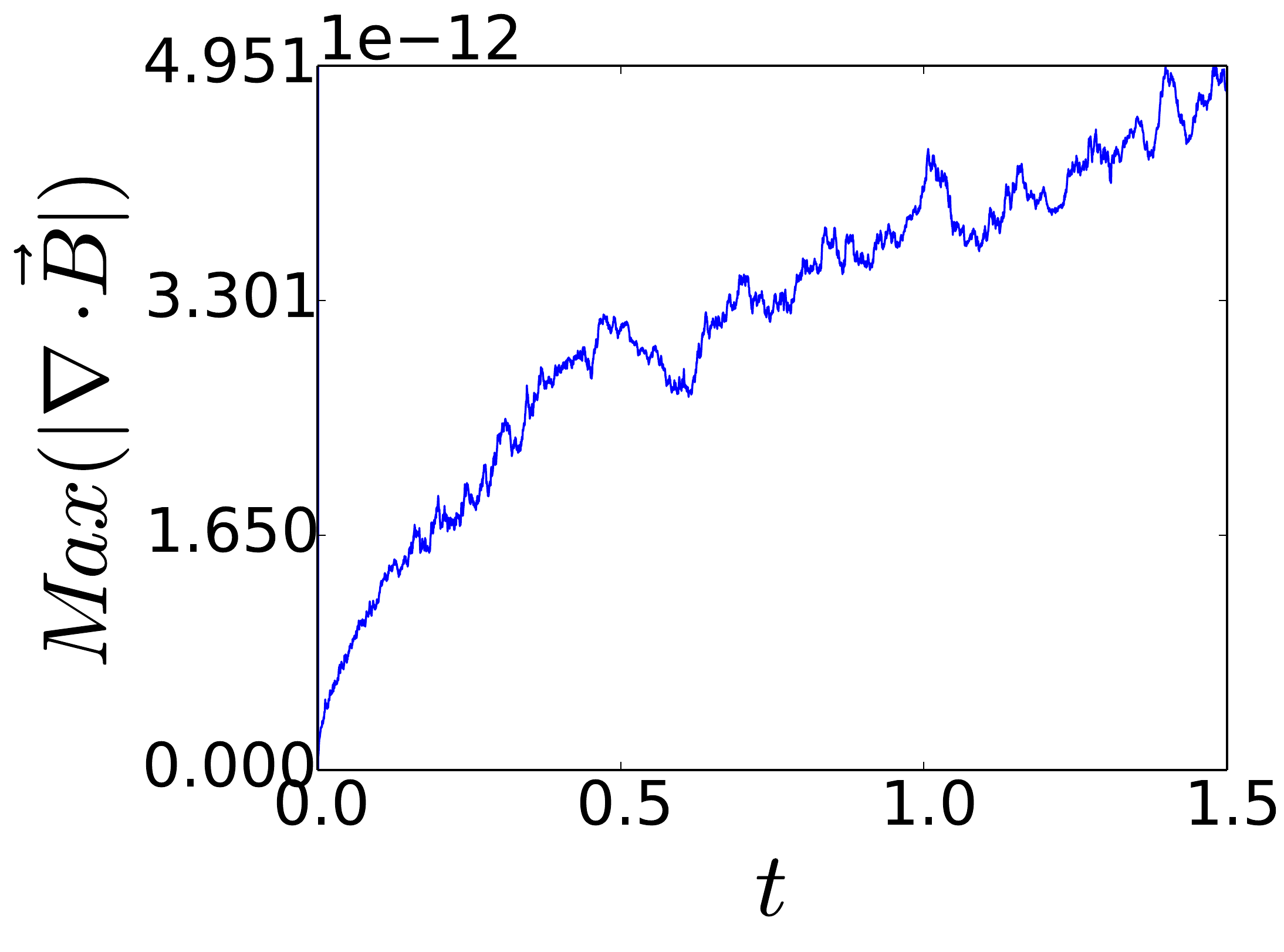}  \\
    \centering\small (d) Test 7: Magnetic Reconnection 
  \end{tabular}\begin{tabular}[b]{@{}p{0.3\textwidth}@{}}
    \centering\includegraphics[height=0.15\textheight]{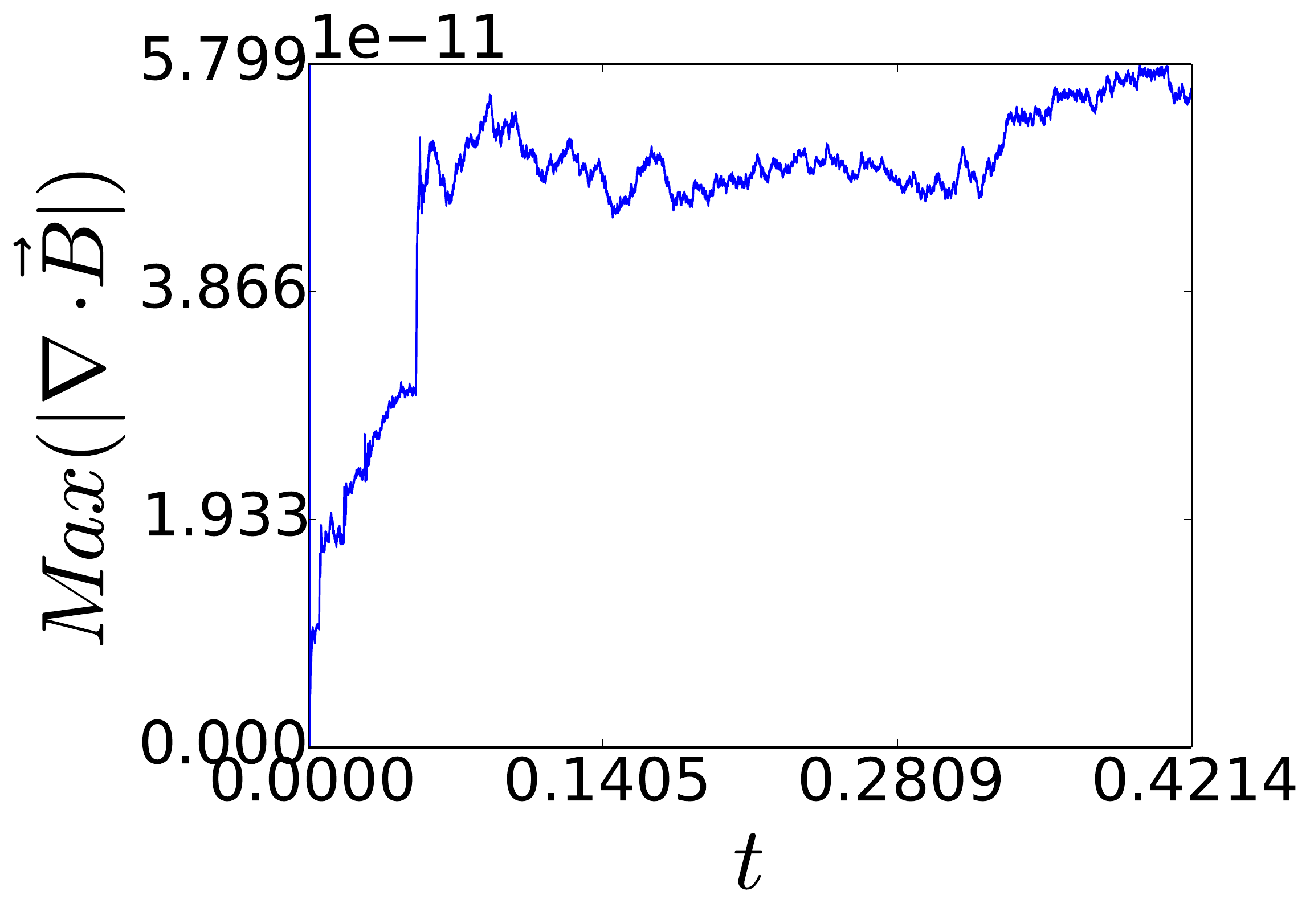}  \\
    \centering\small (e) Test 8: Thermal Conduction 
  \end{tabular}

  \bigskip
    \bigskip

  \begin{tabular}[b]{@{}p{0.3\textwidth}@{}}
    \centering\includegraphics[height=0.15\textheight]{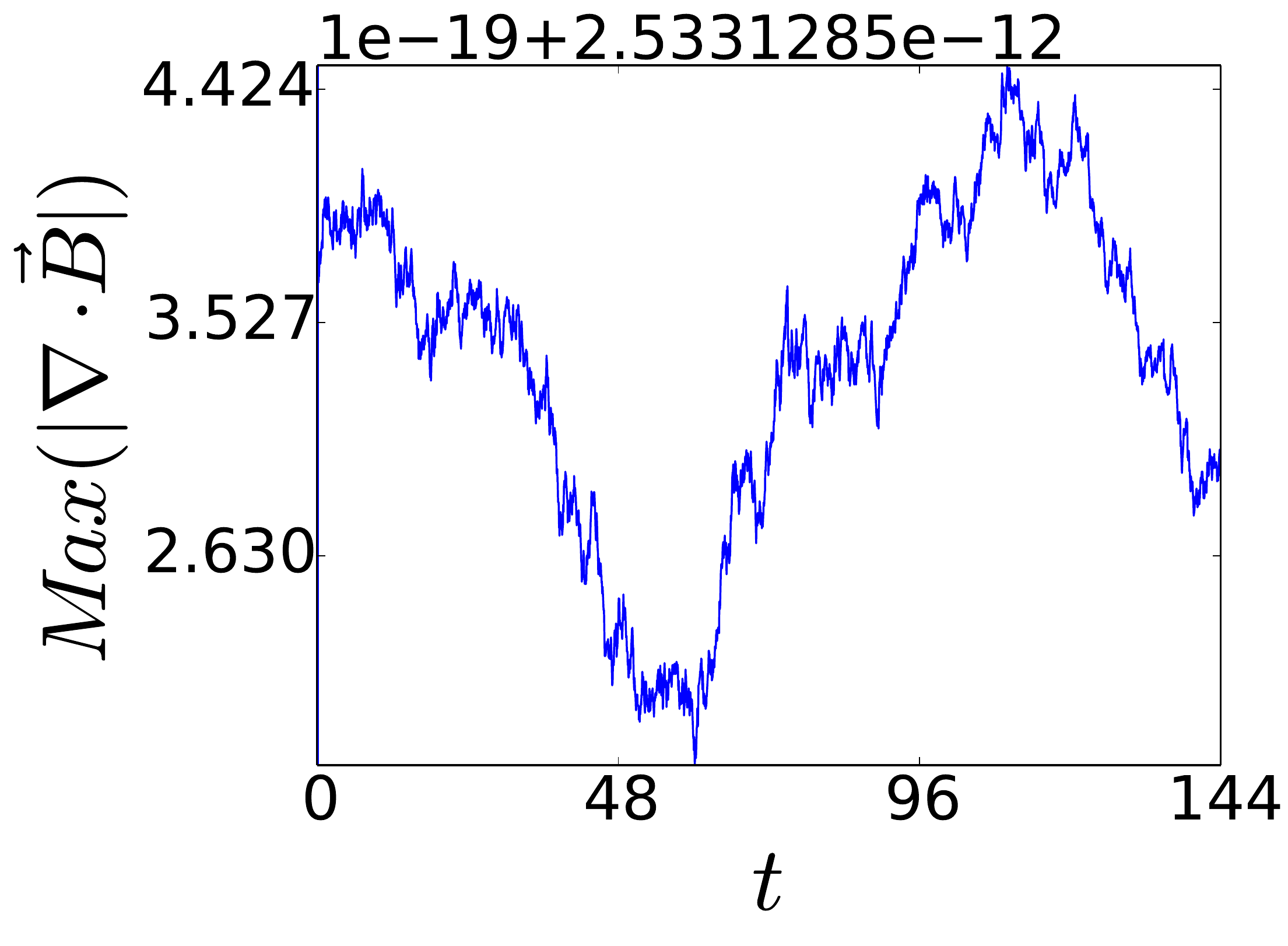}  \\
    \centering\small (f) Test 9: Transverse Oscillations (Tesla/km)
  \end{tabular}    \begin{tabular}[b]{@{}p{0.3\textwidth}@{}}
   \centering\includegraphics[height=0.15\textheight]{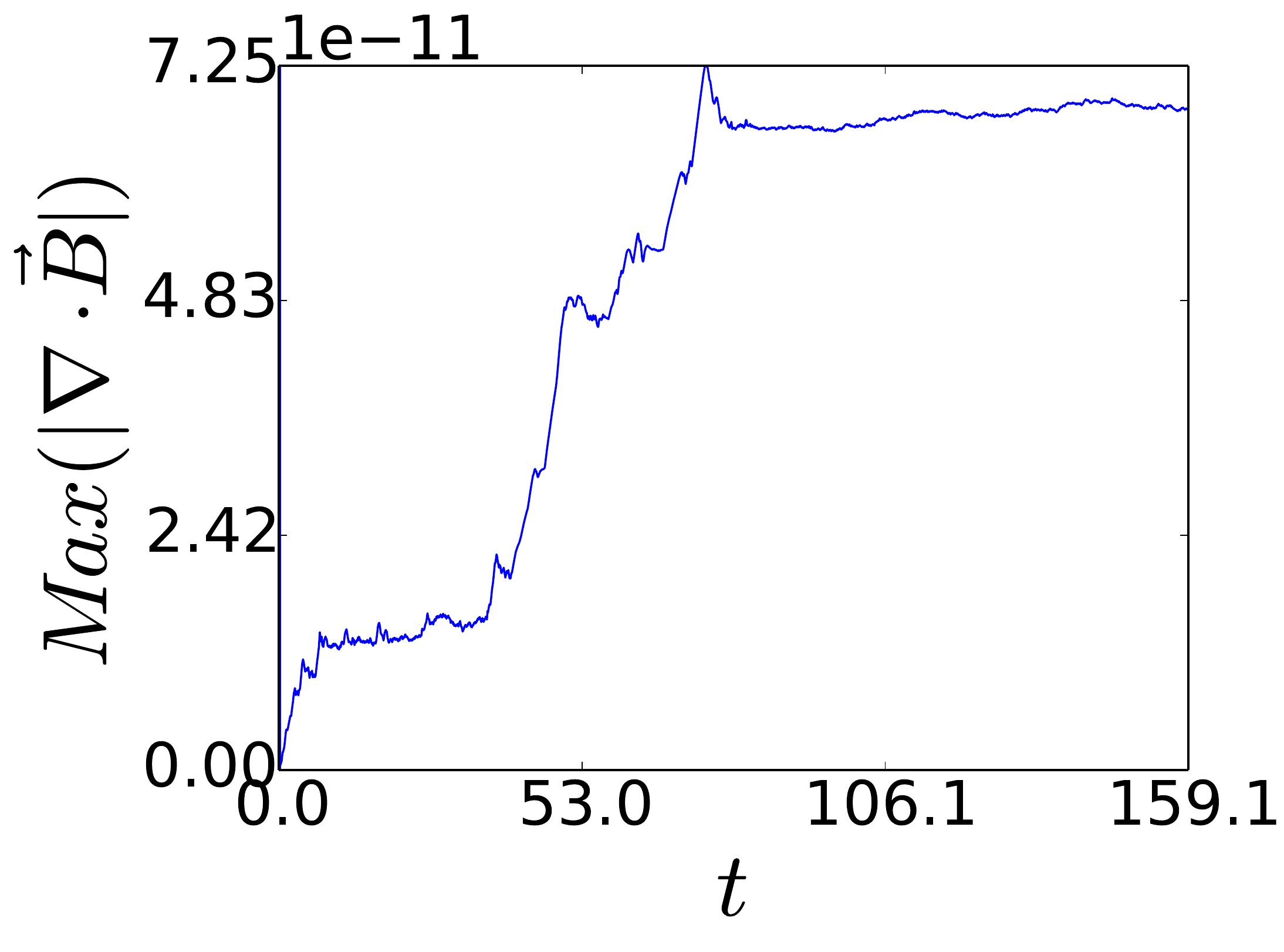}  \\
    \centering\small (g) Test 10: Gravity Waves (Tesla/km)
  \end{tabular}  
   
  \caption{Maximum Divergence of the magnetic field $\nabla \cdot \vec{B}$ in function of the time for each of the 2D and 3D tests \label{fig_DivBmax}}
\end{figure*}

For the simulation, we use the information given by table \ref{tabla_info} with boundary conditions fixed in time at $z=-0.25$ Mm, $z=5.75$ Mm, and outflow in the other faces. In figure \ref{fig_gravity_waves}, we plot the vertical velocity $v_z$ at times $t=35.71$ s, $t=71.43$ s, $t=107.14$ s and $t=157.14$ s in the plane $y=0$. In this plots, we can track the propagation of the longitudinal magnetoacoustic-gravity waves caused by the initial perturbation in the vertical velocity $v_z$. The fast and slow modes are coupled and expanded in a quasi-isotropic way across the z-direction. Finally, the maximum violation in the divergence of the magnetic field (in Tesla/km) is plotted in figure \ref{fig_DivBmax}(g).

\section{Discussion and conclusions}
\label{sec:conclusions}

We have presented in this work a resistive MHD code with heat flux terms. Here we have summarized how the numerical methods were implemented, emphasizing in the HLLC Riemann solver and the slight modification of the Flux Constrained Transport method, insights of the inclusion of the resistivity. We have shown that this code is able to reproduce the standard MHD tests in 1D and 2D since we have obtained the expected behavior of the solutions in all the cases. The resistivity and heat conduction were verified in an independent way and proved to be capable of displaying the expected phenomena, for instance, the magnetic reconnection and the dissipation of the energy. Furthermore, evolutions of MHD waves in solar physics problems were carried out to evaluate the efficiency of the numerical schemes in their description. 

We conclude that one of the bests qualities of our results is the successful implementation of the Flux Constrained Transport method, since we managed to preserve $\nabla\cdot\vec{B}$ in the order of the machine round-off error, $1\times10^{-12}$, as was displayed for the 2D and 3D tests. This is especially important in the cases where resistivity and heat flux were considered, since it shows that the numerical scheme was well adapted for them. We would like to point out that the numerical techniques need to be carefully adjusted to describe the physics laws, and this constraint should not be sacrificed in order to program a faster code. This is the reason why we preferred to build a unigrid code, where no unphysical behaviors could appear when AMR technics are implemented. Finally, it is worth mentioning that the code has been programmed in such a way  that new approximate Riemann solvers, reconstructors, divergence-free methods and time integrators, can be easily added.

\section*{Acknowledgments}

We want to thank professor K. TAKAHASHI for providing us with the exact solution of the Brio-Wu problem.  A. N wants to thanks the financial support from COLCIENCIAS, Colombia, under the program ``Becas Doctorados Nacionales 647'' and Universidad Industrial de Santander.  F.D.L-C gratefully acknowledges the financial support from VIE-UIS, grant number 1822.  G.A.G. was supported in part by VIE-UIS, under Grants No.  1347 and No.  1838, and by COLCIENCIAS, Colombia, under Grant No. 8840.


\begin{thebibliography}{76}
\expandafter\ifx\csname natexlab\endcsname\relax\def\natexlab#1{#1}\fi
\expandafter\ifx\csname href\endcsname\relax
  \def\href#1#2{}\fi
\expandafter\ifx\csname urllinklabel\endcsname\relax
  \def\urllinklabel{[LINK]}\fi
\expandafter\ifx\csname adsurllinklabel\endcsname\relax
  \def\adsurllinklabel{[ADS]}\fi

\bibitem[{{Antolin} {et~al.}(2015){Antolin}, {Okamoto}, {De Pontieu},
  {Uitenbroek}, {Van Doorsselaere}, \& {Yokoyama}}]{AntolinOkamoto2015}
{Antolin}, P., {Okamoto}, T.~J., {De Pontieu}, B., {Uitenbroek}, H., {Van
  Doorsselaere}, T., \& {Yokoyama}, T. 2015, ApJ, 809, 72


\bibitem[{{Arber} {et~al.}(2001){Arber}, {Longbottom}, {Gerrard}, \&
  {Milne}}]{MHDcodeLagrangianEulerian}
{Arber}, T.~D., {Longbottom}, A.~W., {Gerrard}, C.~L., \& {Milne}, A.~M. 2001,
  €ŽJ. Comput. Phys., 171, 151


\bibitem[{Balsara(2004)}]{CT_Balsara}
Balsara, D.~S. 2004, ApJS, 151, 149


\bibitem[{{Balsara} \& {Spicer}(1999)}]{Balsara1999}
{Balsara}, D.~S. \& {Spicer}, D.~S. 1999, J. Comput. Phys., 149, 270


\bibitem[{{Baumann} {et~al.}(2013){Baumann}, {Galsgaard}, \&
  {Nordlund}}]{BaumannGalsgaardNordlund2013}
{Baumann}, G., {Galsgaard}, K., \& {Nordlund}, {\AA}. 2013, Sol. Phys., 284,
  467


\bibitem[{{Brio} \& {Wu}(1988)}]{BrioWu}
{Brio}, M. \& {Wu}, C.~C. 1988, J. Comput. Phys., 75, 400


\bibitem[{{Bryan} {et~al.}(2014){Bryan}, {Norman}, {O'Shea}, {Abel}, {Wise},
  {Turk}, {Reynolds}, {Collins}, {Wang}, {Skillman}, {Smith}, {Harkness},
  {Bordner}, {Kim}, {Kuhlen}, {Xu}, {Goldbaum}, {Hummels}, {Kritsuk}, {Tasker},
  {Skory}, {Simpson}, {Hahn}, {Oishi}, {So}, {Zhao}, {Cen}, {Li}, \& {Enzo
  Collaboration}}]{enzo_code}
{Bryan}, G.~L., {Norman}, M.~L., {O'Shea}, B.~W., {Abel}, T., {Wise}, J.~H.,
  {Turk}, M.~J., {Reynolds}, D.~R., {Collins}, D.~C., {Wang}, P., {Skillman},
  S.~W., {Smith}, B., {Harkness}, R.~P., {Bordner}, J., {Kim}, J.-h., {Kuhlen},
  M., {Xu}, H., {Goldbaum}, N., {Hummels}, C., {Kritsuk}, A.~G., {Tasker}, E.,
  {Skory}, S., {Simpson}, C.~M., {Hahn}, O., {Oishi}, J.~S., {So}, G.~C.,
  {Zhao}, F., {Cen}, R., {Li}, Y., \& {Enzo Collaboration}. 2014, ApJS, 211, 19


\bibitem[{{Chmielewski} {et~al.}(2014){Chmielewski}, {Murawski}, {Musielak}, \&
  {Srivastava}}]{Chmielewski_Murawski_Musielak_Srivastava2014}
{Chmielewski}, P., {Murawski}, K., {Musielak}, Z.~E., \& {Srivastava}, A.~K.
  2014, ApJ, 793, 43


\bibitem[{Dagum \& Menon(1998)}]{openmp}
Dagum, L. \& Menon, R. 1998, IEEE Comput. Sci. Eng., 5, 46


\bibitem[{{Dai} \& {Woodward}(1998)}]{Woodward1998}
{Dai}, W. \& {Woodward}, P.~R. 1998, J. Comput. Phys., 142, 331


\bibitem[{{De Moortel} {et~al.}(2016){De Moortel}, {Pascoe}, {Wright}, \&
  {Hood}}]{DeMoortelPascoeWright2016}
{De Moortel}, I., {Pascoe}, D.~J., {Wright}, A.~N., \& {Hood}, A.~W. 2016,
  Plasma Phys. Contr. F., 58, 014001


\bibitem[{{De Pontieu} {et~al.}(2007){De Pontieu}, {McIntosh}, {Carlsson},
  {Hansteen}, {Tarbell}, {Schrijver}, {Title}, {Shine}, {Tsuneta}, {Katsukawa},
  {Ichimoto}, {Suematsu}, {Shimizu}, \& {Nagata}}]{PontieuEtAl2007}
{De Pontieu}, B., {McIntosh}, S.~W., {Carlsson}, M., {Hansteen}, V.~H.,
  {Tarbell}, T.~D., {Schrijver}, C.~J., {Title}, A.~M., {Shine}, R.~A.,
  {Tsuneta}, S., {Katsukawa}, Y., {Ichimoto}, K., {Suematsu}, Y., {Shimizu},
  T., \& {Nagata}, S. 2007, Science, 318, 1574


\bibitem[{{Del Zanna} {et~al.}(2005){Del Zanna}, {Schaekens}, \&
  {Velli}}]{transverse_waves}
{Del Zanna}, L., {Schaekens}, E., \& {Velli}, M. 2005, \aap, 431, 1095


\bibitem[{{Evans} \& {Hawley}(1988)}]{CT_evans}
{Evans}, C.~R. \& {Hawley}, J.~F. 1988, ApJ, 332, 659


\bibitem[{{Fryxell} {et~al.}(2000){Fryxell}, {Olson}, {Ricker}, {Timmes},
  {Zingale}, {Lamb}, {MacNeice}, {Rosner}, {Truran}, \& {Tufo}}]{flash_code}
{Fryxell}, B., {Olson}, K., {Ricker}, P., {Timmes}, F.~X., {Zingale}, M.,
  {Lamb}, D.~Q., {MacNeice}, P., {Rosner}, R., {Truran}, J.~W., \& {Tufo}, H.
  2000, ApJS, 131, 273


\bibitem[{{Fuchs} {et~al.}(2011){Fuchs}, {McMurry}, {Mishra}, \&
  {Waagan}}]{surya_code}
{Fuchs}, F.~G., {McMurry}, A.~D., {Mishra}, S., \& {Waagan}, K. 2011, ApJ, 732,
  75


\bibitem[{{Fujimura} \& {Tsuneta}(2009)}]{FujimuraTsuneta2009}
{Fujimura}, D. \& {Tsuneta}, S. 2009, ApJ, 702, 1443


\bibitem[{{Gonz{\'a}lez-Avil{\'e}s} {et~al.}(2015){Gonz{\'a}lez-Avil{\'e}s},
  {Cruz-Osorio}, {Lora-Clavijo}, \& {Guzm{\'a}n}}]{CafeNewtoniano}
{Gonz{\'a}lez-Avil{\'e}s}, J.~J., {Cruz-Osorio}, A., {Lora-Clavijo}, F.~D., \&
  {Guzm{\'a}n}, F.~S. 2015, MNRAS, 454, 1871


\bibitem[{{Gudiksen} {et~al.}(2011){Gudiksen}, {Carlsson}, {Hansteen}, {Hayek},
  {Leenaarts}, \& {Mart{\'{\i}}nez-Sykora}}]{bifrost_code}
{Gudiksen}, B.~V., {Carlsson}, M., {Hansteen}, V.~H., {Hayek}, W., {Leenaarts},
  J., \& {Mart{\'{\i}}nez-Sykora}, J. 2011, \aap, 531, A154


\bibitem[{{Hansteen} {et~al.}(2015){Hansteen}, {Guerreiro}, {De Pontieu}, \&
  {Carlsson}}]{HansteenGuerreiro2015}
{Hansteen}, V., {Guerreiro}, N., {De Pontieu}, B., \& {Carlsson}, M. 2015, ApJ,
  811, 106


\bibitem[{{Harten} {et~al.}(1983){Harten}, {Lax}, \& {van
  Leer}}]{Harten_etal_1983}
{Harten}, A., {Lax}, P.~D., \& {van Leer}, B. 1983, SIAM Rev., 25, 35


\bibitem[{{Hawley} \& {Stone}(1995)}]{current_sheet_chapter3_4}
{Hawley}, J.~F. \& {Stone}, J.~M. 1995, Comput. Phys. Commun., 89, 127


\bibitem[{{Janvier} {et~al.}(2015){Janvier}, {Aulanier}, \&
  {D{\'e}moulin}}]{JanvierAulanierDemoulin2015}
{Janvier}, M., {Aulanier}, G., \& {D{\'e}moulin}, P. 2015, Sol. Phys., 290,
  3425


\bibitem[{{Jel{\'{\i}}nek} \& {Murawski}(2013)}]{JelinekMurawski2013}
{Jel{\'{\i}}nek}, P. \& {Murawski}, K. 2013, MNRAS, 434, 2347


\bibitem[{{Jess} {et~al.}(2009){Jess}, {Mathioudakis}, Erd{\'e}lyi, {Crockett},
  {Keenan}, \& {Christian}}]{JessMathioudakisErdelyi2009}
{Jess}, D.~B., {Mathioudakis}, M., Erd{\'e}lyi, R., {Crockett}, P.~J.,
  {Keenan}, F.~P., \& {Christian}, D.~J. 2009, Science, 323, 1582


\bibitem[{Jess {et~al.}(2012)Jess, Pascoe, Christian, Mathioudakis, Keys, \&
  Keenan}]{JessPascoeChristian2012}
Jess, D.~B., Pascoe, D.~J., Christian, D.~J., Mathioudakis, M., Keys, P.~H., \&
  Keenan, F.~P. 2012, Astrophys. J., 744, L5


\bibitem[{{Jess} {et~al.}(2012){Jess}, {Shelyag}, {Mathioudakis}, {Keys},
  {Christian}, \& {Keenan}}]{JessShelyagMathioudakis2012}
{Jess}, D.~B., {Shelyag}, S., {Mathioudakis}, M., {Keys}, P.~H., {Christian},
  D.~J., \& {Keenan}, F.~P. 2012, Apj, 746, 183


\bibitem[{Jiang {et~al.}(2012)Jiang, Fang, \& Chen}]{ResistiveCode_Jiang}
Jiang, R.-L., Fang, C., \& Chen, P.-F. 2012, Comput. Phys. Commun., 183, 1617


\bibitem[{{Jiang} {et~al.}(2012){Jiang}, {Fang}, \& {Chen}}]{JiangFangChen2012}
{Jiang}, R.-L., {Fang}, C., \& {Chen}, P.-F. 2012, ApJ, 751, 152


\bibitem[{{Khomenko} \& {Collados}(2015)}]{Khomenko2015}
{Khomenko}, E. \& {Collados}, M. 2015, Living Rev. Sol. Phys., 12, 6


\bibitem[{{Konkol} {et~al.}(2010){Konkol}, {Murawski}, {Lee}, \&
  {Weide}}]{KonkolMurawskiLeeWeide2010}
{Konkol}, P., {Murawski}, K., {Lee}, D., \& {Weide}, K. 2010, \aap, 521, A34


\bibitem[{{Li}(2005)}]{hllc_mhd_Li}
{Li}, S. 2005, ‎J. Comput. Phys., 203, 344


\bibitem[{{Lora-Clavijo} {et~al.}(2015){Lora-Clavijo}, {Cruz-Osorio}, \&
  {Guzm{\'a}n}}]{CafeRelativista}
{Lora-Clavijo}, F.~D., {Cruz-Osorio}, A., \& {Guzm{\'a}n}, F.~S. 2015, ApJS,
  218, 24


\bibitem[{{Lyon} {et~al.}(2004){Lyon}, {Fedder}, \& {Mobarry}}]{LFM_code}
{Lyon}, J.~G., {Fedder}, J.~A., \& {Mobarry}, C.~M. 2004, JASTP, 66, 1333


\bibitem[{{Mignone} {et~al.}(2007){Mignone}, {Bodo}, {Massaglia}, {Matsakos},
  {Tesileanu}, {Zanni}, \& {Ferrari}}]{pluto_code}
{Mignone}, A., {Bodo}, G., {Massaglia}, S., {Matsakos}, T., {Tesileanu}, O.,
  {Zanni}, C., \& {Ferrari}, A. 2007, ApJS, 170, 228


\bibitem[{{Morton} {et~al.}(2012){Morton}, {Verth}, {McLaughlin}, \&
  Erd{\'e}lyi}]{MortonVerthMcLaughlinErdelyi2012}
{Morton}, R.~J., {Verth}, G., {McLaughlin}, J.~A., \& Erd{\'e}lyi, R. 2012,
  ApJ, 744, 5


\bibitem[{{M{\"o}stl} {et~al.}(2013){M{\"o}stl}, {Temmer}, \&
  {Veronig}}]{MostlTemmerVeronig2013}
{M{\"o}stl}, U.~V., {Temmer}, M., \& {Veronig}, A.~M. 2013, ApJ, 766, L12


\bibitem[{{Murawski} {et~al.}(2013){Murawski}, {Ballai}, {Srivastava}, \&
  {Lee}}]{gravity_waves}
{Murawski}, K., {Ballai}, I., {Srivastava}, A.~K., \& {Lee}, D. 2013, MNRAS,
  436, 1268


\bibitem[{{Murawski} {et~al.}(2016){Murawski}, {Chmielewski}, {Zaqarashvili},
  \& {Khomenko}}]{MurawskiChmielewski2016}
{Murawski}, K., {Chmielewski}, P., {Zaqarashvili}, T.~V., \& {Khomenko}, E.
  2016, MNRAS, 459, 2566


\bibitem[{{Murawski} \& {Musielak}(2010)}]{Murawski_Musielak2010}
{Murawski}, K. \& {Musielak}, Z.~E. 2010, A\&A, 518, A37


\bibitem[{{Murawski} {et~al.}(2011){Murawski}, {Srivastava}, \&
  {Zaqarashvili}}]{Murawski_Srivastava_Zaqarashvili2011}
{Murawski}, K., {Srivastava}, A.~K., \& {Zaqarashvili}, T.~V. 2011, A\&A, 535,
  A58


\bibitem[{{Nakariakov} {et~al.}(1999){Nakariakov}, {Ofman}, {Deluca},
  {Roberts}, \& {Davila}}]{NakariakovOfmanDelucaRobertsDavila1999}
{Nakariakov}, V.~M., {Ofman}, L., {Deluca}, E.~E., {Roberts}, B., \& {Davila},
  J.~M. 1999, Science, 285, 862


\bibitem[{{Nakariakov} \& {Verwichte}(2005)}]{NakariakovVerwichte2005}
{Nakariakov}, V.~M. \& {Verwichte}, E. 2005, Living Rev. Sol. Phys., 2, 3


\bibitem[{Norman(2000)}]{zeus_code}
Norman, M.~L. 2000, Rev. Mex. Astron. Astrof. Ser. Conf., 9, 66


\bibitem[{{Ofman}(2010)}]{Ofman2010}
{Ofman}, L. 2010, Living Rev. Sol. Phys., 7, 4


\bibitem[{{Okamoto} {et~al.}(2007){Okamoto}, {Tsuneta}, {Berger}, {Ichimoto},
  {Katsukawa}, {Lites}, {Nagata}, {Shibata}, {Shimizu}, {Shine}, {Suematsu},
  {Tarbell}, \& {Title}}]{Okamoto2007}
{Okamoto}, T.~J., {Tsuneta}, S., {Berger}, T.~E., {Ichimoto}, K., {Katsukawa},
  Y., {Lites}, B.~W., {Nagata}, S., {Shibata}, K., {Shimizu}, T., {Shine},
  R.~A., {Suematsu}, Y., {Tarbell}, T.~D., \& {Title}, A.~M. 2007, Science,
  318, 1577


\bibitem[{{Pascoe} {et~al.}(2013){Pascoe}, {Nakariakov}, \&
  {Kupriyanova}}]{PascoeNakariakov2013}
{Pascoe}, D.~J., {Nakariakov}, V.~M., \& {Kupriyanova}, E.~G. 2013, A\&A, 560,
  A97


\bibitem[{{Petschek}(1964)}]{Petschek1964}
{Petschek}, H.~E. 1964, NASA Spec. Publ., 50, 425


\bibitem[{Powell(1994)}]{Powell}
Powell, K.~G. 1994, An approximate Riemann solver for magnetohydrodynamics(That
  works in more than one dimension), Tech. rep.


\bibitem[{{Priest}(2014)}]{Priest2014}
{Priest}, E. 2014, {Magnetohydrodynamics of the Sun}


\bibitem[{{Priest} \& {Hood}(1991)}]{Priest1991}
{Priest}, E.~R. \& {Hood}, A.~W., eds. 1991, {Advances in Solar System
  Magnetohydrodynamics, 1991}


\bibitem[{{Roe}(1986)}]{Roe1986}
{Roe}, P.~L. 1986, ‎Annu. Rev. Fluid Mech., 18, 337


\bibitem[{{Shelyag} {et~al.}(2013){Shelyag}, {Cally}, {Reid}, \&
  {Mathioudakis}}]{Shelyag2013}
{Shelyag}, S., {Cally}, P.~S., {Reid}, A., \& {Mathioudakis}, M. 2013, ApJL,
  776, L4


\bibitem[{{Shelyag} {et~al.}(2008){Shelyag}, {Fedun}, \&
  {Erd{\'e}lyi}}]{vac_code}
{Shelyag}, S., {Fedun}, V., \& {Erd{\'e}lyi}, R. 2008, \aap, 486, 655


\bibitem[{{Shelyag} {et~al.}(2011){Shelyag}, {Keys}, {Mathioudakis}, \&
  {Keenan}}]{Shelyag2011}
{Shelyag}, S., {Keys}, P., {Mathioudakis}, M., \& {Keenan}, F.~P. 2011, A\&A,
  526, A5


\bibitem[{{Shibata} \& {Magara}(2011)}]{ShibataMagara2011}
{Shibata}, K. \& {Magara}, T. 2011, Living Rev. Sol. Phys., 8, 6


\bibitem[{Spitzer(2006)}]{Spitzer2006}
Spitzer, L. 2006, Physics of Fully Ionized Gases, Dover Books on Physics (Dover
  Publications)


\bibitem[{{Stone} {et~al.}(2008){Stone}, {Gardiner}, {Teuben}, {Hawley}, \&
  {Simon}}]{athena_code}
{Stone}, J.~M., {Gardiner}, T.~A., {Teuben}, P., {Hawley}, J.~F., \& {Simon},
  J.~B. 2008, ApJS, 178, 137


\bibitem[{Takahashi \& Yamada(2013)}]{takahashi2013}
Takahashi, K. \& Yamada, S. 2013, J. Plasma Phys., 79, 335


\bibitem[{{Terradas} {et~al.}(2015){Terradas}, {Soler}, {Luna}, {Oliver}, \&
  {Ballester}}]{TerradasSolerLunaOliver2015}
{Terradas}, J., {Soler}, R., {Luna}, M., {Oliver}, R., \& {Ballester}, J.~L.
  2015, ApJ, 799, 94


\bibitem[{Titarev \& Toro(2004)}]{TitarevToro2004}
Titarev, V.~A. \& Toro, E.~F. 2004, J. Comput. Phys., 201, 238


\bibitem[{{Titarev} \& {Toro}(2005)}]{2005JCoPh.204..715T}
{Titarev}, V.~A. \& {Toro}, E.~F. 2005, Journal of Computational Physics, 204,
  715


\bibitem[{{Tomczyk} {et~al.}(2007){Tomczyk}, {McIntosh}, {Keil}, {Judge},
  {Schad}, {Seeley}, \&
  {Edmondson}}]{TomczykMcIntoshKeilJudgeSchadSeeleyEdmondson2007}
{Tomczyk}, S., {McIntosh}, S.~W., {Keil}, S.~L., {Judge}, P.~G., {Schad}, T.,
  {Seeley}, D.~H., \& {Edmondson}, J. 2007, Science, 317, 1192


\bibitem[{Toro(2009)}]{Toro}
Toro, E.~F. 2009, Riemann Solvers and Numerical Methods for Fluid Dynamics
  (Springer)


\bibitem[{{Toro} {et~al.}(1994){Toro}, {Spruce}, \& {Speares}}]{Toro_hllc}
{Toro}, E.~F., {Spruce}, M., \& {Speares}, W. 1994, Shock Waves, 4, 25


\bibitem[{{T{\'o}th}(2000)}]{CT_toth}
{T{\'o}th}, G. 2000, ‎J. Comput. Phys., 161, 605


\bibitem[{{Van Doorsselaere} {et~al.}(2008){Van Doorsselaere}, {Nakariakov}, \&
  {Verwichte}}]{DoorsselaereNakariakovVerwichte2008}
{Van Doorsselaere}, T., {Nakariakov}, V.~M., \& {Verwichte}, E. 2008, ApJ, 676,
  L73


\bibitem[{{van Leer}(1977)}]{mc}
{van Leer}, B. 1977, J. Comput. Phys., 23, 263


\bibitem[{{Vernazza} {et~al.}(1981){Vernazza}, {Avrett}, \&
  {Loeser}}]{Vernazza_temperature_profile}
{Vernazza}, J.~E., {Avrett}, E.~H., \& {Loeser}, R. 1981, ApJS, 45, 635


\bibitem[{{Vigeesh} {et~al.}(2012){Vigeesh}, {Fedun}, {Hasan}, \&
  {Erd{\'e}lyi}}]{VigeeshFedun2012}
{Vigeesh}, G., {Fedun}, V., {Hasan}, S.~S., \& {Erd{\'e}lyi}, R. 2012, ApJ,
  755, 18


\bibitem[{{V{\"o}gler} {et~al.}(2005){V{\"o}gler}, {Shelyag}, {Sch{\"u}ssler},
  {Cattaneo}, {Emonet}, \& {Linde}}]{muram_code}
{V{\"o}gler}, A., {Shelyag}, S., {Sch{\"u}ssler}, M., {Cattaneo}, F., {Emonet},
  T., \& {Linde}, T. 2005, A\&A, 429, 335


\bibitem[{Vr{\v s}nak {et~al.}(2013)Vr{\v s}nak, {\v Z}ic, Vrbanec, Temmer,
  Rollett, M{\"o}stl, Veronig, {\v C}alogovi{\'c}, Dumbovi{\'c}, Luli{\'c},
  Moon, \& {Shanmugaraju}}]{Vrsnac2013}
Vr{\v s}nak, B., {\v Z}ic, T., Vrbanec, D., Temmer, M., Rollett, T., M{\"o}stl,
  C., Veronig, A., {\v C}alogovi{\'c}, J., Dumbovi{\'c}, M., Luli{\'c}, S.,
  Moon, Y.-J., \& {Shanmugaraju}, A. 2013, Sol. Phys., 285, 295


\bibitem[{{Wedemeyer} {et~al.}(2013){Wedemeyer}, {Ludwig}, \&
  {Steiner}}]{Wedemeyer2013}
{Wedemeyer}, S., {Ludwig}, H.-G., \& {Steiner}, O. 2013, Astron. Nachr., 334,
  137


\bibitem[{{Wedemeyer-B{\"o}hm} {et~al.}(2012){Wedemeyer-B{\"o}hm}, {Scullion},
  {Steiner}, {Rouppe van der Voort}, {de La Cruz Rodriguez}, {Fedun}, \&
  {Erd{\'e}lyi}}]{Wedemeyer2012}
{Wedemeyer-B{\"o}hm}, S., {Scullion}, E., {Steiner}, O., {Rouppe van der
  Voort}, L., {de La Cruz Rodriguez}, J., {Fedun}, V., \& {Erd{\'e}lyi}, R.
  2012, Nature, 486, 505


\bibitem[{{Yang} {et~al.}(2015){Yang}, {Zhang}, {He}, {Peter}, {Tu}, {Wang},
  {Zhang}, \& {Feng}}]{YangZhangHePeter2015}
{Yang}, L., {Zhang}, L., {He}, J., {Peter}, H., {Tu}, C., {Wang}, L., {Zhang},
  S., \& {Feng}, X. 2015, ApJ, 800, 111


\bibitem[{{Ziegler}(2008)}]{nirvana_code}
{Ziegler}, U. 2008, Comput. Phys. Commun., 179, 227


\end{thebibliography}

\end{document}